\documentclass[prx,amsmath,amssymb,twocolumn,superscriptaddress]{revtex4-2}

\usepackage{graphicx}                           
\usepackage{amsmath}                            
\usepackage{amssymb}                            
\usepackage{tabularx}                           
\usepackage{multirow}
\usepackage{hyperref}
\usepackage{array}                              
\usepackage{tikz}                               
\usepackage{mathrsfs}                           
\usepackage{mwe}                                
\usepackage{csquotes}
\usepackage[normalem]{ulem} 

\usepackage{siunitx}
\newcommand{\overbar}[1]{\mkern 1.5mu\overline{\mkern-1.5mu#1\mkern-1.5mu}\mkern 1.5mu}

\begin{document}
\title{Coherent Phonons and Quasiparticle Renormalization in Semimetals from First Principles}
\author{Christoph Emeis}
\affiliation{Institute of Theoretical Physics and Astrophysics, Kiel University, 24118 Kiel, Germany}
\author{Stephan Jauernik}
\affiliation{Institute of Experimental and Applied Physics, Kiel University, 24118 Kiel, Germany}
\author{Sunil Dahiya}
\affiliation{Institute of Experimental and Applied Physics, Kiel University, 24118 Kiel, Germany}
\author{Yiming Pan}
\affiliation{Institute of Theoretical Physics and Astrophysics, Kiel University, 24118 Kiel, Germany}
\author{Carl E. Jensen}
\affiliation{Institute of Experimental and Applied Physics, Kiel University, 24118 Kiel, Germany}
\author{Petra Hein}
\affiliation{Institute of Experimental and Applied Physics, Kiel University, 24118 Kiel, Germany}
\author{Michael Bauer}
\affiliation{Institute of Experimental and Applied Physics, Kiel University, 24118 Kiel, Germany}
\affiliation{
Kiel Nano, Surface and Interface Science KiNSIS, Kiel University, 24118 Kiel, Germany
}
\author{Fabio Caruso}
\affiliation{Institute of Theoretical Physics and Astrophysics, Kiel University, 24118 Kiel, Germany}
\affiliation{
Kiel Nano, Surface and Interface Science KiNSIS, Kiel University, 24118 Kiel, Germany
}

\begin{abstract}
Coherent phonons, light-induced coherent lattice vibrations in solids, provide
a powerful route to engineer structural and electronic degrees of freedom using
light.  In this manuscript, we formulate an ab initio theory of the displacive
excitation of coherent phonons (DECP), the primary mechanism for light-induced
structural control in semimetals. Our study -- based on the ab initio
simulations of the ultrafast electron and coherent-phonon dynamics in presence
of electron-phonon interactions -- establishes a predictive computational
framework for describing the emergence of light-induced structural changes and
the ensuing transient band-structure renormalization arising from the DECP
mechanism.  We validate this framework via a combined theoretical and
experimental investigation of coherent phonons in the elemental semimetal
antimony. Via a Fourier analysis of time- and angle-resolved photoemission
spectroscopy (tr-ARPES) measurements, we  retrieve information about transient
spectral features and quasiparticle renormalization arising from the coherent
$A_{1g}$ phonon as a function of momentum, energy, time, and fluence.  The
qualitative and quantitative agreement between experiment and theory
corroborates the first-principles approach formulated in this study.  Besides
advancing the fundamental understanding of electron-phonon interactions
mediated by coherent phonons, this study opens new opportunities for
predictively engineering structural and electronic degrees of freedom in
semimetals via the DECP mechanism.
\end{abstract}
 \maketitle
\section{Introduction}

Coherent phonons refer to the damped oscillations experienced by a crystalline
lattice following a time-dependent perturbation, and they are a direct
manifestation of light-induced structural changes in photo-excited solids. They
were first observed in the early 1990s in pump-probe transient-reflectivity
measurements of elemental semimetals
\cite{thomsen1984coherent,cheng1990impulsive} and since then they have been
detected and characterized in various material families using a range of
spectroscopy techniques
\cite{Cheng1991,kuznetsov1995coherent,dekorsy1995emission,hase2005ultrafast,ishioka2008coherent,chatelain2014coherent,waldecker2017coherent,gerber2017femtosecond}.
In addition to time-resolved optical and scattering probes, coherent phonons
have also been observed in tr-ARPES studies in a wide range of different
materials \cite{perfetti2006time, Petersen2011, Hellmann2012,
papalazarou2012coherent, sobota2014distinguishing, rettig2014coherent,
Golias2016, Tang2020, Baldini2023}. This technique stands out for its ability
to study the coupling of coherent phonons to the electronic degrees of freedom
in an exceptionally direct, momentum- and energy-resolved manner.  The recent
development of frequency-domain angle-resolved photoemission spectroscopy
(FD-ARPES) allows for {a mode-resolved} tracking of band-structure
renormalization effects caused by coherent phonons
\cite{hein2020mode,de2020direct,suzuki2021detecting,lee2023layer,ren2023phase}.
Via a Fourier analysis, the contributions of individual modes to changes in the
band structure can be disentangled, providing unprecedented information on
electron-phonon interactions mediated by coherent phonons with energy, momentum, 
and phonon frequency resolution.

\begin{figure}[]
    \centering
    \includegraphics[width=1\linewidth]{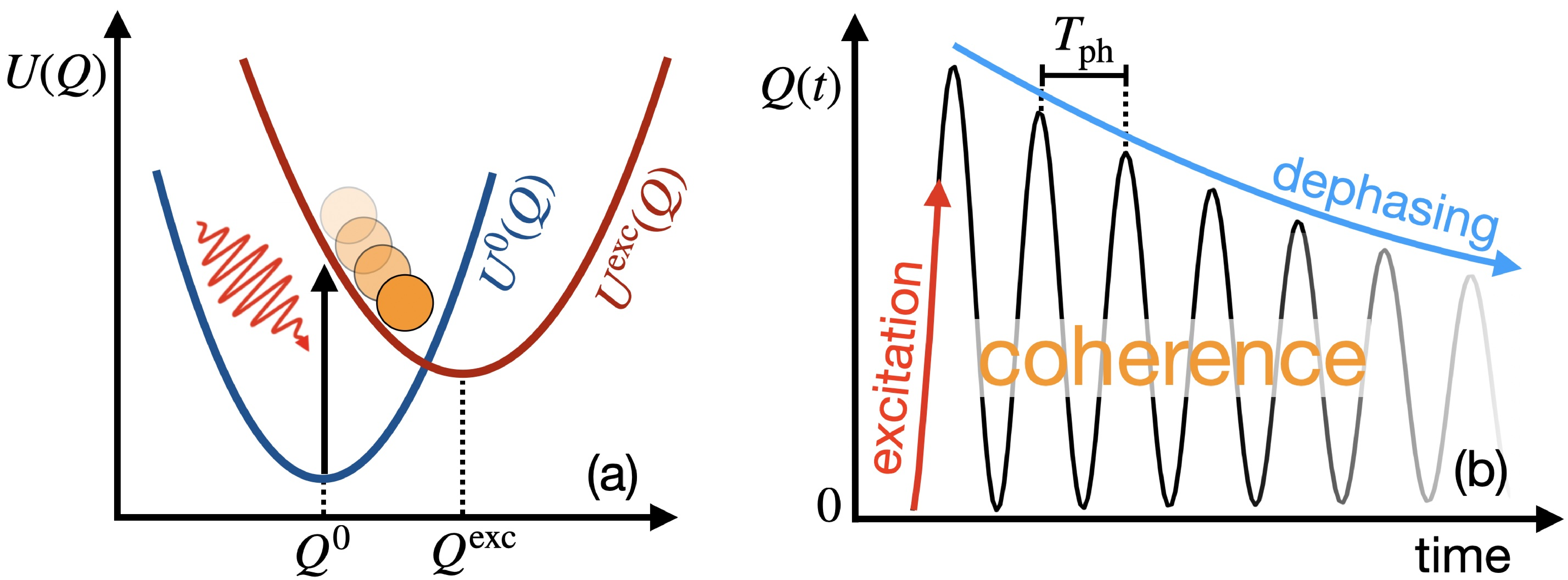}
    \caption{(a) Schematic representation of the DECP mechanism. $U^0$ ($U^{\rm
exc}$) denotes the ground-state (excited-state) potential-energy surface before
(after)  photoexcitation. $Q$ is an adimensional collective displacement
amplitude of the lattice along a fully-symmetric Raman-active phonon. (b)
Characteristic time-dependence of the displacement amplitude $Q$ for the DECP
mechanism. Following excitation, the coherent phonon dynamics is characterized
by damped oscillations with periodicity set by the phonon period $T_{\rm ph} =
2\pi \omega_{\rm ph}^{-1} $.}
    \label{fig0}
\end{figure}

The fundamental interest\sout{s} in coherent phonons is motivated by the wide
possibilities they offer to directly tailor specific materials properties -- as
e.g.  magnetism \cite{wall2009ultrafast,kim2012ultrafast,Nova2017,afanasiev_ultrafast_2021}, ferroelectricity
\cite{bakker1998coherent,Juraschek2017,mankowsky2017ultrafast,fechner2024quenched}, topology
\cite{papalazarou2012coherent,sobota2014distinguishing,sie2019ultrafast}, 
and superconductivity\cite{gerber2017femtosecond,Mankowsky2014,LiuFoerst2020,Mitrano2016}.  Additionally, they can
serve as a precursor to drive light-induced phase transitions
\cite{rettig2014coherent,Juraschek2017b,de2021colloquium}, or to access hidden-metastable
states inaccessible under equilibrium conditions \cite{mocatti2023light,yang2021experimental}.
Coherent phonons can be launched via a variety of excitation pathways,
including inelastic stimulated Raman scattering, ionic Raman scattering,
infrared absorption, and the displacive excitation of coherent phonons
(DECP)\cite{zeiger1992theory,kuznetsov1994theory,dhar1994,Giret2011,Juraschek2018c,lakehal2019microscopic,caruso2023quantum}.
Among these, the DECP mechanism constitutes the primary excitation pathway for
semimetals and it can play an important role in photo-excited semiconductors.

In the DECP mechanism, the coherent motion of the lattice is triggered by
modifications of the electron-density distribution due to photoabsorption
\cite{kuznetsov1994theory}.  Carrier photoexcitation results into a
quasi-adiabatic displacement of the potential energy surface, as depicted in
Fig.~\ref{fig0}~(a), triggering damped lattice oscillations around the new
minimum with the characteristic oscillatory pattern illustrated in
Fig.~\ref{fig0}~(b).  Phenomenological models can be designed to reproduce
targeted experimental fingerprints arising from these phenomena
\cite{Garrett1996}.  Yet, they are inherently dependent on experimental
observation, and they are unsuitable to capture the coherent phonon spectrum in
structurally complex compounds. However, predictive and transferable atomistic
models for the DECP mechanism in semimetals have not yet been presented. The
challenge in developing a first-principles theory of DECP is twofold: First, it
is necessary to describe the coherent phonon excitation induced by dynamical
changes of the electron density.  Second, one must consider the resulting band
structure renormalization, which arises from the electron-phonon interactions
mediated by the driven coherent phonons within a perturbative many-body
framework.

In this manuscript, we formulate an ab initio theory for the DECP mechanism in
semimetals. Our approach specifically targets (i) the emergence of
photo-induced coherent lattice motion following electronic excitation by a pump
pulse and (ii) transient band-structure renormalization effects induced by
coherent phonons.  The former is obtained via the joint solution of the
coherent-phonon equation of motion (EOM) alongside with the time-dependent
Boltzmann equation
(TDBE)\cite{caruso2021nonequilibrium,Tong2021,caruso2022ultrafast,Pan2023} --
the state of the art for ab initio simulations of the joint ultrafast dynamics
of electrons and phonons. Band-structure renormalization effects are accounted
for  within an adiabatic many-body formulation of quasiparticle corrections due
to coherent phonons.  We assess the predictive power of this approach against
tr-ARPES measurements for antimony, whereby we characterize the emergence of
coherent phonons and their impact on the electron band structure as a function
of energy, momentum, time, and pump fluence.  The predictive power of this
theoretical framework, demonstrated by the robust agreement between theory and
experiments, opens up new opportunities to predictively engineer structural and
electronic degrees of freedom in non-equilibrium conditions. 

The manuscript is structured as follows.  In Sec.~\ref{sec:theory}, we
introduce the theoretical framework for the first-principles description of
coherent phonons. Details of the tr-ARPES measurements and the FD-ARPES
technique are reported in Sec.~\ref{sec:exp}.  In Sec.~\ref{sec:eq}, the
equilibrium electronic and vibrational properties of antimony and their
fingerprints in tr-ARPES measurements are reviewed.  Measurements and ab initio
calculations of band-structure renormalization induced by coherent phonons in
antimony are presented in Sec.~\ref{sec:QP}, and their fluence dependence is
analyzed in Sec.~\ref{sec:fluence}.  Discussions and an outlook are presented
in Sec.~\ref{sec:out}. Our concluding remarks are reported in
Sec.~\ref{sec:conc}.

\section{Theory}\label{sec:theory}
In the following, we proceed to introduce a theoretical framework suitable to
capture the emergence of coherent phonons in semimetals based on
first-principles electronic-structure calculations.

A first-principles description of coherent phonons requires tracking the time
dependence of the displacement $\Delta {\tau}_{\kappa p}(t)$ of the $\kappa$-th
nucleus in the $p$-th unit cell from its equilibrium coordinate in presence of
an external time-dependent perturbation (as, e.g., a driving field).  $\Delta
{\tau}_{\kappa p}$ can be quantized by introducing a normal mode basis, leading
to \cite{giustino2017electron}:
\begin{align}\label{eq:Dtau}
    \Delta \hat{\tau}_{\kappa p} = \sum_{{\bf q}\nu} \left(\frac{\hbar}{ 2
\omega_{\mathbf{q} \nu} M_\kappa}\right)^{\frac{1}{2}} e^{i {\bf q \cdot R}_p}
\, {\bf e}^{\kappa}_{{\bf q}\nu} \hat{Q}_{{\bf q}\nu} \quad.  
\end{align}
Here, $\hat{Q}_{{\bf q}\nu} = N_p ^{-\frac{1}{2}}(\hat{a}_{\mathbf{q} \nu} +
\hat{a}^\dagger_{-\mathbf{q} \nu})$ is an adimensional coherent-phonon
amplitude that quantifies the lattice distortion along the ${\bf q}\nu$ normal
mode.  $\hat{a}_{\mathbf{q} \nu}$ and  $\hat{a}^\dagger_{\mathbf{q} \nu}$
denote annihilation and creation bosonic operators, respectively.  $N_p$ is the
number of ${\bf q}$ points, $M_\kappa$  the nuclear mass, $\omega_{\mathbf{q}
\nu}$ the phonon frequency for crystal momentum ${\bf q}$ and mode index $\nu$.
${\bf e}^{\kappa}_{{\bf q}\nu}$ denotes phonon eigenvectors, and ${\bf R}_p$
crystal-lattice vectors.

For the DECP mechanism, this problem can be addressed via the solution of the
Heisenberg EOM for $\hat{Q}_{\mathbf{q} \nu}$, namely $\partial_t^2
\hat{Q}_{\mathbf{q} \nu}(t) = - \hbar^{-2}\big[ [ \hat{Q}_{\mathbf{q} \nu},
\hat{H} ], \hat{H} \big]$ where $\hat{H} = \hat{H}_{\mathrm{ph}} +
\hat{H}_{\mathrm{eph}}$ is the total lattice Hamiltonian, consisting of the
non-interacting phonon Hamiltonian $ \displaystyle \hat{H}_{\mathrm{ph}} = \sum
\hbar \omega_{{\bf q} \nu} ( \hat a^\dagger_{{\bf q} \nu} \hat a_{{\bf q} \nu}
+ 1/2 )$ and the electron-phonon interaction $ \displaystyle
\hat{H}_{\mathrm{eph}} = \sum g^\nu_{mn}({\bf k,q}) [\hat c^\dagger_{m{\bf
k+q}}\hat c_{n{\bf k}} - \delta_{nm}\delta_{{\bf q}0}] \hat{Q}_{{\bf q}\nu}$.
The second term in brackets, recently introduced in
Ref.~\cite{stefanucci2023and}, is critical to correctly capture the coherent
phonon excitation.  Here, $g^\nu_{mn}({\bf k,q})$ is the electron-phonon
coupling matrix element, $\hat c_{n{\bf k}}$ and $\hat c^\dagger_{n{\bf k}}$
are fermionic annihilation and creation operators, respectively, and the sums
extend over all the indices of the arguments.

Making use of standard commutation relations, one promptly arrives at the EOM
for the coherent-phonon amplitude \cite{caruso2023quantum}: 
\begin{align}
    \partial_t^2 Q_{\nu} 
    + \omega^2_{ \nu} Q_{ \nu} &= 
    - {\omega_{\nu}}{\hbar^{-1}} \sum_{n \mathbf{k}} g_{nn}^\nu (\mathbf{k},0)  \Delta f_{n\mathbf{k}}(t) \quad,
    \label{eq:osci}
\end{align}
 $\Delta f_{n{\bf k}}= f_{n{\bf k}} (t)  - f_{n{\bf k}}^{(0)}$,  and $f_{n{\bf
k}}^{(0)}$ is the (equilibrium) Fermi-Dirac distribution before excitation,
whereas $f_{n{\bf k}} (t)$ denotes the non-equilibrium distribution function at
time $t$. The dependence on ${\bf q}$ is omitted, as selection rules require
${\bf q}=0$.  This expression has been derived by neglecting electron coherence
($\langle \hat c^\dagger_{m{\bf k+q}}\hat c_{n{\bf k}} \rangle \simeq f_{n{\bf
k}}$), which is justified for timescales of the order of 50~fs and longer.
{The inclusion of non-diagonal contributions to the density matrix and their
influence on the coherent phonon dynamics can be handled within the framework
of non-equilibrium Green function, as recently demonstrated in the study of
exciton-phonon interactions in monolayer MoS$_2$ \cite{perfetto_theory_2024}. }

Numerical evaluation of Eq.~\eqref{eq:osci} requires knowledge of the
time-dependent occupation function $f_{n{\bf k}} (t)$ whose time-dependence is
governed by the time-dependent Boltzmann equation (TDBE)
\cite{caruso2022ultrafast,caruso2021nonequilibrium}: 
\begin{align}
    \partial_t f_{n \mathbf{k}}   &= \Gamma^{\rm e-ph}\left[f_{n\mathbf{k}},
n_{\mathbf{q} \nu} \right] + \Gamma^{\rm e{-}e} \left[f_{n\mathbf{k}}\right]
\quad, \label{eq:TDBE1} \\ \partial_t n_{\mathbf{q} \nu} &= \Gamma^{\rm
ph-e}\left[f_{n\mathbf{k}}, n_{\mathbf{q} \nu} \right] \quad .
    \label{eq:TDBE2}
\end{align}
Here, $n_{{\bf q}\nu}$ denotes the phonon distribution function, and
$\Gamma^{\rm e-ph}$, $ \Gamma^{\rm e{-}e}$, and  $\Gamma^{\rm ph-e}$ are the
collision integrals due to electron-phonon, electron-electron, and
phonon-electron scattering, respectively, for which explicit expressions can be
found elsewhere \cite{caruso2022ultrafast,rethfeld2002ultrafast}.

Equations~\eqref{eq:osci}-\eqref{eq:TDBE2} form a set of coupled
integro-differential equations, which can be solved by conventional
time-propagation algorithms (e.g., Heun's method or fourth-order Runge-Kutta)
to determine the coherent-phonon amplitude $Q_{\nu}$ alongside with the
electron ($f_{n\mathbf{k}}$) and phonon distribution functions  ($n_{\mathbf{q}
\nu}$).  In this work, we solve Eqs.~\eqref{eq:osci}-\eqref{eq:TDBE2} within
the {\tt EPW} code \cite{lee2023electron} as an initial value problem to
determine the coupled dynamics of electrons and coherent phonons in semimetals
following excitation with a pump pulse.  As initial conditions, we consider
electronic excited states representative of the conditions established in a
pump-probe experiment, and described by a high-temperature Fermi-Dirac
distribution $f_{n{\bf k}}(t=0) = [{\rm exp}(\varepsilon_{n{\bf k}}-\mu)/k_{\rm
B}T_{\rm exc} +1]^{-1}$, where the excitation temperature  $T_{\rm exc}$ is set
to match the estimated incident pump fluence in the tr-ARPES measurements (see
discussion in Appendix~\ref{sec:init}).  All quantities required in the
evaluation of Eqs.~\eqref{eq:osci}-\eqref{eq:TDBE2} are obtained from first
principles via density-functional theory (DFT)
\cite{giannozzi2009quantum,giannozzi2017advanced} and density-function
perturbation theory (DFPT) \cite{baroni2001phonons}.  In particular, we first
obtain electron energies $\varepsilon_{n{\bf k}}$ from DFT, phonon frequencies
$\omega_{\bf q\nu}$ and electron-phonon coupling matrix elements $g_{nm}^\nu
({\bf k, q})$ from DFPT. All quantities are interpolated on dense momentum
grids using maximally-localized Wannier function
\cite{marzari2012maximally,Giustino2007} within the {\tt EPW} and {\tt
Wannier90} codes \cite{pizzi2020wannier90} .  We thus proceed to solve
Eqs.~\eqref{eq:osci}-\eqref{eq:TDBE2} by time-stepping the time derivative via
the second-order Runge-Kutta algorithm (Heun method) for a total duration of
3~ps with time steps of 1~fs.  The last step of this procedure involves the
calculations of quasiparticle corrections due to coherent phonons, which is
detailed in Sec.~\ref{sec:QP}.  A schematic illustration of this workflow is
reported in Fig.~\ref{fig_work}, whereas a detailed description of
computational parameters is provided in Appendix~\ref{sec:comp}. 

\begin{figure}[]
    \centering
    \includegraphics[width=1\linewidth]{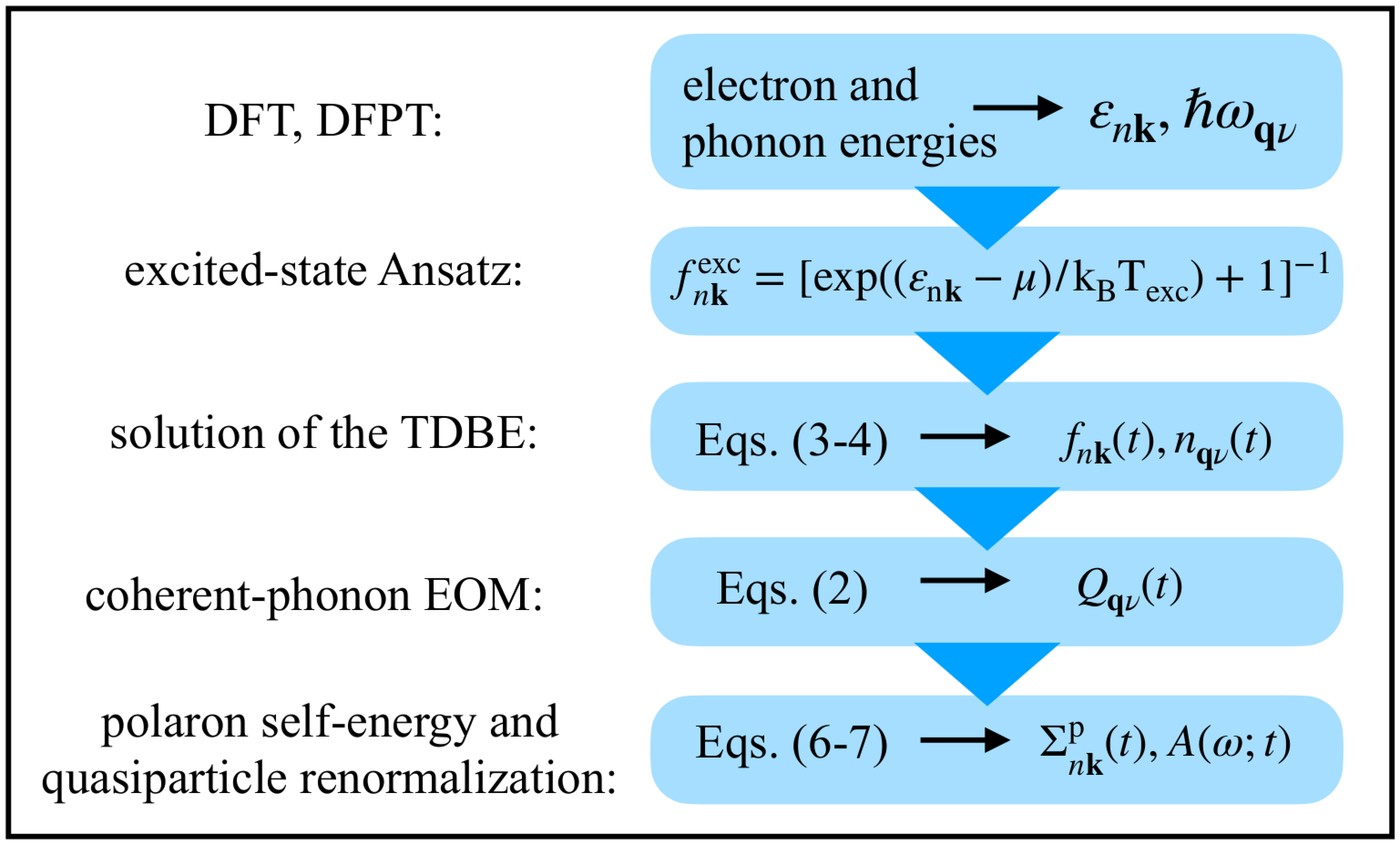}
    \caption{Workflow for the ab initio investigations of coherent phonons in semimetals.}
    \label{fig_work}
\end{figure}

Equation~\eqref{eq:osci} is reminiscent of the EOM for a driven harmonic
oscillator, with a driving term $D_{\rm eph}(t) = - {\omega_{\nu}}{\hbar^{-1}}
\sum_{n \mathbf{k}} g_{nn}^\nu (\mathbf{k},0)  \Delta f_{n\mathbf{k}}(t) $. A
necessary condition for the excitation of coherent phonons is $D_{\rm
eph}(t)\neq 0$.  Equation~\eqref{eq:osci} thus reveals that coherent excitation
of the $\nu$-th zone-center phonon can only occur if $g_{nn}^\nu (\mathbf{k},0)
= \langle u_{n{\bf k}} | \Delta_{\nu} \hat v_{\rm KS} | u_{n{\bf k}} \rangle
\neq 0 $, where $\Delta_{\nu} v_{\rm KS} $ is the linear change of the
Kohn-Sham potential and $ u_{n{\bf k}} $ is the cell-periodic part of a Bloch
state. Simple group-theoretic considerations indicate that this condition can
only be obeyed if the irreducible representation of the $\nu$-th phonon
contains the totally symmetric representation of the lattice (namely,
$\Gamma_{\nu} \supseteq {A}_{1g}$). In short, only totally-symmetry
Raman-active modes -- as, e.g., the $A_{1g}$ phonons in Sb -- can be coherently
excited via the DECP mechanism.  This condition constitutes an important
selection rule for coherent-phonon excitation in semimetals. 

\begin{figure*}[]
    \centering
    \includegraphics[width=1\textwidth]{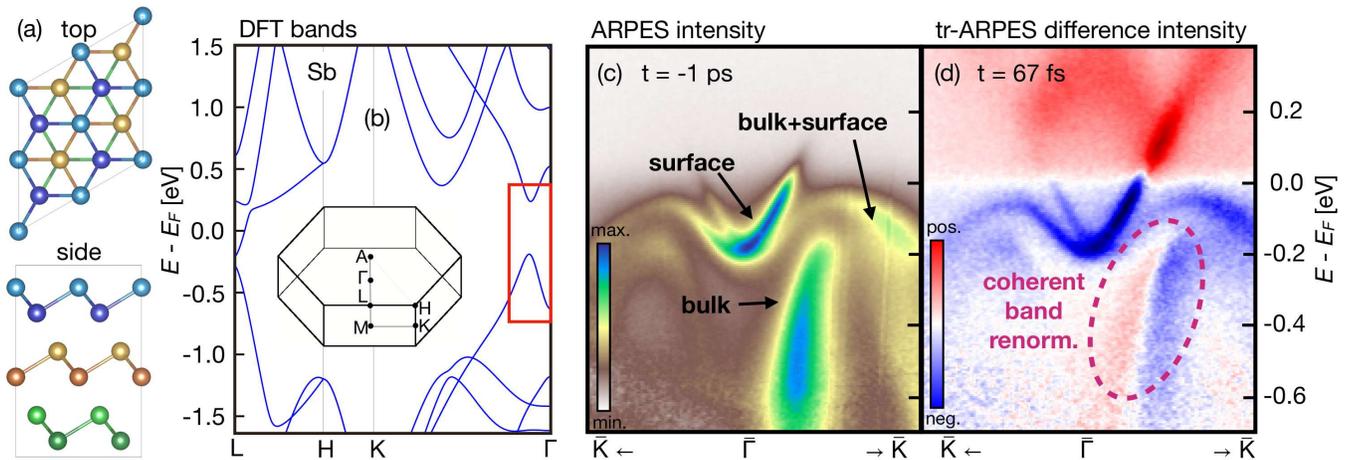}
    \caption{(a) Top and side view of the crystal structure of Sb.  (b)
Electron {bulk} band structure along the L-H-K-$\Gamma$ high-symmetry path in
the BZ as obtained from DFT. The hexagonal non-primitive BZ of Sb is shown as
an inset and energies are relative to the Fermi energy.  (c) Static ARPES
spectrum $I_0$ obtained before photoexcitation for energies and crystal momenta
within the red rectangle in panel (b). In addition to bulk bands, measurements
reveal surface states and bands of hybrid surface-bulk character (not shown in
panel (b)).  (d) Difference-intensity tr-ARPES spectrum $\Delta I(t) =
I(t)-I_0$ for a time delay of $67$~fs and an excitation fluence of
0.26~mJ~cm$^{-2}$. Red (blue) denotes intensity gain (loss).  The dashed oval
marks the energy and momentum region exhibiting pronounced transient
band-structure renormalizations due to coupling with coherent phonons. 
    }
    \label{fig1}
\end{figure*}

\section{Time-resolved and Frequency-domain ARPES} \label{sec:exp}

In the following, we proceed with the details of the tr-ARPES
measurements and the FD-ARPES technique. Details of the experimental setup are
described in Ref.~\cite{Jauernik2018}.  For the experiments, we used Sb single
crystals with a (111) surface orientation. Tr-ARPES measurements were performed
at a time resolution of  $\approx \SI{100}{\femto\second}$ (FWHM of pump-probe
cross-correlation) using \SI{1.48}{\electronvolt}, sub-\SI{30}{\femto\second}
pump pulses and \SI{5.90}{\electronvolt}, \SI{95}{\femto\second} probe pulses
focused on the sample surface at an angle of incidence of 45$^{\circ}$. The
horizontal and vertical size of the pump (probe) beam at the sample surface was
$\SI{237}{} \times \SI{180}{\micro\meter\squared}$ ($\SI{43}{} \times
\SI{70}{\micro\meter\squared}$) in full width at half maximum. The plane of
incident was the $\overbar{\Gamma {\rm K}}$ plane of the Sb(111) surface Brillouin
zone (BZ) and both pump and probe light polarization were oriented parallel to
this plane. The incident pump fluence on the sample was varied between
\SI{0.04}{\milli\joule\per\square\centi\meter} and
\SI{0.26}{\milli\joule\per\square\centi\meter} corresponding to absorbed
fluences between \SI{0.016}{\milli\joule\per\square\centi\meter} and
\SI{0.10}{\milli\joule\per\square\centi\meter} \cite{hass1972optical}.
Photoelectrons were collected with a hemispherical analyzer along the
$\overbar{\Gamma {\rm K}}$ direction with a total energy resolution of
\SI{35}{\milli\electronvolt}. The sample quality was checked by low-energy
electron diffraction and the sample was simultaneously aligned in the
$\overbar{\Gamma {\rm K}}$ direction for the tr-ARPES measurements. In
the ARPES data a sharp signature of the surface band is another indicator for a
good sample preparation. All experiments were performed at room temperature at
a pressure of $<\SI{2e-10}{\milli\bar}$.

The tr-ARPES data were taken in a total (pump-probe) time delay
range of $\approx \SI{2.75}{\pico\second}$ with a sampling rate of
\SI{60}{\tera\hertz} to comply with the Nyquist sampling criterion for probing
the $A_{1g}$ mode. For the energy- and momentum-resolved Fourier analysis, we
divided the tr-ARPES intensity maps into small integration regions of
$\SI{6}{\milli\electronvolt} \times \SI{0.0022}{\per\angstrom}$. Each of these
regions exhibits a photoemission intensity transient showing a superposition of
carrier population dynamics and coherent-phonon-induced modulations. To
separate the two signal contributions, we fitted the transients with a
fifth-order polynomial (starting at a time delay of \SI{33}{\femto\second}) to
model the carrier relaxation dynamics. By subtracting the fitting results from
the transients, the oscillatory part of the signal can be extracted (see also Supplementary Figure~S6).
The resulting transients were zero-padded to triple their lengths before their
amplitude spectra were determined via fast Fourier transformation. This results
in a data set of energy- and momentum-dependent Fourier amplitude spectra (with
a frequency increment of \SI{0.12}{\tera\hertz}). FD-ARPES spectra are cuts
through this data set at a fixed frequency representing the energy- and
momentum-resolved Fourier amplitude at the selected frequency.  

\section{Electronic properties and coherent phonons in antimony}\label{sec:eq}

Antimony (Sb) is an elemental semimetal which crystallizes in a rhombohedral
lattice, illustrated in Fig.~\ref{fig1}~(a), consisting of 2 atoms per unit
cell  and belonging to the R$\overline3$m crystallographic group
\cite{liu1995electronic}.  Owing to its simplicity and the existence of a
well-defined coherent phonon spectrum, it constitutes an ideal test case for
the benchmark and validation of the ab initio approach formulated in this
manuscript.  In particular, at the $\Gamma$ point the phonon dispersion
obtained from DFPT (Supplementary Figure~S2) is characterized by a
non-degenerate totally-symmetric $A_{1g}$ mode and one doubly-degenerate $E_g$
mode with frequencies of 4.4 and 3.2~THz, respectively, which slightly
underestimate the experimental values (4.5 and 3.5 THz).  Coherent phonons of
both $A_{1g}$ and $E_{g}$ characters can be excited in Sb, however, only the
former obeys symmetry selection rules required to undergo excitation via the
DECP mechanism. Conversely, coherent $E_g$ phonons require alternative
excitation pathways (as, e.g., ionic Raman scattering and inelastic stimulated
Raman scattering), which are higher-order in the interaction
\cite{Zijlstra2006}. 

The DFT-PBE band structure along the L-H-K-$\Gamma$ path is reported in
Fig.~\ref{fig1}~(b).  Here and below, we consider the non-primitive hexagonal
unit cell (6 atoms per cell) and BZ (illustrated in the inset of
Fig.~\ref{fig1}~(b)), as this choice enables the straightforward identification
of the high-symmetry path probed by tr-ARPES  on the (111) surface. A
comparison of the primitive and non-primitive structures and BZ is included in
Supplementary Figure~S2  \cite{supp}.  The Fermi surface is located in the
vicinity of the L point and the red rectangle marks the region probed by the
tr-ARPES measurements.  The static ARPES intensity $I_{0} $ along the
K-$\Gamma$-K path is reported in Fig.~\ref{fig1}~(c) before photoexcitation.
The spectral features are in good agreement with previous experimental works
\cite{clark2021observation, sakamoto2022connection} and reveal photoemission
signatures of different origins: one bulk band -- well captured by the bulk DFT
calculation -- as well as surface and hybrid surface-bulk bands.  The
interfacial origin of these spectral features is corroborated by comparison
with DFT simulations for a finite slab consisting of 16 Sb layers (see
Supplementary Figure~S3 \cite{supp}).

\begin{figure*}[]
    \centering
    \includegraphics[width=1\textwidth]{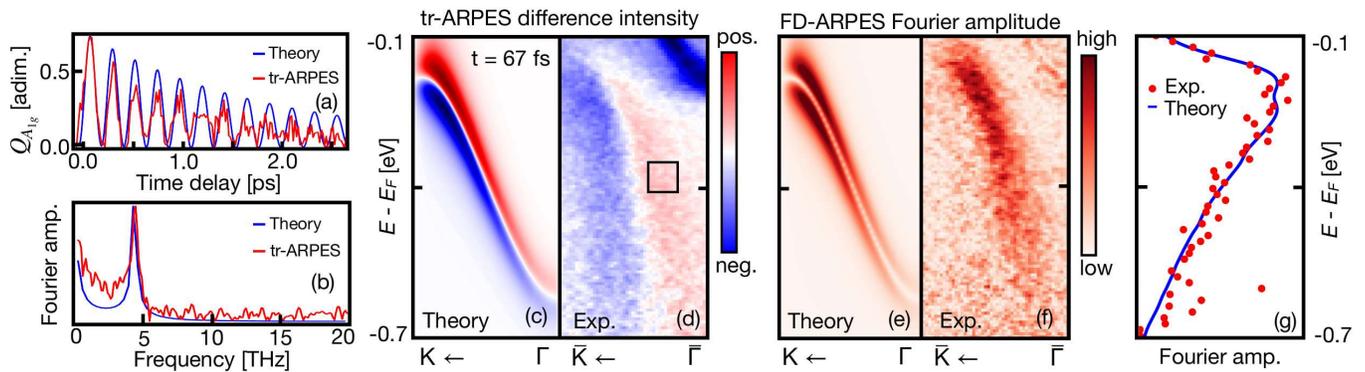}
    \caption{
    (a) Displacement amplitude of the $A_{1g}$ mode as obtained from the
solution of the coherent-phonon EOM [Eq.\,\eqref{eq:osci}]. For comparison, we
report the oscillation of the transient tr-ARPES signal taken within a region
of interest marked by a black rectangle in (d).  (b) Fourier transform of the
tr-ARPES signal and the calculated displacement amplitude in (a).  Simulations
(c) and measurements (d) of changes in the tr-ARPES spectral function at a time
delay of $67$~fs after photoexcitation for crystal momenta in the vicinity of
the $\Gamma$ point. Red (blue) denotes intensity gain (loss). An incident pump
fluence of 0.26~mJ~cm$^{-2}$ was considered. Theoretical (e) and experimental
(f) FD-ARPES map obtained from the Fourier analysis of the time-resolved
spectral functions at the frequency of the $A_{1g}$-mode of Sb.  (e) Comparison
between simulations and measurements for the maximum Fourier amplitude in the
FD-ARPES map, as a function of energy.
    }
    \label{fig2}
\end{figure*}
 
In Fig.~\ref{fig1}~(d),  we report the tr-ARPES difference intensity $\Delta
I(t) = I(t)-I_{0}$  for a time delay of $t=67$~fs after excitation.  Red (blue)
colors denote intensity gain (loss).  For energies larger than $-0.2$~eV, the
gain-loss pattern predominantly reflects carrier photoexcitation above the
Fermi energy, with the ensuing depopulation of lower-energy states.  In the
region within the dashed oval in Fig.~\ref{fig1}~(d) we further observe
transient intensity renormalizations that are indicative of the excitation of
coherent phonons. These gain-loss patterns arise from the upward-downward
shifting of the bulk band, and oscillate with a characteristic frequency that
coincides with the $A_{1g}$ mode frequency. These band oscillations are further
illustrated in Supplementary Figure~S6 and in the Supplementary Video
\cite{supp}.  These spectral fingerprints are a direct manifestation of
electron-phonon interactions involving the bulk Sb band and the
coherently-driven $A_{1g}$ mode, and their dynamics reflects the transient
band-structure renormalization driven by a photo-induced structural change. 

\section{Quasiparticle Renormalization via coherent phonons in
antimony}\label{sec:QP}

To describe the coherent phonon dynamics triggered by an electronic excitation
in Sb, we solve Eqs.~\eqref{eq:osci}-\eqref{eq:TDBE2} from first principles and
determine the time-dependent coherent-phonon amplitude $Q_\nu$. This procedure
has been applied to all zone-center optical phonons of Sb. Only the $A_{1g}$
mode displays coherent dynamics, whereas the $E_g$ modes do not satisfy the
symmetry selection rules outlined in Sec.~\ref{sec:theory}, leading to
vanishing coherent amplitude.  

The electron distribution function $f_{n{\bf k}}$ derived from the solution of
Eqs.~\eqref{eq:TDBE1} and \eqref{eq:TDBE2} is illustrated in Supplementary
Figure~S4 \cite{supp}.  The calculated coherent-phonon amplitude  $Q_{A_{1g}}$
for the $A_{1g}$  mode is marked in {blue} in Fig.~\ref{fig2}~(a). It is
characterized by periodic and damped cosine-like oscillations, with an
oscillation period equal to the phonon period $T_\nu = 2 \pi \omega_\nu^{-1} =
0.22~$ps.  In the simulations, we explicitly account for coherent-phonon
damping due to the phonon-phonon interaction \cite{LAX1964487} via the
scattering rate $\gamma_\nu =0.5$~ps$^{-1}$. This quantity is estimated from ab
initio finite-difference calculations of the third-order force constant tensor
(see Appendix~\ref{sec:comp}) and it is accounted for by inclusion of a term
$\gamma_\nu \partial_t Q_\nu $ in the left-hand side of Eq.~\eqref{eq:osci}.
For comparison, Fig.~\ref{fig2}~(a) further illustrates the time-dependent
modulation of the tr-ARPES signal averaged over the spectral region marked by
the black rectangle in Fig.~\ref{fig2}~(d).  The experimental curves in
Fig.~\ref{fig2}~(a) have been scaled to match the amplitude of the first
maximum of the calculated coherent phonon amplitude. 
The oscillation frequencies and damping of the tr-ARPES data agree well with ab
initio simulations.  Our results are further consistent with timescales of
coherent-phonon damping revealed by earlier experimental studies and further
support the identification of phonon-phonon scattering as the primary source of
damping \cite{hase1998dynamics,ishioka2001ultrafast,fahy2016resonant}.

We further illustrate in Fig.~\ref{fig2}~(b) the Fourier transform of the
calculated and measured coherent-phonon oscillations from panel~(a).  The
Fourier analysis confirms that the frequency of the $A_{\rm 1g}$ mode dominates
the coherent-phonon oscillations. {Additionally, Fig.~\ref{fig2}~(b) indicates
that the peak broadening of the calculated Fourier amplitude agrees well with
the measurements consolidating the interpretation of phonon-phonon scattering
as the primary mechanism for coherent-phonon damping. 

In the following, we proceed to describe band-structure renormalization effects
mediated by coherent phonons within a perturbative many-body framework.  Under
equilibrium conditions, the phonon-assisted renormalization of the
electron-band structures is well understood and it  provides the basis for
understanding the temperature dependence of the electronic properties of
solids. At equilibrium, quasiparticle energies are renormalized by the
inclusion of a self-energy correction \cite{giustino2017electron}: 
\begin{align}
\varepsilon^{\rm QP}_{n{\bf k}} & =
\varepsilon_{n{\bf k}}  + {\rm Re}\Sigma_{n{\bf k}} (\varepsilon^{\rm QP}_{n{\bf k}}) 
 \quad .
\end{align}
Here, $\Sigma_{n \mathbf{k}}$ is the electron self-energy due to the
electron-phonon interaction, which includes the Fan-Migdal  $\Sigma^{\rm FM}$
and Debye-Waller $\Sigma^{\rm DW}$ terms.  In presence of light-induced
structural changes, an additional self-energy term contribute to the band
structure renormalization
\cite{lafuente-bartolome_unified_2022,stefanucci2024semiconductor,stefanucci2023and}:  
\begin{align}
    \Sigma^{p}_{n \mathbf{k}} (t) =  \sum_{{\bf q}\nu} g_{nn}^\nu (\mathbf{k},
\mathbf{q}) Q_{\mathbf{q} \nu}(t) \quad .  \label{eq:CPenre}
\end{align}
This term arises from the perturbative correction of the single-particle
eigenvalues at first-order in the electron-phonon interaction. While it exactly
vanishes at equilibrium ($Q_{\mathbf{q} \nu}=0$), this term is directly
responsible for the band-structure renormalization due coherent phonons.  We
point out that Eq.~\eqref{eq:CPenre} is formally identical to the polaron
self-energy introduced in the many-body polaron theory
\cite{lafuente2022ab,lafuente-bartolome_unified_2022}, and its appearance is
also consistent with a formulation of the coupled electron-phonon dynamics
based on non-equilibrium Green's functions
\cite{stefanucci2023and,stefanucci2024semiconductor}. For consistency with
earlier works, we therefore refer to Eq.~\eqref{eq:CPenre} as the polaron
self-energy.

In order to account for coherent-phonon renormalization of the electronic
properties, we evaluate the time-dependent electron spectral function according
to: 
\begin{align}
     A(\omega,t) = 
     &\sum_{n{\bf k}}\frac{ \pi^{-1} \mathrm{Im}\,\Sigma_{n \mathbf{k}}
}{\left[\hbar\omega - \varepsilon_{n \bf k} - \mathrm{Re}\,\Sigma_{n
\mathbf{k}}  \right]^2 + \left[\mathrm{Im}\,\Sigma_{n \mathbf{k}} \right]^2}
\quad . \label{eq:td-A(w)}
\end{align}
Here, the total time- and frequency-dependent self-energy is defined as
$\Sigma_{n \mathbf{k}}(\omega,t) = \Sigma_{n \mathbf{k}}^{\rm FM}(\omega) +
\Sigma_{n \mathbf{k}}^{p}(t)$. In short, the spectral function $A(\omega,t)$
accounts for static phonon-assisted renormalization of the band structure via
the Fan-Migdal term, whereas time-dependent effects due to coherent phonons are
introduced via the polaron self-energy.  Similar to earlier studies, the
Debye-Waller contribution is  approximately accounted for by enforcing particle
conservation \cite{ponce2016epw}.  We further account for optical selection
rules via inclusion of the dipole matrix elements in the spectral function (see
discussion in Appendix~\ref{sec:fast}), and for finite experimental resolution
by applying a 40~meV broadening to the spectral function.

Figure~\ref{fig2}~(c) displays the calculated transient spectral function
$\Delta A(\omega, t) = A(\omega, t) - A(\omega, 0)$ for a time delay of
$t=67$~fs after photo-excitation, which corresponds to the first maximum of the
displacement amplitude in Fig.~\ref{fig2}~(a). The static spectral function
$A(\omega,0)$ is reported in Supplementary Figure~S5 \cite{supp}.  To determine
initial conditions for simulations compatible with experiments, we set the
initial excited-state energy in our calculations to match the estimated
absorbed pump-energy per unit cell. A detailed discussion of this procedure is
provided in the Appendix~\ref{sec:init}.  The intensity gain and loss patterns
-- marked in red and blue, respectively -- reflect an upward energy
renormalization of the bulk electron band owing to the coupling with the
coherent $A_{1g}$ phonon, and it oscillates following the time profile of
Fig.~\ref{fig2}~(a). 

In Fig.~\ref{fig2}~(d), we report the measured tr-ARPES transient spectral
function at the same time delay $t$ for an incident fluence of
$0.26$~mJ$~$cm$^{-2}$.  The calculated gain-loss pattern as well as its time
dependence are in good agreement with tr-ARPES measurements.  In the vicinity
of the Fermi energy, the tr-ARPES measurements are sensitive to the surface
band, which manifests itself as a high-intensity feature in the top-right
corner of Fig.~\ref{fig2}~(d) due to photoexcitation of the hybrid surface-bulk
band marked in Fig.~\ref{fig1}~(c)}. These features dominate the transient
spectral intensity above $-0.2$~eV below the Fermi energy, overshadowing the
energy renormalization of the bulk band. Surface bands, however, couple weakly
to coherent phonons as compared to bulk states, and they exhibit much weaker
intensity oscillations \cite{sakamoto2022connection}, as illustrated in
Supplementary Figures~S6 and S7.

In Figs.~\ref{fig2}~(e) and \ref{fig2}~(f), we report the theoretical and
experimental FD-ARPES spectra obtained from the Fourier transformation of the
tr-ARPES spectral functions for a fluence of 0.26~mJ~cm$^{-2}$.  The maximum
Fourier amplitude of experiment (theory) is obtained at a frequency
$\omega_{\rm exp} = 4.5$~THz ($\omega_{\rm theo} = 4.4$~THz), which coincides
with the oscillation of the $A_{1g}$ mode.  Overall, FD-ARPES  filters out the
spectral features that do not exhibit an oscillatory component,  thereby
highlighting electronic states that are strongly coupled to the coherent
phonons.  The intensity of the surface states, conversely, is suppressed owing
to their weak coupling to the $A_{1g}$ mode.  In Fig.~\ref{fig2}~(g), we report
the maximum  Fourier amplitude as a function of energy.  The highest Fourier
amplitude is obtained at -0.2~eV and it decreases at lower energies. This trend
is well reproduced by our ab initio calculations across the full energy range
of the measurements. This behaviour can be attributed to a suppression of the
ARPES intensity in the vicinity of the $\Gamma$ point due to the polarization
of the probe pulse and the dipole selection rules for optical transitions (see
also Appendix~\ref{sec:fast}).  

\begin{figure*}[]
    \centering
    \includegraphics[width=1\linewidth]{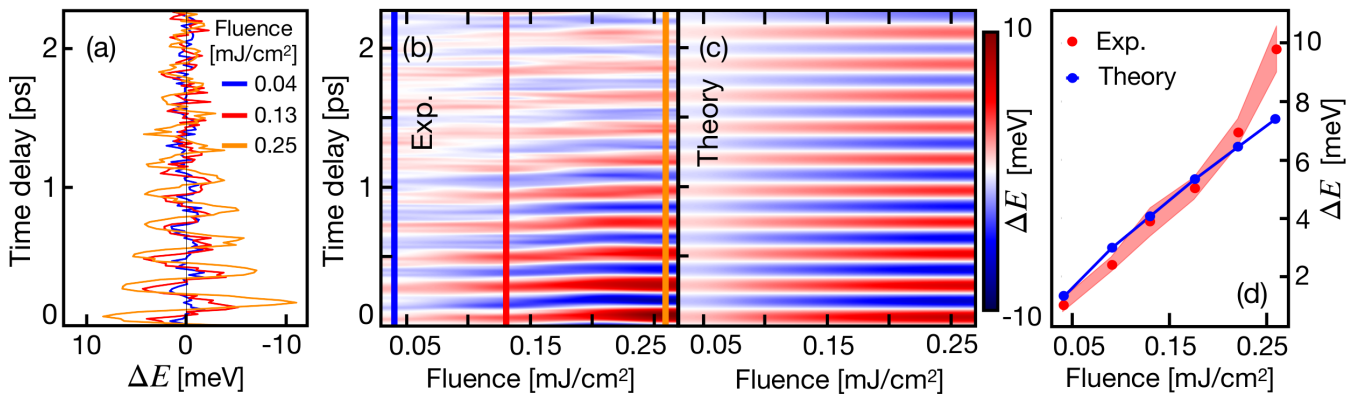}
    \caption{
    (a) Momentum-averaged energy renormalization of the bulk band in the
vicinity of the $\Gamma$ point due to coherent phonon excitation for incident
fluences of $0.04$ (blue), $0.13$ (red) and $0.26$~mJ$~$cm$^{-2}$ (orange).
(b) Momentum-averaged energy renormalization of the bulk band in the vicinity
of the $\Gamma$ point as a function of fluence and time extracted from the
tr-ARPES data. The orange, red and blue lines highlight the time-dependent
energy renormalization at $0.04$, $0.13$ and $0.26$~mJ$~$cm$^{-2}$ presented in
panel (a), respectively.  (c) Calculated energy renormalization of the bulk
band at $\Gamma$ as a function of fluence and time obtained from
Eq.~\eqref{eq:CPenre}.  (d) Maximum energy renormalization versus incident
fluence of the theoretical and fitted experimental oscillations. 
    }
    \label{fig3}
\end{figure*}

\section{Fluence-Dependent Band-structure Renormalization }\label{sec:fluence}

Having established a predictive ab initio framework for the analysis of
coherent phonons in semimetals, we proceed to explore to which extent
quasiparticle renormalization effects can be tailored upon control of the
external driving field.  In particular, we varied the pump-fluence in the
tr-ARPES measurements between 0.04 and 0.26~mJ~cm$^{-2}$, and investigated its
impact on the energy renormalization $\Delta E$ due to coherent phonons.
Figures~\ref{fig3}~(a) illustrates the time-dependent energy renormalization
$\Delta E$ of the bulk band in the vicinity of the $\Gamma$ point for fluences
of $0.04$ (blue), $0.13$ (red) and $0.26$~mJ~cm$^{-2}$ (orange). To focus on
the oscillatory component of the band renormalization, we subtracted the
non-oscillating component of the transient spectral function following the
procedure outlined in the Supplementary Note~S1, and illustrated in
Supplementary Figure~S1 \cite{supp}.  For all fluences, measurements reveal a
periodic renormalization of the band energies and indicate a significant
increase of the maximum energy renormalization with the pump fluence.

The full time- and fluence-dependent measurements of the bulk-band
renormalization is further illustrated in Fig.~\ref{fig3}~(b). The orange, red
and blue lines highlight the fluences reported in Fig.~\ref{fig3}~(a). In
addition to an increase of coherent-phonon energy renormalization with fluence,
Fig.~\ref{fig3}~(b) further reveals a damping of the oscillations that is not
substantially affected by the pump fluence.  Figure~\ref{fig3}~(c) reports ab
initio simulations of fluence- and time-dependent  energy renormalizations of
the bulk band in the vicinity of the $\Gamma$ point.  Overall, we find very
good agreement between experiment and theory across the full range of fluences
and time delays.  Experimental data further exhibit an increase in softening of
the coherent phonon frequency for larger driving fluences. These effects (not
captured in the simulations) are reminiscent of phonon renormalization  arising
in highly-doped semiconductors due to electron-phonon coupling
\cite{caruso_2017,novko_2020}. We thus tentatively attribute these effect to
the dressing of the coherent phonons via electron-phonon interactions mediated
by the photo-excited carriers, in close analogy to the static counterpart of
these effects.  To quantitatively compare theory and experiments, we further
display in Fig.~\ref{fig3}~(d) the measured and calculated maximum energy
renormalization of the bulk band as a function of the incident fluence.
Tr-ARPES indicates a linear dependence of the energy renormalization on
fluence, which is well reproduced by ab initio calculations. This trend
suggests that the driving fluences considered in this work are well captured
within the linear-response regime. 

\begin{figure*}
    \centering
    \includegraphics[width=0.8\textwidth]{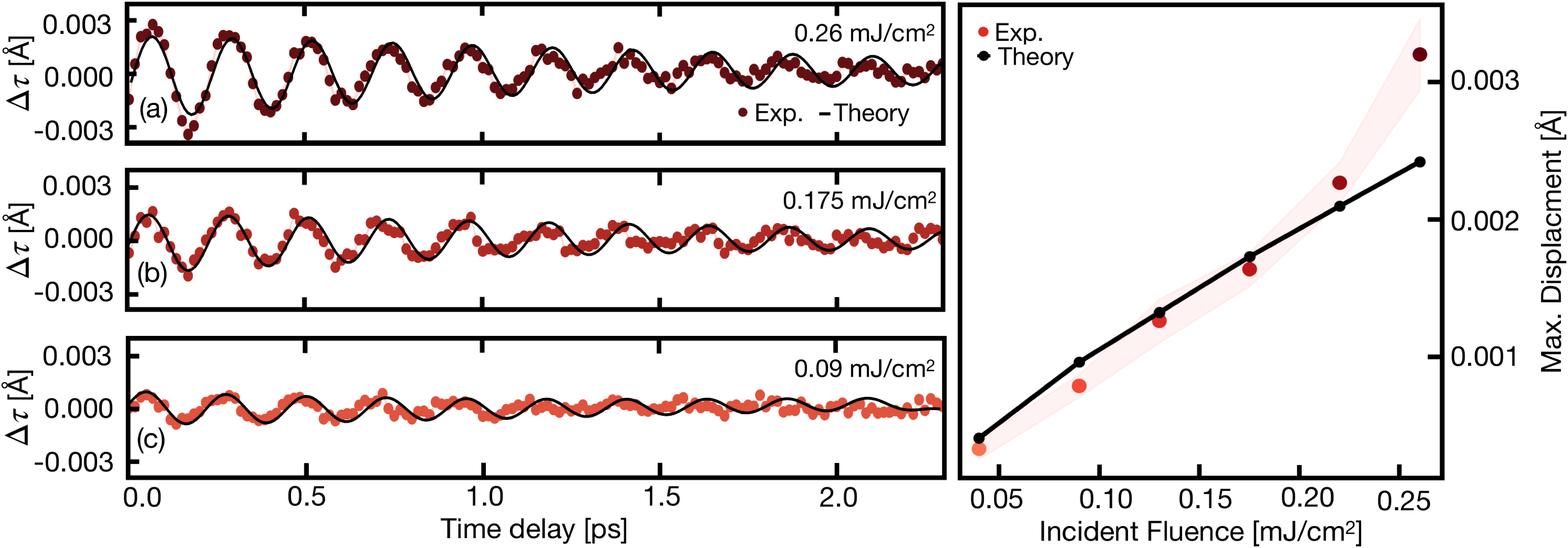}
    \caption{
    (a-c) Comparison between theory and experiment for the
displacement $\Delta \tau$ of Sb atoms from equilibrium due to the excitation
the A$_{1g}$ coherent phonon for fluences as indicated.  (d) Comparison between
theory and experiment for the maximum of the absolute displacement $|\Delta
\tau|$ of Sb as a function of the incident fluence.
    }
    \label{fig:Q}
\end{figure*}

A breakdown of the linear regime has been reported in pump-probe Raman
experiments for antimony at driving fluences exceeding 5~mJ~cm$^{-2}$
\cite{fausti2009ultrafast}. This value constitutes an upper limit for the
maximum fluence to excite and probe coherent-phonon band-structure
renormalization. By extrapolating the linear trend observed in
Fig.~\ref{fig3}~(d), we expect that energy renormalizations exceeding 100~meV
could be realized at high-driving fields without undergoing structural damage. 

\section{Discussion and outlook} \label{sec:out}

Overall, the ab initio formalism presented in this manuscript offers new
opportunities for predictive simulations of light-induced phenomena in
materials and, specifically, for monitoring structural changes and
quasiparticle renormalization effects emerging on ultrafast timescales.  Some
perspectives and opportunities for the application of these concepts beyond the
domain of this manuscript are discussed below.

The fluence-dependent analysis reported in Sec.~\ref{sec:fluence} indicates
that energy renormalization of the order of 10 to 100~meV can be realized in
semimetals via experimentally accessible pump fluences. Energy shifts of this
magnitude may suffice to tailor specific band-structure features via coherent
phonons or even to induce topological features in otherwise topologically
trivial compounds.  Coherent-phonon excitation in Weyl semimetals, for example,
is expected to induce dynamical renormalization of the Weyl points and their
spin textures \cite{hein2020mode, sakano_radial_2020,gatti_radial_2020}, with
likely implications for topological properties \cite{sie2019ultrafast} and
transport anomalies \cite{yan_topological_2017}.  Band engineering could
further offer the opportunity to induced additional band crossings, leading to
the dynamical formation of topological Weyl points in otherwise topologically
trivial semimetals, or insulator-to-metal transitions in narrow-gap
semiconductors \cite{Ning_2022_light}.  More generally, the ab initio modelling
of DECP in combination with tr-ARPES measurements can likely guide the
identification of unexplored opportunities for engineering the band structure
of semimetals via coherent phonons.

Besides band-structure engineering, the ab initio description of coherent
phonons constitutes a powerful opportunity to monitor light-induced structural
changes, which could shed light on the microscopic origin of light-induced
phase transitions and symmetry breaking \cite{guan2022optical,
mocatti2023light}. By combining information from tr-ARPES and ab initio
simulations, for example, one may directly retrieve the nuclear trajectories
induced by coherent phonons. We illustrate this point in
Fig.~\ref{fig:Q}~(a-c), where we  reconstructed the time-dependent displacement
of Sb from equilibrium from tr-ARPES measurements. Here, the experimental
displacement has been directly estimated from experiments via
Eq.~\eqref{eq:Dtau} as $\Delta \tau = (\hbar/2M_{\rm Sb}
\omega_\nu)^{\frac{1}{2}} e^\nu Q_\nu^{\rm exp}(t)$ using   the experimental
coherent phonon amplitude $Q_\nu^{\rm exp}(t) \simeq \langle \Delta
\varepsilon^{\rm exp}_{n{\bf k}}(t) \rangle / \langle g^{\rm theo}_{nn} ({\bf
k},0)\rangle$. In this expression, $\Delta \varepsilon_{n{\bf k}}^{\rm exp}(t)$
denotes the measured band shift due to coherent phonons, $ g^{\rm theo}_{nn}
({\bf k},0)$ is the momentum-averaged coupling matrix elements obtained from ab
initio calculations, and $\langle\cdots\rangle$ indicates averaging over
momentum. The calculated displacement, obtained from Eq.~\eqref{eq:osci} is
reported for comparison in Fig.~\ref{fig:Q}, where we accounted for the
experimental time resolution of $\Delta t = 100\,$fs FWHM via a Gaussian
convolution.  We further illustrate in Fig.~\ref{fig:Q}~(d) experimental and
calculated maximum displacement as a function of pump fluence.  This analysis
indicates the possibility to directly determine the displacement of the nuclei
from the experimental data when the magnitude of the electron-phonon
interaction is known from theory. 

Another tr-ARPES study of Sb(111) provides additional experimental reference
data for comparison with the results of our ab initio theory
\cite{sakamoto2022connection}. The paper reports a maximum oscillation
amplitude in the Sb bulk band of about 3.3 meV (see Fig. 2(c) in Ref.
\cite{sakamoto2022connection}). For the incident fluence of
\SI{0.17}~mJ~cm$^{-2}$ and considering the other experimental parameters
\cite{gauthier2020tuning}, we estimate an absorbed fluence of 0.05~mJ~cm$^{-2}$
and calculate an oscillation amplitude of $\approx$~\SI{4.7}{\milli eV} for a
sample temperature of \SI{20}{K} in reasonable agreement with the experimental
value.   Similarly, our calculations for the displacement amplitude $\Delta
\tau$ yield \SI{0.15}{\pico m} which slightly overestimates the value
$\approx$~\SI{0.1}{\pico m} reported in Ref.~\cite{sakamoto2022connection}. 

While the Raman-active phonons excited via the DECP mechanism preserve the
symmetry of the lattice, symmetry breaking could occur at high fluences via
non-linear (ternary and quartic) phonon-phonon coupling mechanisms, such as,
e.g.,  ionic Raman scattering.  Overall, the ab initio approach developed here
can be straightforwardly extended to account for these phenomena. It may thus
offer new opportunities for crystal-structure control via non-linear phononics.

Finally, we emphasize that the solution of the coherent-phonon EOM
(Eq.~\eqref{eq:osci}) and the estimation of dynamical band-structure
renormalization effects (Eqs.~\eqref{eq:CPenre} and \eqref{eq:td-A(w)}) do not
entail a significant increase in  computational costs as compared to the
solution of the TDBE (Eqs.~\eqref{eq:TDBE1} and \eqref{eq:TDBE2}).  Thus, the
workflow presented in Fig.~\ref{fig_work} for the simulation of coherent
phonons can be promptly integrated into existing ab initio codes for
electron-phonon coupling calculations, whenever Eqs.~\eqref{eq:TDBE1} and
\eqref{eq:TDBE2} are already available. 

\section{Conclusion} \label{sec:conc}

In conclusion, we reported a combined theoretical and experimental
investigation of the light-induced structural dynamics and electron-phonon
interactions in driven semimetals.  We formulated a transferable \textit{ab
initio} theory of  the displacive excitation of coherent phonons, which
combines the time-dependent Boltzmann equation, the EOM for the coherent-phonon
amplitude, and quasiparticle renormalization effects due to the polaron
self-energy.  Our approach specifically targets (i) the emergence of
photo-induced coherent lattice motion following electronic excitation by a pump
pulse and (ii) transient band-structure renormalization effects induced by
coherent phonons.  The points (i-ii) are a direct manifestation of the
electron-phonon interactions occurring under non-equilibrium conditions, and
their description eludes the capability of the equilibrium many-body approaches
to the electron-phonon coupling.  

We validated and benchmarked this approach against tr-ARPES experiments for the
elemental semimetal antimony.  To enable a quantitative comparison between
experiment and theory, we detected and characterized the transient oscillations
of the measured spectral function induced by coherent phonons for a wide range
of pump fluences. We further conducted a frequency-domain analysis of coherent
phonon oscillation (FD-ARPES), to deduce momentum- and mode-resolved
information on the electron-phonon interactions.  The good qualitative and
quantitative agreement between measurements and simulations corroborate our ab
initio approach. We further demonstrate the possibility to directly retrieve,
via a combined experimental-theory approach, information about the
electron-phonon coupling matrix elements and the nuclear trajectories from the
tr-ARPES data.  

Overall, the combination of ab initio theory and tr-ARPES emerges as a powerful
diagnostic framework to directly characterize the light-induced structural
changes and quasiparticle-energy renormalization in driven semimetals. We are
confident that these advancements will open up opportunities for
dynamic\sout{al} structural control and band-structure engineering. 

\section*{ACKNOWLEDGMENTS}
This work was funded by the Deutsche Forschungsgemeinschaft (DFG), Projects No.
499426961 and 443988403. The authors gratefully acknowledge the computing time
provided by the high-performance computer Lichtenberg at the NHR Centers
NHR4CES at TU Darmstadt (Project p0021280). F.C. acknowledges discussions with
Elia Stocco. 

\section*{Appendix}

\appendix

\section{Computational details}\label{sec:comp}

\subsection{Ground-state properties}
DFT calculations are conducted with the plane-wave pseudopotential code {\tt
Quantum ESPRESSO}  \cite{giannozzi2009quantum,giannozzi2017advanced}.
Fully-relativistic projector augmented wave (PAW) pseudopotentials
\cite{dal2014pseudopotentials} and the Perdew-Burke-Ernzerhof (PBE)
generalized-gradient approximation \cite{perdew1996generalized} for the
exchange-correlation functional are used. The plane-wave kinetic-energy cutoff
is set to 50~Ry, and the BZ is sampled with a $10\times 10 \times 8$
Monkhorst-Pack grid.  The phonon dispersion is obtained from DFPT on a $5\times
5 \times 4$ q-point mesh \cite{baroni2001phonons}. Spin-orbit coupling (SOC) is
included throughout all steps of the calculations.  The electron and phonon
eigenvalues, as well as the electron-phonon coupling matrix, are interpolated
on a dense $40\times 40 \times 15$ grid via maximally-localized Wannier
functions \cite{marzari2012maximally} within {\tt EPW } code
\cite{lee2023electron}, which uses {\tt Wannier90} as a library
\cite{pizzi2020wannier90}. 

\subsection{Ultrafast dynamics from the TDBE}
\label{sec:fast}

Equations~\eqref{eq:osci}-\eqref{eq:TDBE2} have been implemented in a modified
version of the {\tt EPW} code,  and solved within an energy window of 1.5~eV
around the Fermi energy. Time propagation is achieved by discretization of the
time derivative via Heun's method for a total simulation time of 10~ps with a
time step of $1$~fs. The electron and phonon collision integrals due to the
electron-phonon interaction ($\Gamma^{\rm e-ph}_{n{\bf k}}$ and $\Gamma^{\rm
ph-e}_{{\bf q}\nu}$) are evaluated at each time-step on dense ${\bf k}$- and
${\bf q}$-grids of $40\times 40 \times 15$, using a broadening of 10~meV for
the numerical evaluation of the Dirac $\delta$ function. Explicit expressions
for the collision integrals $\Gamma^{\rm e-ph}_{n{\bf k}}$ and $\Gamma^{\rm
ph-e}_{{\bf q}\nu}$ and a detailed description of the computational procedure
for their solution is reported elsewhere
\cite{caruso2021nonequilibrium,caruso2022ultrafast}.  We account for
electron-electron scattering within the relaxation-time approximation via the
inclusion of the collision integral:
\begin{align*}
    \Gamma^{\rm e{-}e} \left[ f_{n\mathbf{k}}\right] = \frac{f_{n\mathbf{k}}(t)
- f_{n\mathbf{k}}^{\rm eq}(t)}{\tau^{\rm e{-}e}} \quad . 
\end{align*}
Here, $f_{n\mathbf{k}}^{\rm eq}(t)$ is the equilibrium (Fermi-Dirac)
distribution function at time $t$. To ensure energy conservation,
$f_{n\mathbf{k}}^{\rm eq}(t)$ is determined at each time step via the
requirement that it yields an electronic energy $E_{\rm el} = N_p ^{-1} \sum_{n
{\bf k}} \varepsilon_{n \bf k} f^{\rm eq}_{n \bf k}$ which coincides with the
one corresponding to the non-equilibrium distribution function
$f_{n\mathbf{k}}(t)$.  We set the average electron lifetime due to
electron-electron interaction to $\tau^{\rm e{-}e}\simeq10$~meV. This value was
deduced by averaging the imaginary part of the $GW$ self-energy, as obtained
from the {\tt yambo} code \cite{sangalli2019many}, in an energy window of
1.5~eV around the Fermi energy.  {Simulations of the coherent-phonon dynamics
are robust against changes of $\tau^{\rm e{-}e}$. }

The phonon decoherence time of the $A_{1g}$-mode was obtained estimated via the
following expression:\cite{caruso2021nonequilibrium}
\begin{widetext}
\begin{align}
    \frac{1}{ \tau^{\rm pp}_{{\bf q}\nu }} = \frac{\pi\hbar }{4} \sum_{\nu'
\nu''} \int \frac{d{\bf q'}}{\Omega_{\rm BZ}} \left| \Psi^{\nu  \nu'
\nu''}_{\bf q q' q''} \right|^2 \nonumber &[(n_{{\bf q'} \nu'} - n_{{\bf q''}
\nu''}) \delta(\omega_{{\bf q} \nu} + \omega_{{\bf q'} \nu'} - \omega_{{\bf
q''} \nu''})\delta^{\bf G}_{\bf q q' - q''} \\ & + (n_{{\bf q'} \nu'} + n_{{\bf
q'' +1} \nu''}) \delta(\omega_{{\bf q} \nu} - \omega_{{\bf q'} \nu'} -
\omega_{{\bf q''} \nu''})\delta^{\bf G}_{\bf q -q' - q''}] \quad .
\label{eq:tau}
\end{align}
\end{widetext}
Here $\Psi^{\nu  \nu' \nu''}_{\bf q q' q''}$ is the phonon-phonon coupling
matrix element, which we evaluated for the $A_{1g}$-mode at $\Gamma$. The
third-order force constant required for the estimation of $\Psi^{\nu  \nu'
\nu''}_{\bf q q' q''}$ has evaluated from difference based on the utility {\tt
third-order.py} \cite{ShengBTE_2014}.  Evaluation of Eq.~\eqref{eq:tau} yields
$\tau^{\rm pp}_{\rm A_{1g}}= 2$~ps for antimony at room temperature. This value
is kept constant throughout the simulations. 

\begin{figure}[t]
    \centering
    \includegraphics[width=0.4\textwidth]{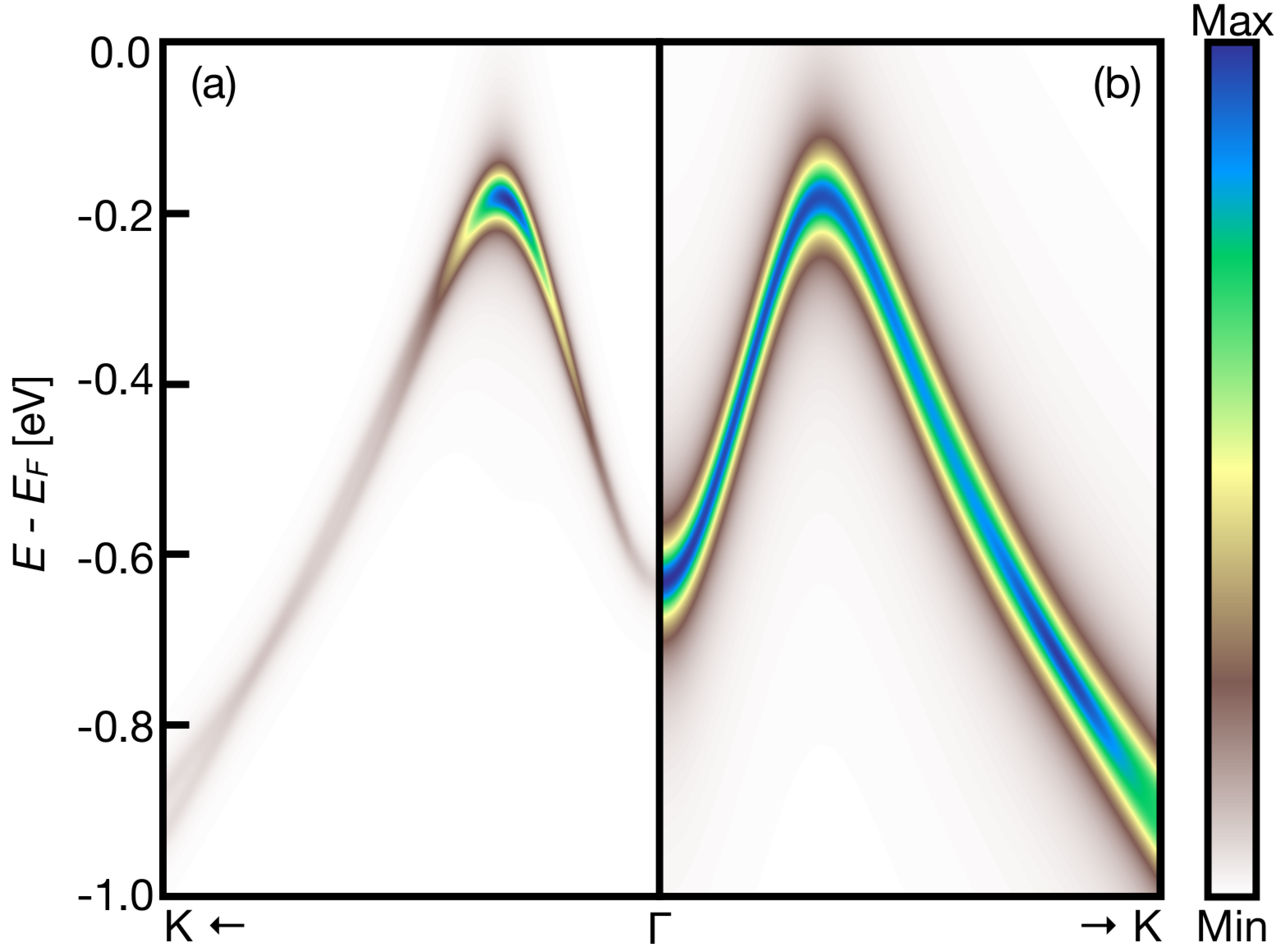}
    \caption{Static electron spectral function for the bulk band of Sb due to
the electron-phonon interaction in the Fan-Migdal approximation evaluated (a)
with and (b) without dipole matrix elements for p-polarized light.}
\label{fig:spec_funcs}
\end{figure}

To account for the effects of optical selection rules arising from the
polarization of the probe pulse, we evaluate the electron spectral function via
\cite{moser2017experimentalist}:
\begin{align*}
     A(\omega,t) = 
     &\sum_{n{\bf k}}^{\rm occ} \sum_{m{\bf k}}^{\rm unocc}  \frac{ \pi^{-1}
\left| d_{nm{\bf k}} \right|^2 \mathrm{Im}\,\Sigma_{n \mathbf{k}}
}{\left[\hbar\omega - \varepsilon_{n \bf k} - \mathrm{Re}\,\Sigma_{n
\mathbf{k}}  \right]^2 + \left[\mathrm{Im}\,\Sigma_{n \mathbf{k}} \right]^2}
\quad, \end{align*}
where $d_{nm\bf k}  = \hat{\boldsymbol{\epsilon}}\cdot\langle u_{m{\bf k}} |
\hat{\bf p}| u_{n{\bf k}} \rangle$ are the dipole transition probabilities,
which we obtained from a modified version of  the utility {\tt epsilon.x}, part
of {\tt Quantum Espresso}. Here, $ \hat{\boldsymbol{\epsilon}}$ denotes the
light-polarization vector, $u_{n{\bf k}}$ the cell-periodic part of Kohn-Sham
Bloch eigenstate, and $\hat{\bf p}$ the momentum operator.  The effect of
dipole transition probabilities for p-polarized light on the spectral function
is illustrated in Fig.~\ref{fig:spec_funcs}. The left (right) panel shows the
spectral function with (without) dipole transition probabilities. The dipole
transition probabilities suppress intensity away from the $\Gamma$ point. 

\section{Initial conditions of ultrafast dynamics simulations}\label{sec:init}

The initial condition for the numerical solution of
Eqs.~\eqref{eq:osci}-\eqref{eq:TDBE2} is a Fermi-Dirac distribution $f^{\rm
FD}_{n \bf k} (T_{\rm exc})= \{{\rm exp} [(\varepsilon_{n \bf k} - \mu) / k_B
T_{\rm exc}]+1\}^{-1}$  evaluated at the effective electronic temperature
$T_{\rm exc}$, where $\mu$ denotes the chemical potential, and $k_B$ the
Boltzmann constant.  The lattice is initially at thermal equilibrium at room
temperature ($T_0 = 300$~K), which  coincides with the experimental conditions.
Correspondingly,  the initial ($t=0$) phonon occupations are set by the
Bose-Einstein statistics according to $n_{\mathbf{q} \nu}(t=0) = \left[ {\rm
exp}({ \hbar \omega_{\mathbf{q} \nu} / k_B T_0 }) - 1 \right]^{-1} $. 
 
The initial electronic temperature $T_{\rm exc}$ is determined as a function of
the incident fluence employed in the experiments on the basis of the following
rationale.  The relation between the incident fluence $F$ and the average
increase of electronic energy per unit cell $\Delta E_{\rm el}$ due to the pump
absorption is given by: 
\begin{align}
    \alpha F =  {d_z }{\Omega^{-1}} \Delta E_{\rm el}  \quad . 
    \label{eq:F2T}
\end{align}
Here, $\Omega$ is the volume of the unit cell, $d_z =15$~nm is the penetration
depth of the pump pulse, which we estimate from the extinction coefficient
$\kappa$ via:
\begin{align}
    d_z = \frac{\lambda}{4 \pi \kappa} \quad .
    \label{eq:dz}
\end{align}
With a wavelength of the laser light of $\lambda=800~$nm and an extinction
coefficient of $\kappa = 4.0803$ \cite{hass1972optical}.  Furthermore,
$\alpha=0.4$ denotes the ratio between incident and absorbed fluence, which we
obtain from the experimental absorption coefficient of Sb
\cite{hass1972optical}. 

\begin{figure}[]
    \centering
    \includegraphics[width=0.4\textwidth]{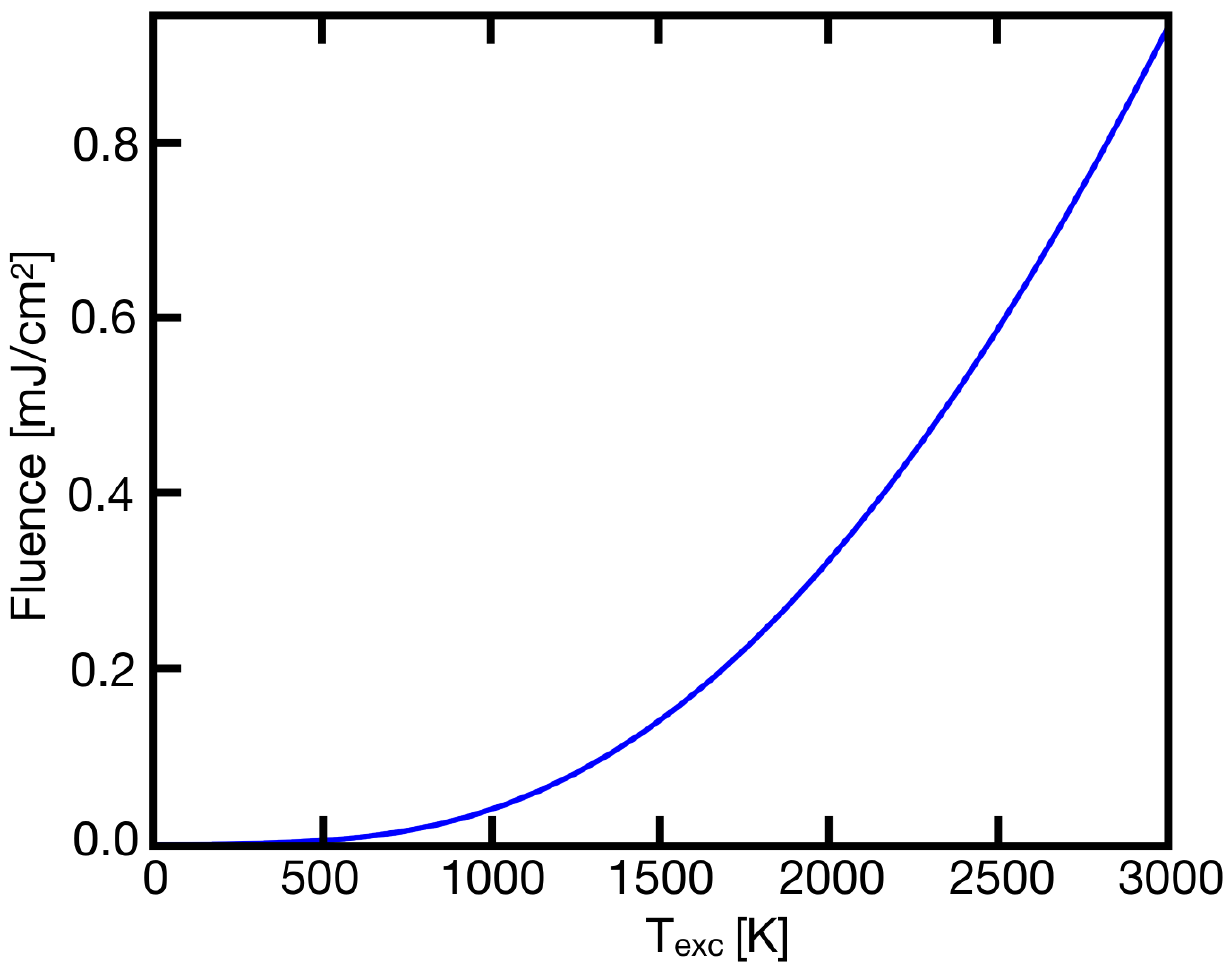}
    \caption{Relation between the incident fluence and the effective electronic
temperature $T_{\rm exc}$ obtained from numerical solution of
Eqs.~\eqref{eq:F2T}  \eqref{eq:Eexc} .} \label{fig:F2T}
\end{figure}

For the numerical solution of Eqs.~\eqref{eq:osci}-\eqref{eq:TDBE2}, we define
the initial electronic occupations $f_{n{\bf k}} (t=0)$ such that they yield an
excess electronic energy per unit cell coinciding with $\Delta E_{\rm el} =
\alpha \Omega F / d_z $ and are thus in compliance with Eq.~\eqref{eq:F2T}. 
In particular, we express the electron excited-state energy as 
 \begin{align}\label{eq:Eexc}
 \Delta E_{\rm el} = N_p ^{-1} \sum_{n {\bf k}} \varepsilon_{n \bf k} [f^{\rm
FD}_{n \bf k} (T_{\rm exc}) - f^{\rm FD}_{n \bf k} (T_{0})] \quad .
\end{align}
  Combining Eqs.~\eqref{eq:F2T} and \eqref{eq:Eexc} we obtain an explicit
relation between $T_{\rm exc}$ and the fluence $F$, which we solve numerically
to determine the effective temperature $T_{\rm exc}$ and the electronic
distribution function $f^{\rm FD}_{n \bf k} (T_{\rm exc})$ corresponding to the
experimental incident fluence.  Figure~\ref{fig:F2T} illustrates the relation
between incident fluence and effective electron temperature $T_{\rm exc}$
obtained via this condition for antimony, considering an equilibrium
temperature $T_0 = 300$~K.

We note that this Ansatz for the initial excited state is unsuitable to capture
electron coherence and the emergence of nonthermal electronic states. These
phenomena typically persist for 50-100~fs and can be systematically accounted
for within a correlated description of the coupled electron-phonon dynamics
based on non-equilibrium Green's
functions.\cite{stefanucci2024semiconductor,stefanucci2023and,PerfettoStefanucci2023,Karlsson2021,Sanchez2017}


\begin{thebibliography}{97}%
\makeatletter
\providecommand \@ifxundefined [1]{%
 \@ifx{#1\undefined}
}%
\providecommand \@ifnum [1]{%
 \ifnum #1\expandafter \@firstoftwo
 \else \expandafter \@secondoftwo
 \fi
}%
\providecommand \@ifx [1]{%
 \ifx #1\expandafter \@firstoftwo
 \else \expandafter \@secondoftwo
 \fi
}%
\providecommand \natexlab [1]{#1}%
\providecommand \enquote  [1]{``#1''}%
\providecommand \bibnamefont  [1]{#1}%
\providecommand \bibfnamefont [1]{#1}%
\providecommand \citenamefont [1]{#1}%
\providecommand \href@noop [0]{\@secondoftwo}%
\providecommand \href [0]{\begingroup \@sanitize@url \@href}%
\providecommand \@href[1]{\@@startlink{#1}\@@href}%
\providecommand \@@href[1]{\endgroup#1\@@endlink}%
\providecommand \@sanitize@url [0]{\catcode `\\12\catcode `\$12\catcode
  `\&12\catcode `\#12\catcode `\^12\catcode `\_12\catcode `\%12\relax}%
\providecommand \@@startlink[1]{}%
\providecommand \@@endlink[0]{}%
\providecommand \url  [0]{\begingroup\@sanitize@url \@url }%
\providecommand \@url [1]{\endgroup\@href {#1}{\urlprefix }}%
\providecommand \urlprefix  [0]{URL }%
\providecommand \Eprint [0]{\href }%
\providecommand \doibase [0]{https://doi.org/}%
\providecommand \selectlanguage [0]{\@gobble}%
\providecommand \bibinfo  [0]{\@secondoftwo}%
\providecommand \bibfield  [0]{\@secondoftwo}%
\providecommand \translation [1]{[#1]}%
\providecommand \BibitemOpen [0]{}%
\providecommand \bibitemStop [0]{}%
\providecommand \bibitemNoStop [0]{.\EOS\space}%
\providecommand \EOS [0]{\spacefactor3000\relax}%
\providecommand \BibitemShut  [1]{\csname bibitem#1\endcsname}%
\let\auto@bib@innerbib\@empty
\bibitem [{\citenamefont {Thomsen}\ \emph {et~al.}(1984)\citenamefont
  {Thomsen}, \citenamefont {Strait}, \citenamefont {Vardeny}, \citenamefont
  {Maris}, \citenamefont {Tauc},\ and\ \citenamefont
  {Hauser}}]{thomsen1984coherent}%
  \BibitemOpen
  \bibfield  {author} {\bibinfo {author} {\bibfnamefont {C.}~\bibnamefont
  {Thomsen}}, \bibinfo {author} {\bibfnamefont {J.}~\bibnamefont {Strait}},
  \bibinfo {author} {\bibfnamefont {Z.}~\bibnamefont {Vardeny}}, \bibinfo
  {author} {\bibfnamefont {H.~J.}\ \bibnamefont {Maris}}, \bibinfo {author}
  {\bibfnamefont {J.}~\bibnamefont {Tauc}},\ and\ \bibinfo {author}
  {\bibfnamefont {J.}~\bibnamefont {Hauser}},\ }\bibfield  {title} {\bibinfo
  {title} {Coherent phonon generation and detection by picosecond light
  pulses},\ }\href@noop {} {\bibfield  {journal} {\bibinfo  {journal} {Phys.
  Rev. Lett.}\ }\textbf {\bibinfo {volume} {53}},\ \bibinfo {pages} {989}
  (\bibinfo {year} {1984})}\BibitemShut {NoStop}%
\bibitem [{\citenamefont {Cheng}\ \emph {et~al.}(1990)\citenamefont {Cheng},
  \citenamefont {Brorson}, \citenamefont {Kazeroonian}, \citenamefont
  {Moodera}, \citenamefont {Dresselhaus}, \citenamefont {Dresselhaus},\ and\
  \citenamefont {Ippen}}]{cheng1990impulsive}%
  \BibitemOpen
  \bibfield  {author} {\bibinfo {author} {\bibfnamefont {T.}~\bibnamefont
  {Cheng}}, \bibinfo {author} {\bibfnamefont {S.}~\bibnamefont {Brorson}},
  \bibinfo {author} {\bibfnamefont {A.}~\bibnamefont {Kazeroonian}}, \bibinfo
  {author} {\bibfnamefont {J.}~\bibnamefont {Moodera}}, \bibinfo {author}
  {\bibfnamefont {G.}~\bibnamefont {Dresselhaus}}, \bibinfo {author}
  {\bibfnamefont {M.}~\bibnamefont {Dresselhaus}},\ and\ \bibinfo {author}
  {\bibfnamefont {E.}~\bibnamefont {Ippen}},\ }\bibfield  {title} {\bibinfo
  {title} {Impulsive excitation of coherent phonons observed in reflection in
  bismuth and antimony},\ }\href@noop {} {\bibfield  {journal} {\bibinfo
  {journal} {Appl. Phys. Lett.}\ }\textbf {\bibinfo {volume} {57}},\ \bibinfo
  {pages} {1004} (\bibinfo {year} {1990})}\BibitemShut {NoStop}%
\bibitem [{\citenamefont {Cheng}\ \emph {et~al.}(1991)\citenamefont {Cheng},
  \citenamefont {Vidal}, \citenamefont {Zeiger}, \citenamefont {Dresselhaus},
  \citenamefont {Dresselhaus},\ and\ \citenamefont {Ippen}}]{Cheng1991}%
  \BibitemOpen
  \bibfield  {author} {\bibinfo {author} {\bibfnamefont {T.~K.}\ \bibnamefont
  {Cheng}}, \bibinfo {author} {\bibfnamefont {J.}~\bibnamefont {Vidal}},
  \bibinfo {author} {\bibfnamefont {H.~J.}\ \bibnamefont {Zeiger}}, \bibinfo
  {author} {\bibfnamefont {G.}~\bibnamefont {Dresselhaus}}, \bibinfo {author}
  {\bibfnamefont {M.~S.}\ \bibnamefont {Dresselhaus}},\ and\ \bibinfo {author}
  {\bibfnamefont {E.~P.}\ \bibnamefont {Ippen}},\ }\bibfield  {title} {\bibinfo
  {title} {Mechanism for displacive excitation of coherent phonons in
  $\mathrm{Sb}$, $\mathrm{Bi}$, $\mathrm{Te}$, and $\mathrm{Ti_2O_3}$},\ }\href
  {https://doi.org/10.1063/1.106187} {\bibfield  {journal} {\bibinfo  {journal}
  {Appl. Phys. Lett.}\ }\textbf {\bibinfo {volume} {59}},\ \bibinfo {pages}
  {1923} (\bibinfo {year} {1991})}\BibitemShut {NoStop}%
\bibitem [{\citenamefont {Kuznetsov}\ and\ \citenamefont
  {Stanton}(1995)}]{kuznetsov1995coherent}%
  \BibitemOpen
  \bibfield  {author} {\bibinfo {author} {\bibfnamefont {A.}~\bibnamefont
  {Kuznetsov}}\ and\ \bibinfo {author} {\bibfnamefont {C.}~\bibnamefont
  {Stanton}},\ }\bibfield  {title} {\bibinfo {title} {Coherent phonon
  oscillations in $\mathrm{GaAs}$},\ }\href@noop {} {\bibfield  {journal}
  {\bibinfo  {journal} {Phys. Rev. B}\ }\textbf {\bibinfo {volume} {51}},\
  \bibinfo {pages} {7555} (\bibinfo {year} {1995})}\BibitemShut {NoStop}%
\bibitem [{\citenamefont {Dekorsy}\ \emph {et~al.}(1995)\citenamefont
  {Dekorsy}, \citenamefont {Auer}, \citenamefont {Waschke}, \citenamefont
  {Bakker}, \citenamefont {Roskos}, \citenamefont {Kurz}, \citenamefont
  {Wagner},\ and\ \citenamefont {Grosse}}]{dekorsy1995emission}%
  \BibitemOpen
  \bibfield  {author} {\bibinfo {author} {\bibfnamefont {T.}~\bibnamefont
  {Dekorsy}}, \bibinfo {author} {\bibfnamefont {H.}~\bibnamefont {Auer}},
  \bibinfo {author} {\bibfnamefont {C.}~\bibnamefont {Waschke}}, \bibinfo
  {author} {\bibfnamefont {H.~J.}\ \bibnamefont {Bakker}}, \bibinfo {author}
  {\bibfnamefont {H.~G.}\ \bibnamefont {Roskos}}, \bibinfo {author}
  {\bibfnamefont {H.}~\bibnamefont {Kurz}}, \bibinfo {author} {\bibfnamefont
  {V.}~\bibnamefont {Wagner}},\ and\ \bibinfo {author} {\bibfnamefont
  {P.}~\bibnamefont {Grosse}},\ }\bibfield  {title} {\bibinfo {title} {Emission
  of submillimeter electromagnetic waves by coherent phonons},\ }\href@noop {}
  {\bibfield  {journal} {\bibinfo  {journal} {Phys. Rev. Lett.}\ }\textbf
  {\bibinfo {volume} {74}},\ \bibinfo {pages} {738} (\bibinfo {year}
  {1995})}\BibitemShut {NoStop}%
\bibitem [{\citenamefont {Hase}\ \emph {et~al.}(2005)\citenamefont {Hase},
  \citenamefont {Ishioka}, \citenamefont {Demsar}, \citenamefont {Ushida},\
  and\ \citenamefont {Kitajima}}]{hase2005ultrafast}%
  \BibitemOpen
  \bibfield  {author} {\bibinfo {author} {\bibfnamefont {M.}~\bibnamefont
  {Hase}}, \bibinfo {author} {\bibfnamefont {K.}~\bibnamefont {Ishioka}},
  \bibinfo {author} {\bibfnamefont {J.}~\bibnamefont {Demsar}}, \bibinfo
  {author} {\bibfnamefont {K.}~\bibnamefont {Ushida}},\ and\ \bibinfo {author}
  {\bibfnamefont {M.}~\bibnamefont {Kitajima}},\ }\bibfield  {title} {\bibinfo
  {title} {Ultrafast dynamics of coherent optical phonons and nonequilibrium
  electrons in transition metals},\ }\href@noop {} {\bibfield  {journal}
  {\bibinfo  {journal} {Phys. Rev. B}\ }\textbf {\bibinfo {volume} {71}},\
  \bibinfo {pages} {184301} (\bibinfo {year} {2005})}\BibitemShut {NoStop}%
\bibitem [{\citenamefont {Ishioka}\ \emph {et~al.}(2008)\citenamefont
  {Ishioka}, \citenamefont {Kitajima},\ and\ \citenamefont
  {Misochko}}]{ishioka2008coherent}%
  \BibitemOpen
  \bibfield  {author} {\bibinfo {author} {\bibfnamefont {K.}~\bibnamefont
  {Ishioka}}, \bibinfo {author} {\bibfnamefont {M.}~\bibnamefont {Kitajima}},\
  and\ \bibinfo {author} {\bibfnamefont {O.~V.}\ \bibnamefont {Misochko}},\
  }\bibfield  {title} {\bibinfo {title} {Coherent $\mathrm{A_{1g}}$ and
  $\mathrm{E_g}$ phonons of antimony},\ }\href@noop {} {\bibfield  {journal}
  {\bibinfo  {journal} {J. Appl. Phys.}\ }\textbf {\bibinfo {volume} {103}},\
  \bibinfo {pages} {123505} (\bibinfo {year} {2008})}\BibitemShut {NoStop}%
\bibitem [{\citenamefont {Chatelain}\ \emph {et~al.}(2014)\citenamefont
  {Chatelain}, \citenamefont {Morrison}, \citenamefont {Klarenaar},\ and\
  \citenamefont {Siwick}}]{chatelain2014coherent}%
  \BibitemOpen
  \bibfield  {author} {\bibinfo {author} {\bibfnamefont {R.~P.}\ \bibnamefont
  {Chatelain}}, \bibinfo {author} {\bibfnamefont {V.~R.}\ \bibnamefont
  {Morrison}}, \bibinfo {author} {\bibfnamefont {B.~L.}\ \bibnamefont
  {Klarenaar}},\ and\ \bibinfo {author} {\bibfnamefont {B.~J.}\ \bibnamefont
  {Siwick}},\ }\bibfield  {title} {\bibinfo {title} {Coherent and incoherent
  electron-phonon coupling in graphite observed with radio-frequency compressed
  ultrafast electron diffraction},\ }\href@noop {} {\bibfield  {journal}
  {\bibinfo  {journal} {Phys. Rev. Lett.}\ }\textbf {\bibinfo {volume} {113}},\
  \bibinfo {pages} {235502} (\bibinfo {year} {2014})}\BibitemShut {NoStop}%
\bibitem [{\citenamefont {Waldecker}\ \emph {et~al.}(2017)\citenamefont
  {Waldecker}, \citenamefont {Vasileiadis}, \citenamefont {Bertoni},
  \citenamefont {Ernstorfer}, \citenamefont {Zier}, \citenamefont {Valencia},
  \citenamefont {Garcia},\ and\ \citenamefont
  {Zijlstra}}]{waldecker2017coherent}%
  \BibitemOpen
  \bibfield  {author} {\bibinfo {author} {\bibfnamefont {L.}~\bibnamefont
  {Waldecker}}, \bibinfo {author} {\bibfnamefont {T.}~\bibnamefont
  {Vasileiadis}}, \bibinfo {author} {\bibfnamefont {R.}~\bibnamefont
  {Bertoni}}, \bibinfo {author} {\bibfnamefont {R.}~\bibnamefont {Ernstorfer}},
  \bibinfo {author} {\bibfnamefont {T.}~\bibnamefont {Zier}}, \bibinfo {author}
  {\bibfnamefont {F.~H.}\ \bibnamefont {Valencia}}, \bibinfo {author}
  {\bibfnamefont {M.~E.}\ \bibnamefont {Garcia}},\ and\ \bibinfo {author}
  {\bibfnamefont {E.~S.}\ \bibnamefont {Zijlstra}},\ }\bibfield  {title}
  {\bibinfo {title} {Coherent and incoherent structural dynamics in
  laser-excited antimony},\ }\href@noop {} {\bibfield  {journal} {\bibinfo
  {journal} {Phys. Rev. B}\ }\textbf {\bibinfo {volume} {95}},\ \bibinfo
  {pages} {054302} (\bibinfo {year} {2017})}\BibitemShut {NoStop}%
\bibitem [{\citenamefont {Gerber}\ \emph {et~al.}(2017)\citenamefont {Gerber},
  \citenamefont {Yang}, \citenamefont {Zhu}, \citenamefont {Soifer},
  \citenamefont {Sobota}, \citenamefont {Rebec}, \citenamefont {Lee},
  \citenamefont {Jia}, \citenamefont {Moritz}, \citenamefont {Jia},
  \citenamefont {Gauthier}, \citenamefont {Li}, \citenamefont {Leuenberger},
  \citenamefont {Zhang}, \citenamefont {Chaix}, \citenamefont {Li},
  \citenamefont {Jang}, \citenamefont {Lee}, \citenamefont {Yi}, \citenamefont
  {Dakovski}, \citenamefont {Song}, \citenamefont {Glownia}, \citenamefont
  {Nelson}, \citenamefont {Kim}, \citenamefont {Chuang}, \citenamefont
  {Hussain}, \citenamefont {Moore}, \citenamefont {Devereaux}, \citenamefont
  {Lee}, \citenamefont {Kirchmann},\ and\ \citenamefont
  {Shen}}]{gerber2017femtosecond}%
  \BibitemOpen
  \bibfield  {author} {\bibinfo {author} {\bibfnamefont {S.}~\bibnamefont
  {Gerber}}, \bibinfo {author} {\bibfnamefont {S.-L.}\ \bibnamefont {Yang}},
  \bibinfo {author} {\bibfnamefont {D.}~\bibnamefont {Zhu}}, \bibinfo {author}
  {\bibfnamefont {H.}~\bibnamefont {Soifer}}, \bibinfo {author} {\bibfnamefont
  {J.~A.}\ \bibnamefont {Sobota}}, \bibinfo {author} {\bibfnamefont
  {S.}~\bibnamefont {Rebec}}, \bibinfo {author} {\bibfnamefont {J.~J.}\
  \bibnamefont {Lee}}, \bibinfo {author} {\bibfnamefont {T.}~\bibnamefont
  {Jia}}, \bibinfo {author} {\bibfnamefont {B.}~\bibnamefont {Moritz}},
  \bibinfo {author} {\bibfnamefont {C.}~\bibnamefont {Jia}}, \bibinfo {author}
  {\bibfnamefont {A.}~\bibnamefont {Gauthier}}, \bibinfo {author}
  {\bibfnamefont {Y.}~\bibnamefont {Li}}, \bibinfo {author} {\bibfnamefont
  {D.}~\bibnamefont {Leuenberger}}, \bibinfo {author} {\bibfnamefont
  {Y.}~\bibnamefont {Zhang}}, \bibinfo {author} {\bibfnamefont
  {L.}~\bibnamefont {Chaix}}, \bibinfo {author} {\bibfnamefont
  {W.}~\bibnamefont {Li}}, \bibinfo {author} {\bibfnamefont {H.}~\bibnamefont
  {Jang}}, \bibinfo {author} {\bibfnamefont {J.-S.}\ \bibnamefont {Lee}},
  \bibinfo {author} {\bibfnamefont {M.}~\bibnamefont {Yi}}, \bibinfo {author}
  {\bibfnamefont {G.~L.}\ \bibnamefont {Dakovski}}, \bibinfo {author}
  {\bibfnamefont {S.}~\bibnamefont {Song}}, \bibinfo {author} {\bibfnamefont
  {J.~M.}\ \bibnamefont {Glownia}}, \bibinfo {author} {\bibfnamefont
  {S.}~\bibnamefont {Nelson}}, \bibinfo {author} {\bibfnamefont {K.~W.}\
  \bibnamefont {Kim}}, \bibinfo {author} {\bibfnamefont {Y.-D.}\ \bibnamefont
  {Chuang}}, \bibinfo {author} {\bibfnamefont {Z.}~\bibnamefont {Hussain}},
  \bibinfo {author} {\bibfnamefont {R.~G.}\ \bibnamefont {Moore}}, \bibinfo
  {author} {\bibfnamefont {T.~P.}\ \bibnamefont {Devereaux}}, \bibinfo {author}
  {\bibfnamefont {W.-S.}\ \bibnamefont {Lee}}, \bibinfo {author} {\bibfnamefont
  {P.~S.}\ \bibnamefont {Kirchmann}},\ and\ \bibinfo {author} {\bibfnamefont
  {Z.-X.}\ \bibnamefont {Shen}},\ }\bibfield  {title} {\bibinfo {title}
  {Femtosecond electron-phonon lock-in by photoemission and x-ray free-electron
  laser},\ }\href {https://doi.org/10.1126/science.aak9946} {\bibfield
  {journal} {\bibinfo  {journal} {Science}\ }\textbf {\bibinfo {volume}
  {357}},\ \bibinfo {pages} {71} (\bibinfo {year} {2017})}\BibitemShut
  {NoStop}%
\bibitem [{\citenamefont {Perfetti}\ \emph {et~al.}(2006)\citenamefont
  {Perfetti}, \citenamefont {Loukakos}, \citenamefont {Lisowski}, \citenamefont
  {Bovensiepen}, \citenamefont {Berger}, \citenamefont {Biermann},
  \citenamefont {Cornaglia}, \citenamefont {Georges},\ and\ \citenamefont
  {Wolf}}]{perfetti2006time}%
  \BibitemOpen
  \bibfield  {author} {\bibinfo {author} {\bibfnamefont {L.}~\bibnamefont
  {Perfetti}}, \bibinfo {author} {\bibfnamefont {P.}~\bibnamefont {Loukakos}},
  \bibinfo {author} {\bibfnamefont {M.}~\bibnamefont {Lisowski}}, \bibinfo
  {author} {\bibfnamefont {U.}~\bibnamefont {Bovensiepen}}, \bibinfo {author}
  {\bibfnamefont {H.}~\bibnamefont {Berger}}, \bibinfo {author} {\bibfnamefont
  {S.}~\bibnamefont {Biermann}}, \bibinfo {author} {\bibfnamefont
  {P.}~\bibnamefont {Cornaglia}}, \bibinfo {author} {\bibfnamefont
  {A.}~\bibnamefont {Georges}},\ and\ \bibinfo {author} {\bibfnamefont
  {M.}~\bibnamefont {Wolf}},\ }\bibfield  {title} {\bibinfo {title} {Time
  evolution of the electronic structure of 1\textit{T}-\textrm{TaS}$_{2}$
  through the insulator-metal transition},\ }\href@noop {} {\bibfield
  {journal} {\bibinfo  {journal} {Phys. Rev. Lett.}\ }\textbf {\bibinfo
  {volume} {97}},\ \bibinfo {pages} {067402} (\bibinfo {year}
  {2006})}\BibitemShut {NoStop}%
\bibitem [{\citenamefont {Petersen}\ \emph {et~al.}(2011)\citenamefont
  {Petersen}, \citenamefont {Kaiser}, \citenamefont {Dean}, \citenamefont
  {Simoncig}, \citenamefont {Liu}, \citenamefont {Cavalieri}, \citenamefont
  {Cacho}, \citenamefont {Turcu}, \citenamefont {Springate}, \citenamefont
  {Frassetto}, \citenamefont {Poletto}, \citenamefont {Dhesi}, \citenamefont
  {Berger},\ and\ \citenamefont {Cavalleri}}]{Petersen2011}%
  \BibitemOpen
  \bibfield  {author} {\bibinfo {author} {\bibfnamefont {J.}~\bibnamefont
  {Petersen}}, \bibinfo {author} {\bibfnamefont {S.}~\bibnamefont {Kaiser}},
  \bibinfo {author} {\bibfnamefont {N.}~\bibnamefont {Dean}}, \bibinfo {author}
  {\bibfnamefont {a.}~\bibnamefont {Simoncig}}, \bibinfo {author}
  {\bibfnamefont {H.}~\bibnamefont {Liu}}, \bibinfo {author} {\bibfnamefont
  {a.}~\bibnamefont {Cavalieri}}, \bibinfo {author} {\bibfnamefont
  {C.}~\bibnamefont {Cacho}}, \bibinfo {author} {\bibfnamefont
  {I.}~\bibnamefont {Turcu}}, \bibinfo {author} {\bibfnamefont
  {E.}~\bibnamefont {Springate}}, \bibinfo {author} {\bibfnamefont
  {F.}~\bibnamefont {Frassetto}}, \bibinfo {author} {\bibfnamefont
  {L.}~\bibnamefont {Poletto}}, \bibinfo {author} {\bibfnamefont
  {S.}~\bibnamefont {Dhesi}}, \bibinfo {author} {\bibfnamefont
  {H.}~\bibnamefont {Berger}},\ and\ \bibinfo {author} {\bibfnamefont
  {a.}~\bibnamefont {Cavalleri}},\ }\bibfield  {title} {\bibinfo {title}
  {{Clocking the Melting Transition of Charge and Lattice Order in
  1\textit{T}-\textrm{TaS}$_{2}$ with Ultrafast Extreme-Ultraviolet
  Angle-Resolved Photoemission Spectroscopy}},\ }\href
  {https://doi.org/10.1103/PhysRevLett.107.177402} {\bibfield  {journal}
  {\bibinfo  {journal} {Phys. Rev. Lett.}\ }\textbf {\bibinfo {volume} {107}},\
  \bibinfo {pages} {1} (\bibinfo {year} {2011})}\BibitemShut {NoStop}%
\bibitem [{\citenamefont {Hellmann}\ \emph {et~al.}(2012)\citenamefont
  {Hellmann}, \citenamefont {Rohwer}, \citenamefont {Kall{\"{a}}ne},
  \citenamefont {Hanff}, \citenamefont {Sohrt}, \citenamefont {Stange},
  \citenamefont {Carr}, \citenamefont {Murnane}, \citenamefont {Kapteyn},
  \citenamefont {Kipp}, \citenamefont {Bauer},\ and\ \citenamefont
  {Rossnagel}}]{Hellmann2012}%
  \BibitemOpen
  \bibfield  {author} {\bibinfo {author} {\bibfnamefont {S.}~\bibnamefont
  {Hellmann}}, \bibinfo {author} {\bibfnamefont {T.}~\bibnamefont {Rohwer}},
  \bibinfo {author} {\bibfnamefont {M.}~\bibnamefont {Kall{\"{a}}ne}}, \bibinfo
  {author} {\bibfnamefont {K.}~\bibnamefont {Hanff}}, \bibinfo {author}
  {\bibfnamefont {C.}~\bibnamefont {Sohrt}}, \bibinfo {author} {\bibfnamefont
  {A.}~\bibnamefont {Stange}}, \bibinfo {author} {\bibfnamefont
  {A.}~\bibnamefont {Carr}}, \bibinfo {author} {\bibfnamefont {M.}~\bibnamefont
  {Murnane}}, \bibinfo {author} {\bibfnamefont {H.}~\bibnamefont {Kapteyn}},
  \bibinfo {author} {\bibfnamefont {L.}~\bibnamefont {Kipp}}, \bibinfo {author}
  {\bibfnamefont {M.}~\bibnamefont {Bauer}},\ and\ \bibinfo {author}
  {\bibfnamefont {K.}~\bibnamefont {Rossnagel}},\ }\bibfield  {title} {\bibinfo
  {title} {{Time-domain classification of charge-density-wave insulators}},\
  }\href@noop {} {\bibfield  {journal} {\bibinfo  {journal} {Nat. Commun.}\
  }\textbf {\bibinfo {volume} {3}},\ \bibinfo {pages} {1069} (\bibinfo {year}
  {2012})}\BibitemShut {NoStop}%
\bibitem [{\citenamefont {Papalazarou}\ \emph {et~al.}(2012)\citenamefont
  {Papalazarou}, \citenamefont {Faure}, \citenamefont {Mauchain}, \citenamefont
  {Marsi}, \citenamefont {Taleb-Ibrahimi}, \citenamefont {Reshetnyak},
  \citenamefont {Van~Roekeghem}, \citenamefont {Timrov}, \citenamefont {Vast},
  \citenamefont {Arnaud},\ and\ \citenamefont
  {Perfetti}}]{papalazarou2012coherent}%
  \BibitemOpen
  \bibfield  {author} {\bibinfo {author} {\bibfnamefont {E.}~\bibnamefont
  {Papalazarou}}, \bibinfo {author} {\bibfnamefont {J.}~\bibnamefont {Faure}},
  \bibinfo {author} {\bibfnamefont {J.}~\bibnamefont {Mauchain}}, \bibinfo
  {author} {\bibfnamefont {M.}~\bibnamefont {Marsi}}, \bibinfo {author}
  {\bibfnamefont {A.}~\bibnamefont {Taleb-Ibrahimi}}, \bibinfo {author}
  {\bibfnamefont {I.}~\bibnamefont {Reshetnyak}}, \bibinfo {author}
  {\bibfnamefont {A.}~\bibnamefont {Van~Roekeghem}}, \bibinfo {author}
  {\bibfnamefont {I.}~\bibnamefont {Timrov}}, \bibinfo {author} {\bibfnamefont
  {N.}~\bibnamefont {Vast}}, \bibinfo {author} {\bibfnamefont {B.}~\bibnamefont
  {Arnaud}},\ and\ \bibinfo {author} {\bibfnamefont {L.}~\bibnamefont
  {Perfetti}},\ }\bibfield  {title} {\bibinfo {title} {Coherent phonon coupling
  to individual bloch states in photoexcited bismuth},\ }\href@noop {}
  {\bibfield  {journal} {\bibinfo  {journal} {Phys. Rev. Lett.}\ }\textbf
  {\bibinfo {volume} {108}},\ \bibinfo {pages} {256808} (\bibinfo {year}
  {2012})}\BibitemShut {NoStop}%
\bibitem [{\citenamefont {Sobota}\ \emph {et~al.}(2014)\citenamefont {Sobota},
  \citenamefont {Yang}, \citenamefont {Leuenberger}, \citenamefont {Kemper},
  \citenamefont {Analytis}, \citenamefont {Fisher}, \citenamefont {Kirchmann},
  \citenamefont {Devereaux},\ and\ \citenamefont
  {Shen}}]{sobota2014distinguishing}%
  \BibitemOpen
  \bibfield  {author} {\bibinfo {author} {\bibfnamefont {J.~A.}\ \bibnamefont
  {Sobota}}, \bibinfo {author} {\bibfnamefont {S.-L.}\ \bibnamefont {Yang}},
  \bibinfo {author} {\bibfnamefont {D.}~\bibnamefont {Leuenberger}}, \bibinfo
  {author} {\bibfnamefont {A.~F.}\ \bibnamefont {Kemper}}, \bibinfo {author}
  {\bibfnamefont {J.~G.}\ \bibnamefont {Analytis}}, \bibinfo {author}
  {\bibfnamefont {I.~R.}\ \bibnamefont {Fisher}}, \bibinfo {author}
  {\bibfnamefont {P.~S.}\ \bibnamefont {Kirchmann}}, \bibinfo {author}
  {\bibfnamefont {T.~P.}\ \bibnamefont {Devereaux}},\ and\ \bibinfo {author}
  {\bibfnamefont {Z.-X.}\ \bibnamefont {Shen}},\ }\bibfield  {title} {\bibinfo
  {title} {Distinguishing bulk and surface electron-phonon coupling in the
  topological insulator $\mathrm{Bi_2Se_3}$ using time-resolved photoemission
  spectroscopy},\ }\href@noop {} {\bibfield  {journal} {\bibinfo  {journal}
  {Phys. Rev. Lett.}\ }\textbf {\bibinfo {volume} {113}},\ \bibinfo {pages}
  {157401} (\bibinfo {year} {2014})}\BibitemShut {NoStop}%
\bibitem [{\citenamefont {Rettig}\ \emph {et~al.}(2014)\citenamefont {Rettig},
  \citenamefont {Chu}, \citenamefont {Fisher}, \citenamefont {Bovensiepen},\
  and\ \citenamefont {Wolf}}]{rettig2014coherent}%
  \BibitemOpen
  \bibfield  {author} {\bibinfo {author} {\bibfnamefont {L.}~\bibnamefont
  {Rettig}}, \bibinfo {author} {\bibfnamefont {J.-H.}\ \bibnamefont {Chu}},
  \bibinfo {author} {\bibfnamefont {I.}~\bibnamefont {Fisher}}, \bibinfo
  {author} {\bibfnamefont {U.}~\bibnamefont {Bovensiepen}},\ and\ \bibinfo
  {author} {\bibfnamefont {M.}~\bibnamefont {Wolf}},\ }\bibfield  {title}
  {\bibinfo {title} {Coherent dynamics of the charge density wave gap in
  tritellurides},\ }\href@noop {} {\bibfield  {journal} {\bibinfo  {journal}
  {Faraday Discuss.}\ }\textbf {\bibinfo {volume} {171}},\ \bibinfo {pages}
  {299} (\bibinfo {year} {2014})}\BibitemShut {NoStop}%
\bibitem [{\citenamefont {Golias}\ and\ \citenamefont
  {S{\'{a}}nchez-Barriga}(2016)}]{Golias2016}%
  \BibitemOpen
  \bibfield  {author} {\bibinfo {author} {\bibfnamefont {E.}~\bibnamefont
  {Golias}}\ and\ \bibinfo {author} {\bibfnamefont {J.}~\bibnamefont
  {S{\'{a}}nchez-Barriga}},\ }\bibfield  {title} {\bibinfo {title} {Observation
  of antiphase coherent phonons in the warped {D}irac cone of
  $\mathrm{Bi_2Te_3}$},\ }\href@noop {} {\bibfield  {journal} {\bibinfo
  {journal} {Phys. Rev. B}\ }\textbf {\bibinfo {volume} {94}},\ \bibinfo
  {pages} {161113(R)} (\bibinfo {year} {2016})}\BibitemShut {NoStop}%
\bibitem [{\citenamefont {Tang}\ \emph {et~al.}(2020)\citenamefont {Tang},
  \citenamefont {Wang}, \citenamefont {Duan}, \citenamefont {Yang},
  \citenamefont {Huang}, \citenamefont {Guo}, \citenamefont {Qian},\ and\
  \citenamefont {Zhang}}]{Tang2020}%
  \BibitemOpen
  \bibfield  {author} {\bibinfo {author} {\bibfnamefont {T.}~\bibnamefont
  {Tang}}, \bibinfo {author} {\bibfnamefont {H.}~\bibnamefont {Wang}}, \bibinfo
  {author} {\bibfnamefont {S.}~\bibnamefont {Duan}}, \bibinfo {author}
  {\bibfnamefont {Y.}~\bibnamefont {Yang}}, \bibinfo {author} {\bibfnamefont
  {C.}~\bibnamefont {Huang}}, \bibinfo {author} {\bibfnamefont
  {Y.}~\bibnamefont {Guo}}, \bibinfo {author} {\bibfnamefont {D.}~\bibnamefont
  {Qian}},\ and\ \bibinfo {author} {\bibfnamefont {W.}~\bibnamefont {Zhang}},\
  }\bibfield  {title} {\bibinfo {title} {{Non-{Coulomb} strong electron-hole
  binding in $\mathrm{Ta_2NiSe_5}$ revealed by time- and angle-resolved
  photoemission spectroscopy}},\ }\href
  {https://doi.org/10.1103/PhysRevB.101.235148} {\bibfield  {journal} {\bibinfo
   {journal} {Phys. Rev. B}\ }\textbf {\bibinfo {volume} {101}},\ \bibinfo
  {pages} {235148} (\bibinfo {year} {2020})}\BibitemShut {NoStop}%
\bibitem [{\citenamefont {Baldini}\ \emph {et~al.}(2023)\citenamefont
  {Baldini}, \citenamefont {Zong}, \citenamefont {Choi}, \citenamefont {Lee},
  \citenamefont {Michael}, \citenamefont {Windgaetter}, \citenamefont {Mazin},
  \citenamefont {Latini}, \citenamefont {Azoury}, \citenamefont {Lv},
  \citenamefont {Kogar}, \citenamefont {Su}, \citenamefont {Wang},
  \citenamefont {Lu}, \citenamefont {Takayama}, \citenamefont {Takagi},
  \citenamefont {Millis}, \citenamefont {Rubio}, \citenamefont {Demler},\ and\
  \citenamefont {Gedik}}]{Baldini2023}%
  \BibitemOpen
  \bibfield  {author} {\bibinfo {author} {\bibfnamefont {E.}~\bibnamefont
  {Baldini}}, \bibinfo {author} {\bibfnamefont {A.}~\bibnamefont {Zong}},
  \bibinfo {author} {\bibfnamefont {D.}~\bibnamefont {Choi}}, \bibinfo {author}
  {\bibfnamefont {C.}~\bibnamefont {Lee}}, \bibinfo {author} {\bibfnamefont
  {M.~H.}\ \bibnamefont {Michael}}, \bibinfo {author} {\bibfnamefont
  {L.}~\bibnamefont {Windgaetter}}, \bibinfo {author} {\bibfnamefont {I.~I.}\
  \bibnamefont {Mazin}}, \bibinfo {author} {\bibfnamefont {S.}~\bibnamefont
  {Latini}}, \bibinfo {author} {\bibfnamefont {D.}~\bibnamefont {Azoury}},
  \bibinfo {author} {\bibfnamefont {B.}~\bibnamefont {Lv}}, \bibinfo {author}
  {\bibfnamefont {A.}~\bibnamefont {Kogar}}, \bibinfo {author} {\bibfnamefont
  {Y.}~\bibnamefont {Su}}, \bibinfo {author} {\bibfnamefont {Y.}~\bibnamefont
  {Wang}}, \bibinfo {author} {\bibfnamefont {Y.}~\bibnamefont {Lu}}, \bibinfo
  {author} {\bibfnamefont {T.}~\bibnamefont {Takayama}}, \bibinfo {author}
  {\bibfnamefont {H.}~\bibnamefont {Takagi}}, \bibinfo {author} {\bibfnamefont
  {A.~J.}\ \bibnamefont {Millis}}, \bibinfo {author} {\bibfnamefont
  {A.}~\bibnamefont {Rubio}}, \bibinfo {author} {\bibfnamefont
  {E.}~\bibnamefont {Demler}},\ and\ \bibinfo {author} {\bibfnamefont
  {N.}~\bibnamefont {Gedik}},\ }\bibfield  {title} {\bibinfo {title} {{The
  spontaneous symmetry breaking in $\mathrm{Ta_2NiSe_5}$ is structural in
  nature}},\ }\href {https://doi.org/10.1073/pnas.2221688120} {\bibfield
  {journal} {\bibinfo  {journal} {Proc. Natl. Acad. Sci.}\ }\textbf {\bibinfo
  {volume} {120}},\ \bibinfo {pages} {1} (\bibinfo {year} {2023})}\BibitemShut
  {NoStop}%
\bibitem [{\citenamefont {Hein}\ \emph {et~al.}(2020)\citenamefont {Hein},
  \citenamefont {Jauernik}, \citenamefont {Erk}, \citenamefont {Yang},
  \citenamefont {Qi}, \citenamefont {Sun}, \citenamefont {Felser},\ and\
  \citenamefont {Bauer}}]{hein2020mode}%
  \BibitemOpen
  \bibfield  {author} {\bibinfo {author} {\bibfnamefont {P.}~\bibnamefont
  {Hein}}, \bibinfo {author} {\bibfnamefont {S.}~\bibnamefont {Jauernik}},
  \bibinfo {author} {\bibfnamefont {H.}~\bibnamefont {Erk}}, \bibinfo {author}
  {\bibfnamefont {L.}~\bibnamefont {Yang}}, \bibinfo {author} {\bibfnamefont
  {Y.}~\bibnamefont {Qi}}, \bibinfo {author} {\bibfnamefont {Y.}~\bibnamefont
  {Sun}}, \bibinfo {author} {\bibfnamefont {C.}~\bibnamefont {Felser}},\ and\
  \bibinfo {author} {\bibfnamefont {M.}~\bibnamefont {Bauer}},\ }\bibfield
  {title} {\bibinfo {title} {Mode-resolved reciprocal space mapping of
  electron-phonon interaction in the {Weyl} semimetal candidate
  \textit{Td}-$\mathrm{WTe_2}$},\ }\href@noop {} {\bibfield  {journal}
  {\bibinfo  {journal} {Nat. Commun.}\ }\textbf {\bibinfo {volume} {11}},\
  \bibinfo {pages} {2613} (\bibinfo {year} {2020})}\BibitemShut {NoStop}%
\bibitem [{\citenamefont {De~Giovannini}\ \emph {et~al.}(2020)\citenamefont
  {De~Giovannini}, \citenamefont {H{\"u}bener}, \citenamefont {Sato},\ and\
  \citenamefont {Rubio}}]{de2020direct}%
  \BibitemOpen
  \bibfield  {author} {\bibinfo {author} {\bibfnamefont {U.}~\bibnamefont
  {De~Giovannini}}, \bibinfo {author} {\bibfnamefont {H.}~\bibnamefont
  {H{\"u}bener}}, \bibinfo {author} {\bibfnamefont {S.~A.}\ \bibnamefont
  {Sato}},\ and\ \bibinfo {author} {\bibfnamefont {A.}~\bibnamefont {Rubio}},\
  }\bibfield  {title} {\bibinfo {title} {Direct measurement of electron-phonon
  coupling with time-resolved {ARPES}},\ }\href@noop {} {\bibfield  {journal}
  {\bibinfo  {journal} {Phys. Rev. Lett.}\ }\textbf {\bibinfo {volume} {125}},\
  \bibinfo {pages} {136401} (\bibinfo {year} {2020})}\BibitemShut {NoStop}%
\bibitem [{\citenamefont {Suzuki}\ \emph {et~al.}(2021)\citenamefont {Suzuki},
  \citenamefont {Shinohara}, \citenamefont {Lu}, \citenamefont {Watanabe},
  \citenamefont {Xu}, \citenamefont {Ishikawa}, \citenamefont {Takagi},
  \citenamefont {Nohara}, \citenamefont {Katayama}, \citenamefont {Sawa},
  \citenamefont {Fujisawa}, \citenamefont {Kanai}, \citenamefont {Itatani},
  \citenamefont {Mizokawa}, \citenamefont {Shin},\ and\ \citenamefont
  {Okazaki}}]{suzuki2021detecting}%
  \BibitemOpen
  \bibfield  {author} {\bibinfo {author} {\bibfnamefont {T.}~\bibnamefont
  {Suzuki}}, \bibinfo {author} {\bibfnamefont {Y.}~\bibnamefont {Shinohara}},
  \bibinfo {author} {\bibfnamefont {Y.}~\bibnamefont {Lu}}, \bibinfo {author}
  {\bibfnamefont {M.}~\bibnamefont {Watanabe}}, \bibinfo {author}
  {\bibfnamefont {J.}~\bibnamefont {Xu}}, \bibinfo {author} {\bibfnamefont
  {K.~L.}\ \bibnamefont {Ishikawa}}, \bibinfo {author} {\bibfnamefont
  {H.}~\bibnamefont {Takagi}}, \bibinfo {author} {\bibfnamefont
  {M.}~\bibnamefont {Nohara}}, \bibinfo {author} {\bibfnamefont
  {N.}~\bibnamefont {Katayama}}, \bibinfo {author} {\bibfnamefont
  {H.}~\bibnamefont {Sawa}}, \bibinfo {author} {\bibfnamefont {M.}~\bibnamefont
  {Fujisawa}}, \bibinfo {author} {\bibfnamefont {T.}~\bibnamefont {Kanai}},
  \bibinfo {author} {\bibfnamefont {J.}~\bibnamefont {Itatani}}, \bibinfo
  {author} {\bibfnamefont {T.}~\bibnamefont {Mizokawa}}, \bibinfo {author}
  {\bibfnamefont {S.}~\bibnamefont {Shin}},\ and\ \bibinfo {author}
  {\bibfnamefont {K.}~\bibnamefont {Okazaki}},\ }\bibfield  {title} {\bibinfo
  {title} {Detecting electron-phonon coupling during photoinduced phase
  transition},\ }\href@noop {} {\bibfield  {journal} {\bibinfo  {journal}
  {Phys. Rev. B}\ }\textbf {\bibinfo {volume} {103}},\ \bibinfo {pages}
  {L121105} (\bibinfo {year} {2021})}\BibitemShut {NoStop}%
\bibitem [{\citenamefont {Lee}\ \emph {et~al.}(2023{\natexlab{a}})\citenamefont
  {Lee}, \citenamefont {Fernandez-Mulligan}, \citenamefont {Tan}, \citenamefont
  {Yan}, \citenamefont {Guan}, \citenamefont {Lee}, \citenamefont {Mei},
  \citenamefont {Liu}, \citenamefont {Yan}, \citenamefont {Mao},\ and\
  \citenamefont {Yang}}]{lee2023layer}%
  \BibitemOpen
  \bibfield  {author} {\bibinfo {author} {\bibfnamefont {W.}~\bibnamefont
  {Lee}}, \bibinfo {author} {\bibfnamefont {S.}~\bibnamefont
  {Fernandez-Mulligan}}, \bibinfo {author} {\bibfnamefont {H.}~\bibnamefont
  {Tan}}, \bibinfo {author} {\bibfnamefont {C.}~\bibnamefont {Yan}}, \bibinfo
  {author} {\bibfnamefont {Y.}~\bibnamefont {Guan}}, \bibinfo {author}
  {\bibfnamefont {S.~H.}\ \bibnamefont {Lee}}, \bibinfo {author} {\bibfnamefont
  {R.}~\bibnamefont {Mei}}, \bibinfo {author} {\bibfnamefont {C.}~\bibnamefont
  {Liu}}, \bibinfo {author} {\bibfnamefont {B.}~\bibnamefont {Yan}}, \bibinfo
  {author} {\bibfnamefont {Z.}~\bibnamefont {Mao}},\ and\ \bibinfo {author}
  {\bibfnamefont {S.}~\bibnamefont {Yang}},\ }\bibfield  {title} {\bibinfo
  {title} {Layer-by-layer disentanglement of {Bloch} states},\ }\href@noop {}
  {\bibfield  {journal} {\bibinfo  {journal} {Nat. Phys.}\ }\textbf {\bibinfo
  {volume} {19}},\ \bibinfo {pages} {950} (\bibinfo {year}
  {2023}{\natexlab{a}})}\BibitemShut {NoStop}%
\bibitem [{\citenamefont {Ren}\ \emph {et~al.}(2023)\citenamefont {Ren},
  \citenamefont {Suzuki}, \citenamefont {Kanai}, \citenamefont {Itatani},
  \citenamefont {Shin},\ and\ \citenamefont {Okazaki}}]{ren2023phase}%
  \BibitemOpen
  \bibfield  {author} {\bibinfo {author} {\bibfnamefont {Q.}~\bibnamefont
  {Ren}}, \bibinfo {author} {\bibfnamefont {T.}~\bibnamefont {Suzuki}},
  \bibinfo {author} {\bibfnamefont {T.}~\bibnamefont {Kanai}}, \bibinfo
  {author} {\bibfnamefont {J.}~\bibnamefont {Itatani}}, \bibinfo {author}
  {\bibfnamefont {S.}~\bibnamefont {Shin}},\ and\ \bibinfo {author}
  {\bibfnamefont {K.}~\bibnamefont {Okazaki}},\ }\bibfield  {title} {\bibinfo
  {title} {Phase-resolved frequency-domain analysis of the photoemission
  spectra for photoexcited 1\textit{T}-\textrm{TaS}$_{2}$ in the {Mott}
  insulating charge density wave state},\ }\href@noop {} {\bibfield  {journal}
  {\bibinfo  {journal} {Appl. Phys. Lett.}\ }\textbf {\bibinfo {volume}
  {122}},\ \bibinfo {pages} {221902} (\bibinfo {year} {2023})}\BibitemShut
  {NoStop}%
\bibitem [{\citenamefont {Wall}\ \emph {et~al.}(2009)\citenamefont {Wall},
  \citenamefont {Prabhakaran}, \citenamefont {Boothroyd},\ and\ \citenamefont
  {Cavalleri}}]{wall2009ultrafast}%
  \BibitemOpen
  \bibfield  {author} {\bibinfo {author} {\bibfnamefont {S.}~\bibnamefont
  {Wall}}, \bibinfo {author} {\bibfnamefont {D.}~\bibnamefont {Prabhakaran}},
  \bibinfo {author} {\bibfnamefont {A.}~\bibnamefont {Boothroyd}},\ and\
  \bibinfo {author} {\bibfnamefont {A.}~\bibnamefont {Cavalleri}},\ }\bibfield
  {title} {\bibinfo {title} {Ultrafast coupling between light, coherent lattice
  vibrations, and the magnetic structure of semicovalent $\mathrm{LaMnO_3}$},\
  }\href@noop {} {\bibfield  {journal} {\bibinfo  {journal} {Phys. Rev. Lett.}\
  }\textbf {\bibinfo {volume} {103}},\ \bibinfo {pages} {097402} (\bibinfo
  {year} {2009})}\BibitemShut {NoStop}%
\bibitem [{\citenamefont {Kim}\ \emph {et~al.}(2012)\citenamefont {Kim},
  \citenamefont {Pashkin}, \citenamefont {Sch{\"a}fer}, \citenamefont {Beyer},
  \citenamefont {Porer}, \citenamefont {Wolf}, \citenamefont {Bernhard},
  \citenamefont {Demsar}, \citenamefont {Huber},\ and\ \citenamefont
  {Leitenstorfer}}]{kim2012ultrafast}%
  \BibitemOpen
  \bibfield  {author} {\bibinfo {author} {\bibfnamefont {K.~W.}\ \bibnamefont
  {Kim}}, \bibinfo {author} {\bibfnamefont {A.}~\bibnamefont {Pashkin}},
  \bibinfo {author} {\bibfnamefont {H.}~\bibnamefont {Sch{\"a}fer}}, \bibinfo
  {author} {\bibfnamefont {M.}~\bibnamefont {Beyer}}, \bibinfo {author}
  {\bibfnamefont {M.}~\bibnamefont {Porer}}, \bibinfo {author} {\bibfnamefont
  {T.}~\bibnamefont {Wolf}}, \bibinfo {author} {\bibfnamefont {C.}~\bibnamefont
  {Bernhard}}, \bibinfo {author} {\bibfnamefont {J.}~\bibnamefont {Demsar}},
  \bibinfo {author} {\bibfnamefont {R.}~\bibnamefont {Huber}},\ and\ \bibinfo
  {author} {\bibfnamefont {A.}~\bibnamefont {Leitenstorfer}},\ }\bibfield
  {title} {\bibinfo {title} {Ultrafast transient generation of
  spin-density-wave order in the normal state of $\mathrm{BaFe_2As_2}$ driven
  by coherent lattice vibrations},\ }\href@noop {} {\bibfield  {journal}
  {\bibinfo  {journal} {Nat. Mater.}\ }\textbf {\bibinfo {volume} {11}},\
  \bibinfo {pages} {497} (\bibinfo {year} {2012})}\BibitemShut {NoStop}%
\bibitem [{\citenamefont {Nova}\ \emph {et~al.}(2017)\citenamefont {Nova},
  \citenamefont {Cartella}, \citenamefont {Cantaluppi}, \citenamefont
  {F{\"o}rst}, \citenamefont {Bossini}, \citenamefont {Mikhaylovskiy},
  \citenamefont {Kimel}, \citenamefont {Merlin},\ and\ \citenamefont
  {Cavalleri}}]{Nova2017}%
  \BibitemOpen
  \bibfield  {author} {\bibinfo {author} {\bibfnamefont {T.~F.}\ \bibnamefont
  {Nova}}, \bibinfo {author} {\bibfnamefont {A.}~\bibnamefont {Cartella}},
  \bibinfo {author} {\bibfnamefont {A.}~\bibnamefont {Cantaluppi}}, \bibinfo
  {author} {\bibfnamefont {M.}~\bibnamefont {F{\"o}rst}}, \bibinfo {author}
  {\bibfnamefont {D.}~\bibnamefont {Bossini}}, \bibinfo {author} {\bibfnamefont
  {R.~V.}\ \bibnamefont {Mikhaylovskiy}}, \bibinfo {author} {\bibfnamefont
  {A.~V.}\ \bibnamefont {Kimel}}, \bibinfo {author} {\bibfnamefont
  {R.}~\bibnamefont {Merlin}},\ and\ \bibinfo {author} {\bibfnamefont
  {A.}~\bibnamefont {Cavalleri}},\ }\bibfield  {title} {\bibinfo {title} {An
  effective magnetic field from optically driven phonons},\ }\href
  {https://doi.org/10.1038/nphys3925} {\bibfield  {journal} {\bibinfo
  {journal} {Nat. Phys.}\ }\textbf {\bibinfo {volume} {13}},\ \bibinfo {pages}
  {132} (\bibinfo {year} {2017})}\BibitemShut {NoStop}%
\bibitem [{\citenamefont {Afanasiev}\ \emph {et~al.}(2021)\citenamefont
  {Afanasiev}, \citenamefont {Hortensius}, \citenamefont {Ivanov},
  \citenamefont {Sasani}, \citenamefont {Bousquet}, \citenamefont {Blanter},
  \citenamefont {Mikhaylovskiy}, \citenamefont {Kimel},\ and\ \citenamefont
  {Caviglia}}]{afanasiev_ultrafast_2021}%
  \BibitemOpen
  \bibfield  {author} {\bibinfo {author} {\bibfnamefont {D.}~\bibnamefont
  {Afanasiev}}, \bibinfo {author} {\bibfnamefont {J.~R.}\ \bibnamefont
  {Hortensius}}, \bibinfo {author} {\bibfnamefont {B.~A.}\ \bibnamefont
  {Ivanov}}, \bibinfo {author} {\bibfnamefont {A.}~\bibnamefont {Sasani}},
  \bibinfo {author} {\bibfnamefont {E.}~\bibnamefont {Bousquet}}, \bibinfo
  {author} {\bibfnamefont {Y.~M.}\ \bibnamefont {Blanter}}, \bibinfo {author}
  {\bibfnamefont {R.~V.}\ \bibnamefont {Mikhaylovskiy}}, \bibinfo {author}
  {\bibfnamefont {A.~V.}\ \bibnamefont {Kimel}},\ and\ \bibinfo {author}
  {\bibfnamefont {A.~D.}\ \bibnamefont {Caviglia}},\ }\bibfield  {title}
  {\bibinfo {title} {Ultrafast control of magnetic interactions via
  light-driven phonons},\ }\href {https://doi.org/10.1038/s41563-021-00922-7}
  {\bibfield  {journal} {\bibinfo  {journal} {Nat. Mater.}\ }\textbf {\bibinfo
  {volume} {20}},\ \bibinfo {pages} {607} (\bibinfo {year} {2021})}\BibitemShut
  {NoStop}%
\bibitem [{\citenamefont {Bakker}\ \emph {et~al.}(1998)\citenamefont {Bakker},
  \citenamefont {Hunsche},\ and\ \citenamefont {Kurz}}]{bakker1998coherent}%
  \BibitemOpen
  \bibfield  {author} {\bibinfo {author} {\bibfnamefont {H.}~\bibnamefont
  {Bakker}}, \bibinfo {author} {\bibfnamefont {S.}~\bibnamefont {Hunsche}},\
  and\ \bibinfo {author} {\bibfnamefont {H.}~\bibnamefont {Kurz}},\ }\bibfield
  {title} {\bibinfo {title} {Coherent phonon polaritons as probes of anharmonic
  phonons in ferroelectrics},\ }\href@noop {} {\bibfield  {journal} {\bibinfo
  {journal} {Rev. Mod. Phys.}\ }\textbf {\bibinfo {volume} {70}},\ \bibinfo
  {pages} {523} (\bibinfo {year} {1998})}\BibitemShut {NoStop}%
\bibitem [{\citenamefont {Juraschek}\ \emph
  {et~al.}(2017{\natexlab{a}})\citenamefont {Juraschek}, \citenamefont
  {Fechner}, \citenamefont {Balatsky},\ and\ \citenamefont
  {Spaldin}}]{Juraschek2017}%
  \BibitemOpen
  \bibfield  {author} {\bibinfo {author} {\bibfnamefont {D.~M.}\ \bibnamefont
  {Juraschek}}, \bibinfo {author} {\bibfnamefont {M.}~\bibnamefont {Fechner}},
  \bibinfo {author} {\bibfnamefont {A.~V.}\ \bibnamefont {Balatsky}},\ and\
  \bibinfo {author} {\bibfnamefont {N.~A.}\ \bibnamefont {Spaldin}},\
  }\bibfield  {title} {\bibinfo {title} {Dynamical multiferroicity},\ }\href
  {https://doi.org/10.1103/PhysRevMaterials.1.014401} {\bibfield  {journal}
  {\bibinfo  {journal} {Phys. Rev. Mater.}\ }\textbf {\bibinfo {volume} {1}},\
  \bibinfo {pages} {014401} (\bibinfo {year} {2017}{\natexlab{a}})}\BibitemShut
  {NoStop}%
\bibitem [{\citenamefont {Mankowsky}\ \emph {et~al.}(2017)\citenamefont
  {Mankowsky}, \citenamefont {von Hoegen}, \citenamefont {F{\"o}rst},\ and\
  \citenamefont {Cavalleri}}]{mankowsky2017ultrafast}%
  \BibitemOpen
  \bibfield  {author} {\bibinfo {author} {\bibfnamefont {R.}~\bibnamefont
  {Mankowsky}}, \bibinfo {author} {\bibfnamefont {A.}~\bibnamefont {von
  Hoegen}}, \bibinfo {author} {\bibfnamefont {M.}~\bibnamefont {F{\"o}rst}},\
  and\ \bibinfo {author} {\bibfnamefont {A.}~\bibnamefont {Cavalleri}},\
  }\bibfield  {title} {\bibinfo {title} {Ultrafast reversal of the
  ferroelectric polarization},\ }\href@noop {} {\bibfield  {journal} {\bibinfo
  {journal} {Phys. Rev. Lett.}\ }\textbf {\bibinfo {volume} {118}},\ \bibinfo
  {pages} {197601} (\bibinfo {year} {2017})}\BibitemShut {NoStop}%
\bibitem [{\citenamefont {Fechner}\ \emph {et~al.}(2024)\citenamefont
  {Fechner}, \citenamefont {Först}, \citenamefont {Orenstein}, \citenamefont
  {Krapivin}, \citenamefont {Disa}, \citenamefont {Buzzi}, \citenamefont
  {Von~Hoegen}, \citenamefont {De~La~Pena}, \citenamefont {Nguyen},
  \citenamefont {Mankowsky}, \citenamefont {Sander}, \citenamefont {Lemke},
  \citenamefont {Deng}, \citenamefont {Trigo},\ and\ \citenamefont
  {Cavalleri}}]{fechner2024quenched}%
  \BibitemOpen
  \bibfield  {author} {\bibinfo {author} {\bibfnamefont {M.}~\bibnamefont
  {Fechner}}, \bibinfo {author} {\bibfnamefont {M.}~\bibnamefont {Först}},
  \bibinfo {author} {\bibfnamefont {G.}~\bibnamefont {Orenstein}}, \bibinfo
  {author} {\bibfnamefont {V.}~\bibnamefont {Krapivin}}, \bibinfo {author}
  {\bibfnamefont {A.~S.}\ \bibnamefont {Disa}}, \bibinfo {author}
  {\bibfnamefont {M.}~\bibnamefont {Buzzi}}, \bibinfo {author} {\bibfnamefont
  {A.}~\bibnamefont {Von~Hoegen}}, \bibinfo {author} {\bibfnamefont
  {G.}~\bibnamefont {De~La~Pena}}, \bibinfo {author} {\bibfnamefont {Q.~L.}\
  \bibnamefont {Nguyen}}, \bibinfo {author} {\bibfnamefont {R.}~\bibnamefont
  {Mankowsky}}, \bibinfo {author} {\bibfnamefont {M.}~\bibnamefont {Sander}},
  \bibinfo {author} {\bibfnamefont {H.}~\bibnamefont {Lemke}}, \bibinfo
  {author} {\bibfnamefont {Y.}~\bibnamefont {Deng}}, \bibinfo {author}
  {\bibfnamefont {M.}~\bibnamefont {Trigo}},\ and\ \bibinfo {author}
  {\bibfnamefont {A.}~\bibnamefont {Cavalleri}},\ }\bibfield  {title} {\bibinfo
  {title} {Quenched lattice fluctuations in optically driven
  $\mathrm{SrTiO_3}$},\ }\href@noop {} {\bibfield  {journal} {\bibinfo
  {journal} {Nat. Mater.}\ }\textbf {\bibinfo {volume} {23}},\ \bibinfo {pages}
  {363} (\bibinfo {year} {2024})}\BibitemShut {NoStop}%
\bibitem [{\citenamefont {Sie}\ \emph {et~al.}(2019)\citenamefont {Sie},
  \citenamefont {Nyby}, \citenamefont {Pemmaraju}, \citenamefont {Park},
  \citenamefont {Shen}, \citenamefont {Yang}, \citenamefont {Hoffmann},
  \citenamefont {Ofori-Okai}, \citenamefont {Li}, \citenamefont {Reid},
  \citenamefont {Weathersby}, \citenamefont {Mannebach}, \citenamefont
  {Finney}, \citenamefont {Rhodes}, \citenamefont {Chenet}, \citenamefont
  {Antony}, \citenamefont {Balicas}, \citenamefont {Hone}, \citenamefont
  {Devereaux}, \citenamefont {Heinz}, \citenamefont {Wang},\ and\ \citenamefont
  {Lindenberg}}]{sie2019ultrafast}%
  \BibitemOpen
  \bibfield  {author} {\bibinfo {author} {\bibfnamefont {E.~J.}\ \bibnamefont
  {Sie}}, \bibinfo {author} {\bibfnamefont {C.~M.}\ \bibnamefont {Nyby}},
  \bibinfo {author} {\bibfnamefont {C.~D.}\ \bibnamefont {Pemmaraju}}, \bibinfo
  {author} {\bibfnamefont {S.~J.}\ \bibnamefont {Park}}, \bibinfo {author}
  {\bibfnamefont {X.}~\bibnamefont {Shen}}, \bibinfo {author} {\bibfnamefont
  {J.}~\bibnamefont {Yang}}, \bibinfo {author} {\bibfnamefont {M.~C.}\
  \bibnamefont {Hoffmann}}, \bibinfo {author} {\bibfnamefont {B.~K.}\
  \bibnamefont {Ofori-Okai}}, \bibinfo {author} {\bibfnamefont
  {R.}~\bibnamefont {Li}}, \bibinfo {author} {\bibfnamefont {A.~H.}\
  \bibnamefont {Reid}}, \bibinfo {author} {\bibfnamefont {S.}~\bibnamefont
  {Weathersby}}, \bibinfo {author} {\bibfnamefont {E.}~\bibnamefont
  {Mannebach}}, \bibinfo {author} {\bibfnamefont {N.}~\bibnamefont {Finney}},
  \bibinfo {author} {\bibfnamefont {D.}~\bibnamefont {Rhodes}}, \bibinfo
  {author} {\bibfnamefont {D.}~\bibnamefont {Chenet}}, \bibinfo {author}
  {\bibfnamefont {A.}~\bibnamefont {Antony}}, \bibinfo {author} {\bibfnamefont
  {L.}~\bibnamefont {Balicas}}, \bibinfo {author} {\bibfnamefont
  {J.}~\bibnamefont {Hone}}, \bibinfo {author} {\bibfnamefont {T.~P.}\
  \bibnamefont {Devereaux}}, \bibinfo {author} {\bibfnamefont {T.~F.}\
  \bibnamefont {Heinz}}, \bibinfo {author} {\bibfnamefont {X.}~\bibnamefont
  {Wang}},\ and\ \bibinfo {author} {\bibfnamefont {A.~M.}\ \bibnamefont
  {Lindenberg}},\ }\bibfield  {title} {\bibinfo {title} {An ultrafast symmetry
  switch in a {Weyl} semimetal},\ }\href@noop {} {\bibfield  {journal}
  {\bibinfo  {journal} {Nature}\ }\textbf {\bibinfo {volume} {565}},\ \bibinfo
  {pages} {61} (\bibinfo {year} {2019})}\BibitemShut {NoStop}%
\bibitem [{\citenamefont {Mankowsky}\ \emph {et~al.}(2014)\citenamefont
  {Mankowsky}, \citenamefont {Subedi}, \citenamefont {F{\"o}rst}, \citenamefont
  {Mariager}, \citenamefont {Chollet}, \citenamefont {Lemke}, \citenamefont
  {Robinson}, \citenamefont {Glownia}, \citenamefont {Minitti}, \citenamefont
  {Frano}, \citenamefont {Fechner}, \citenamefont {Spaldin}, \citenamefont
  {Loew}, \citenamefont {Keimer}, \citenamefont {Georges},\ and\ \citenamefont
  {Cavalleri}}]{Mankowsky2014}%
  \BibitemOpen
  \bibfield  {author} {\bibinfo {author} {\bibfnamefont {R.}~\bibnamefont
  {Mankowsky}}, \bibinfo {author} {\bibfnamefont {A.}~\bibnamefont {Subedi}},
  \bibinfo {author} {\bibfnamefont {M.}~\bibnamefont {F{\"o}rst}}, \bibinfo
  {author} {\bibfnamefont {S.~O.}\ \bibnamefont {Mariager}}, \bibinfo {author}
  {\bibfnamefont {M.}~\bibnamefont {Chollet}}, \bibinfo {author} {\bibfnamefont
  {H.~T.}\ \bibnamefont {Lemke}}, \bibinfo {author} {\bibfnamefont {J.~S.}\
  \bibnamefont {Robinson}}, \bibinfo {author} {\bibfnamefont {J.~M.}\
  \bibnamefont {Glownia}}, \bibinfo {author} {\bibfnamefont {M.~P.}\
  \bibnamefont {Minitti}}, \bibinfo {author} {\bibfnamefont {A.}~\bibnamefont
  {Frano}}, \bibinfo {author} {\bibfnamefont {M.}~\bibnamefont {Fechner}},
  \bibinfo {author} {\bibfnamefont {N.~A.}\ \bibnamefont {Spaldin}}, \bibinfo
  {author} {\bibfnamefont {T.}~\bibnamefont {Loew}}, \bibinfo {author}
  {\bibfnamefont {B.}~\bibnamefont {Keimer}}, \bibinfo {author} {\bibfnamefont
  {A.}~\bibnamefont {Georges}},\ and\ \bibinfo {author} {\bibfnamefont
  {A.}~\bibnamefont {Cavalleri}},\ }\bibfield  {title} {\bibinfo {title}
  {Nonlinear lattice dynamics as a basis for enhanced superconductivity in
  {YBa}$_2${Cu}$_3${O}$_{6.5}$},\ }\href {https://doi.org/10.1038/nature13875}
  {\bibfield  {journal} {\bibinfo  {journal} {Nature}\ }\textbf {\bibinfo
  {volume} {516}},\ \bibinfo {pages} {71} (\bibinfo {year} {2014})}\BibitemShut
  {NoStop}%
\bibitem [{\citenamefont {Liu}\ \emph {et~al.}(2020)\citenamefont {Liu},
  \citenamefont {F\"orst}, \citenamefont {Fechner}, \citenamefont {Nicoletti},
  \citenamefont {Porras}, \citenamefont {Loew}, \citenamefont {Keimer},\ and\
  \citenamefont {Cavalleri}}]{LiuFoerst2020}%
  \BibitemOpen
  \bibfield  {author} {\bibinfo {author} {\bibfnamefont {B.}~\bibnamefont
  {Liu}}, \bibinfo {author} {\bibfnamefont {M.}~\bibnamefont {F\"orst}},
  \bibinfo {author} {\bibfnamefont {M.}~\bibnamefont {Fechner}}, \bibinfo
  {author} {\bibfnamefont {D.}~\bibnamefont {Nicoletti}}, \bibinfo {author}
  {\bibfnamefont {J.}~\bibnamefont {Porras}}, \bibinfo {author} {\bibfnamefont
  {T.}~\bibnamefont {Loew}}, \bibinfo {author} {\bibfnamefont {B.}~\bibnamefont
  {Keimer}},\ and\ \bibinfo {author} {\bibfnamefont {A.}~\bibnamefont
  {Cavalleri}},\ }\bibfield  {title} {\bibinfo {title} {Pump frequency
  resonances for light-induced incipient superconductivity in
  ${\mathrm{yba}}_{2}{\mathrm{cu}}_{3}{\mathrm{o}}_{6.5}$},\ }\href
  {https://doi.org/10.1103/PhysRevX.10.011053} {\bibfield  {journal} {\bibinfo
  {journal} {Phys. Rev. X}\ }\textbf {\bibinfo {volume} {10}},\ \bibinfo
  {pages} {011053} (\bibinfo {year} {2020})}\BibitemShut {NoStop}%
\bibitem [{\citenamefont {Mitrano}\ \emph {et~al.}(2016)\citenamefont
  {Mitrano}, \citenamefont {Cantaluppi}, \citenamefont {Nicoletti},
  \citenamefont {Kaiser}, \citenamefont {Perucchi}, \citenamefont {Lupi},
  \citenamefont {Di~Pietro}, \citenamefont {Pontiroli}, \citenamefont
  {Ricc{\`o}}, \citenamefont {Clark}, \citenamefont {Jaksch},\ and\
  \citenamefont {Cavalleri}}]{Mitrano2016}%
  \BibitemOpen
  \bibfield  {author} {\bibinfo {author} {\bibfnamefont {M.}~\bibnamefont
  {Mitrano}}, \bibinfo {author} {\bibfnamefont {A.}~\bibnamefont {Cantaluppi}},
  \bibinfo {author} {\bibfnamefont {D.}~\bibnamefont {Nicoletti}}, \bibinfo
  {author} {\bibfnamefont {S.}~\bibnamefont {Kaiser}}, \bibinfo {author}
  {\bibfnamefont {A.}~\bibnamefont {Perucchi}}, \bibinfo {author}
  {\bibfnamefont {S.}~\bibnamefont {Lupi}}, \bibinfo {author} {\bibfnamefont
  {P.}~\bibnamefont {Di~Pietro}}, \bibinfo {author} {\bibfnamefont
  {D.}~\bibnamefont {Pontiroli}}, \bibinfo {author} {\bibfnamefont
  {M.}~\bibnamefont {Ricc{\`o}}}, \bibinfo {author} {\bibfnamefont {S.~R.}\
  \bibnamefont {Clark}}, \bibinfo {author} {\bibfnamefont {D.}~\bibnamefont
  {Jaksch}},\ and\ \bibinfo {author} {\bibfnamefont {A.}~\bibnamefont
  {Cavalleri}},\ }\bibfield  {title} {\bibinfo {title} {Possible light-induced
  superconductivity in {K}$_3${C}$_{60}$ at high temperature},\ }\href
  {https://doi.org/10.1038/nature16522} {\bibfield  {journal} {\bibinfo
  {journal} {Nature}\ }\textbf {\bibinfo {volume} {530}},\ \bibinfo {pages}
  {461} (\bibinfo {year} {2016})}\BibitemShut {NoStop}%
\bibitem [{\citenamefont {Juraschek}\ \emph
  {et~al.}(2017{\natexlab{b}})\citenamefont {Juraschek}, \citenamefont
  {Fechner},\ and\ \citenamefont {Spaldin}}]{Juraschek2017b}%
  \BibitemOpen
  \bibfield  {author} {\bibinfo {author} {\bibfnamefont {D.~M.}\ \bibnamefont
  {Juraschek}}, \bibinfo {author} {\bibfnamefont {M.}~\bibnamefont {Fechner}},\
  and\ \bibinfo {author} {\bibfnamefont {N.~A.}\ \bibnamefont {Spaldin}},\
  }\bibfield  {title} {\bibinfo {title} {Ultrafast structure switching through
  nonlinear phononics},\ }\href
  {https://doi.org/10.1103/PhysRevLett.118.054101} {\bibfield  {journal}
  {\bibinfo  {journal} {Phys. Rev. Lett.}\ }\textbf {\bibinfo {volume} {118}},\
  \bibinfo {pages} {054101} (\bibinfo {year} {2017}{\natexlab{b}})}\BibitemShut
  {NoStop}%
\bibitem [{\citenamefont {De~La~Torre}\ \emph {et~al.}(2021)\citenamefont
  {De~La~Torre}, \citenamefont {Kennes}, \citenamefont {Claassen},
  \citenamefont {Gerber}, \citenamefont {McIver},\ and\ \citenamefont
  {Sentef}}]{de2021colloquium}%
  \BibitemOpen
  \bibfield  {author} {\bibinfo {author} {\bibfnamefont {A.}~\bibnamefont
  {De~La~Torre}}, \bibinfo {author} {\bibfnamefont {D.~M.}\ \bibnamefont
  {Kennes}}, \bibinfo {author} {\bibfnamefont {M.}~\bibnamefont {Claassen}},
  \bibinfo {author} {\bibfnamefont {S.}~\bibnamefont {Gerber}}, \bibinfo
  {author} {\bibfnamefont {J.~W.}\ \bibnamefont {McIver}},\ and\ \bibinfo
  {author} {\bibfnamefont {M.~A.}\ \bibnamefont {Sentef}},\ }\bibfield  {title}
  {\bibinfo {title} {Colloquium: {Nonthermal} pathways to ultrafast control in
  quantum materials},\ }\href@noop {} {\bibfield  {journal} {\bibinfo
  {journal} {Rev. Mod. Phys.}\ }\textbf {\bibinfo {volume} {93}},\ \bibinfo
  {pages} {041002} (\bibinfo {year} {2021})}\BibitemShut {NoStop}%
\bibitem [{\citenamefont {Mocatti}\ \emph {et~al.}(2023)\citenamefont
  {Mocatti}, \citenamefont {Marini},\ and\ \citenamefont
  {Calandra}}]{mocatti2023light}%
  \BibitemOpen
  \bibfield  {author} {\bibinfo {author} {\bibfnamefont {S.}~\bibnamefont
  {Mocatti}}, \bibinfo {author} {\bibfnamefont {G.}~\bibnamefont {Marini}},\
  and\ \bibinfo {author} {\bibfnamefont {M.}~\bibnamefont {Calandra}},\
  }\bibfield  {title} {\bibinfo {title} {Light-induced nonthermal phase
  transition to the topological crystalline insulator state in {SnSe}},\
  }\href@noop {} {\bibfield  {journal} {\bibinfo  {journal} {J. Phys. Chem.
  Lett.}\ }\textbf {\bibinfo {volume} {14}},\ \bibinfo {pages} {9329} (\bibinfo
  {year} {2023})}\BibitemShut {NoStop}%
\bibitem [{\citenamefont {Yang}\ \emph {et~al.}(2021)\citenamefont {Yang},
  \citenamefont {Rohde}, \citenamefont {Chen}, \citenamefont {Shi},
  \citenamefont {Liu}, \citenamefont {Chen}, \citenamefont {Chen},
  \citenamefont {Rossnagel},\ and\ \citenamefont
  {Bauer}}]{yang2021experimental}%
  \BibitemOpen
  \bibfield  {author} {\bibinfo {author} {\bibfnamefont {L.}~\bibnamefont
  {Yang}}, \bibinfo {author} {\bibfnamefont {G.}~\bibnamefont {Rohde}},
  \bibinfo {author} {\bibfnamefont {Y.}~\bibnamefont {Chen}}, \bibinfo {author}
  {\bibfnamefont {W.}~\bibnamefont {Shi}}, \bibinfo {author} {\bibfnamefont
  {Z.}~\bibnamefont {Liu}}, \bibinfo {author} {\bibfnamefont {F.}~\bibnamefont
  {Chen}}, \bibinfo {author} {\bibfnamefont {Y.}~\bibnamefont {Chen}}, \bibinfo
  {author} {\bibfnamefont {K.}~\bibnamefont {Rossnagel}},\ and\ \bibinfo
  {author} {\bibfnamefont {M.}~\bibnamefont {Bauer}},\ }\bibfield  {title}
  {\bibinfo {title} {Experimental evidence for a metastable state in
  $\mathrm{FeTe_{1-x}Se_x}$ following coherent-phonon excitation},\ }\href@noop
  {} {\bibfield  {journal} {\bibinfo  {journal} {J. Electron Spectros. Relat.
  Phenomena}\ }\textbf {\bibinfo {volume} {250}},\ \bibinfo {pages} {147085}
  (\bibinfo {year} {2021})}\BibitemShut {NoStop}%
\bibitem [{\citenamefont {Zeiger}\ \emph {et~al.}(1992)\citenamefont {Zeiger},
  \citenamefont {Vidal}, \citenamefont {Cheng}, \citenamefont {Ippen},
  \citenamefont {Dresselhaus},\ and\ \citenamefont
  {Dresselhaus}}]{zeiger1992theory}%
  \BibitemOpen
  \bibfield  {author} {\bibinfo {author} {\bibfnamefont {H.}~\bibnamefont
  {Zeiger}}, \bibinfo {author} {\bibfnamefont {J.}~\bibnamefont {Vidal}},
  \bibinfo {author} {\bibfnamefont {T.}~\bibnamefont {Cheng}}, \bibinfo
  {author} {\bibfnamefont {E.}~\bibnamefont {Ippen}}, \bibinfo {author}
  {\bibfnamefont {G.}~\bibnamefont {Dresselhaus}},\ and\ \bibinfo {author}
  {\bibfnamefont {M.}~\bibnamefont {Dresselhaus}},\ }\bibfield  {title}
  {\bibinfo {title} {Theory for displacive excitation of coherent phonons},\
  }\href@noop {} {\bibfield  {journal} {\bibinfo  {journal} {Phys. Rev. B}\
  }\textbf {\bibinfo {volume} {45}},\ \bibinfo {pages} {768} (\bibinfo {year}
  {1992})}\BibitemShut {NoStop}%
\bibitem [{\citenamefont {Kuznetsov}\ and\ \citenamefont
  {Stanton}(1994)}]{kuznetsov1994theory}%
  \BibitemOpen
  \bibfield  {author} {\bibinfo {author} {\bibfnamefont {A.~V.}\ \bibnamefont
  {Kuznetsov}}\ and\ \bibinfo {author} {\bibfnamefont {C.~J.}\ \bibnamefont
  {Stanton}},\ }\bibfield  {title} {\bibinfo {title} {Theory of coherent phonon
  oscillations in semiconductors},\ }\href@noop {} {\bibfield  {journal}
  {\bibinfo  {journal} {Phys. Rev. Lett.}\ }\textbf {\bibinfo {volume} {73}},\
  \bibinfo {pages} {3243} (\bibinfo {year} {1994})}\BibitemShut {NoStop}%
\bibitem [{\citenamefont {Dhar}\ \emph {et~al.}(1994)\citenamefont {Dhar},
  \citenamefont {Rogers},\ and\ \citenamefont {Nelson}}]{dhar1994}%
  \BibitemOpen
  \bibfield  {author} {\bibinfo {author} {\bibfnamefont {L.}~\bibnamefont
  {Dhar}}, \bibinfo {author} {\bibfnamefont {J.~A.}\ \bibnamefont {Rogers}},\
  and\ \bibinfo {author} {\bibfnamefont {K.~A.}\ \bibnamefont {Nelson}},\
  }\bibfield  {title} {\bibinfo {title} {Time-resolved vibrational spectroscopy
  in the impulsive limit},\ }\href {https://doi.org/10.1021/cr00025a006}
  {\bibfield  {journal} {\bibinfo  {journal} {Chem. Rev.}\ }\textbf {\bibinfo
  {volume} {94}},\ \bibinfo {pages} {157} (\bibinfo {year} {1994})}\BibitemShut
  {NoStop}%
\bibitem [{\citenamefont {Giret}\ \emph {et~al.}(2011)\citenamefont {Giret},
  \citenamefont {Gell\'e},\ and\ \citenamefont {Arnaud}}]{Giret2011}%
  \BibitemOpen
  \bibfield  {author} {\bibinfo {author} {\bibfnamefont {Y.}~\bibnamefont
  {Giret}}, \bibinfo {author} {\bibfnamefont {A.}~\bibnamefont {Gell\'e}},\
  and\ \bibinfo {author} {\bibfnamefont {B.}~\bibnamefont {Arnaud}},\
  }\bibfield  {title} {\bibinfo {title} {Entropy driven atomic motion in
  laser-excited bismuth},\ }\href
  {https://doi.org/10.1103/PhysRevLett.106.155503} {\bibfield  {journal}
  {\bibinfo  {journal} {Phys. Rev. Lett.}\ }\textbf {\bibinfo {volume} {106}},\
  \bibinfo {pages} {155503} (\bibinfo {year} {2011})}\BibitemShut {NoStop}%
\bibitem [{\citenamefont {Juraschek}\ and\ \citenamefont
  {Maehrlein}(2018)}]{Juraschek2018c}%
  \BibitemOpen
  \bibfield  {author} {\bibinfo {author} {\bibfnamefont {D.~M.}\ \bibnamefont
  {Juraschek}}\ and\ \bibinfo {author} {\bibfnamefont {S.~F.}\ \bibnamefont
  {Maehrlein}},\ }\bibfield  {title} {\bibinfo {title} {Sum-frequency ionic
  raman scattering},\ }\href {https://doi.org/10.1103/PhysRevB.97.174302}
  {\bibfield  {journal} {\bibinfo  {journal} {Phys. Rev. B}\ }\textbf {\bibinfo
  {volume} {97}},\ \bibinfo {pages} {174302} (\bibinfo {year}
  {2018})}\BibitemShut {NoStop}%
\bibitem [{\citenamefont {Lakehal}\ and\ \citenamefont
  {Paul}(2019)}]{lakehal2019microscopic}%
  \BibitemOpen
  \bibfield  {author} {\bibinfo {author} {\bibfnamefont {M.}~\bibnamefont
  {Lakehal}}\ and\ \bibinfo {author} {\bibfnamefont {I.}~\bibnamefont {Paul}},\
  }\bibfield  {title} {\bibinfo {title} {Microscopic description of displacive
  coherent phonons},\ }\href@noop {} {\bibfield  {journal} {\bibinfo  {journal}
  {Phys. Rev. B}\ }\textbf {\bibinfo {volume} {99}},\ \bibinfo {pages} {035131}
  (\bibinfo {year} {2019})}\BibitemShut {NoStop}%
\bibitem [{\citenamefont {Caruso}\ and\ \citenamefont
  {Zacharias}(2023)}]{caruso2023quantum}%
  \BibitemOpen
  \bibfield  {author} {\bibinfo {author} {\bibfnamefont {F.}~\bibnamefont
  {Caruso}}\ and\ \bibinfo {author} {\bibfnamefont {M.}~\bibnamefont
  {Zacharias}},\ }\bibfield  {title} {\bibinfo {title} {Quantum theory of
  light-driven coherent lattice dynamics},\ }\href@noop {} {\bibfield
  {journal} {\bibinfo  {journal} {Phys. Rev. B}\ }\textbf {\bibinfo {volume}
  {107}},\ \bibinfo {pages} {054102} (\bibinfo {year} {2023})}\BibitemShut
  {NoStop}%
\bibitem [{\citenamefont {Garrett}\ \emph {et~al.}(1996)\citenamefont
  {Garrett}, \citenamefont {Albrecht}, \citenamefont {Whitaker},\ and\
  \citenamefont {Merlin}}]{Garrett1996}%
  \BibitemOpen
  \bibfield  {author} {\bibinfo {author} {\bibfnamefont {G.~A.}\ \bibnamefont
  {Garrett}}, \bibinfo {author} {\bibfnamefont {T.~F.}\ \bibnamefont
  {Albrecht}}, \bibinfo {author} {\bibfnamefont {J.~F.}\ \bibnamefont
  {Whitaker}},\ and\ \bibinfo {author} {\bibfnamefont {R.}~\bibnamefont
  {Merlin}},\ }\bibfield  {title} {\bibinfo {title} {Coherent thz phonons
  driven by light pulses and the $\mathrm{Sb}$ problem: What is the
  mechanism?},\ }\href {https://doi.org/10.1103/PhysRevLett.77.3661} {\bibfield
   {journal} {\bibinfo  {journal} {Phys. Rev. Lett.}\ }\textbf {\bibinfo
  {volume} {77}},\ \bibinfo {pages} {3661} (\bibinfo {year}
  {1996})}\BibitemShut {NoStop}%
\bibitem [{\citenamefont {Caruso}(2021)}]{caruso2021nonequilibrium}%
  \BibitemOpen
  \bibfield  {author} {\bibinfo {author} {\bibfnamefont {F.}~\bibnamefont
  {Caruso}},\ }\bibfield  {title} {\bibinfo {title} {Nonequilibrium lattice
  dynamics in monolayer $\mathrm{MoS_2}$},\ }\href@noop {} {\bibfield
  {journal} {\bibinfo  {journal} {J. Phys. Chem. Lett.}\ }\textbf {\bibinfo
  {volume} {12}},\ \bibinfo {pages} {1734} (\bibinfo {year}
  {2021})}\BibitemShut {NoStop}%
\bibitem [{\citenamefont {Tong}\ and\ \citenamefont
  {Bernardi}(2021)}]{Tong2021}%
  \BibitemOpen
  \bibfield  {author} {\bibinfo {author} {\bibfnamefont {X.}~\bibnamefont
  {Tong}}\ and\ \bibinfo {author} {\bibfnamefont {M.}~\bibnamefont
  {Bernardi}},\ }\bibfield  {title} {\bibinfo {title} {Toward precise
  simulations of the coupled ultrafast dynamics of electrons and atomic
  vibrations in materials},\ }\href
  {https://doi.org/10.1103/PhysRevResearch.3.023072} {\bibfield  {journal}
  {\bibinfo  {journal} {Phys. Rev. Res.}\ }\textbf {\bibinfo {volume} {3}},\
  \bibinfo {pages} {023072} (\bibinfo {year} {2021})}\BibitemShut {NoStop}%
\bibitem [{\citenamefont {Caruso}\ and\ \citenamefont
  {Novko}(2022)}]{caruso2022ultrafast}%
  \BibitemOpen
  \bibfield  {author} {\bibinfo {author} {\bibfnamefont {F.}~\bibnamefont
  {Caruso}}\ and\ \bibinfo {author} {\bibfnamefont {D.}~\bibnamefont {Novko}},\
  }\bibfield  {title} {\bibinfo {title} {Ultrafast dynamics of electrons and
  phonons: from the two-temperature model to the time-dependent {Boltzmann}
  equation},\ }\href@noop {} {\bibfield  {journal} {\bibinfo  {journal} {Adv.
  Phys.: X}\ }\textbf {\bibinfo {volume} {7}},\ \bibinfo {pages} {2095925}
  (\bibinfo {year} {2022})}\BibitemShut {NoStop}%
\bibitem [{\citenamefont {Pan}\ and\ \citenamefont {Caruso}(2023)}]{Pan2023}%
  \BibitemOpen
  \bibfield  {author} {\bibinfo {author} {\bibfnamefont {Y.}~\bibnamefont
  {Pan}}\ and\ \bibinfo {author} {\bibfnamefont {F.}~\bibnamefont {Caruso}},\
  }\bibfield  {title} {\bibinfo {title} {Vibrational dichroism of chiral valley
  phonons},\ }\href {https://doi.org/10.1021/acs.nanolett.3c01904} {\bibfield
  {journal} {\bibinfo  {journal} {Nano Lett.}\ }\textbf {\bibinfo {volume}
  {23}},\ \bibinfo {pages} {7463} (\bibinfo {year} {2023})}\BibitemShut
  {NoStop}%
\bibitem [{\citenamefont {Giustino}(2017)}]{giustino2017electron}%
  \BibitemOpen
  \bibfield  {author} {\bibinfo {author} {\bibfnamefont {F.}~\bibnamefont
  {Giustino}},\ }\bibfield  {title} {\bibinfo {title} {Electron-phonon
  interactions from first principles},\ }\href@noop {} {\bibfield  {journal}
  {\bibinfo  {journal} {Rev. Mod. Phys.}\ }\textbf {\bibinfo {volume} {89}},\
  \bibinfo {pages} {015003} (\bibinfo {year} {2017})}\BibitemShut {NoStop}%
\bibitem [{\citenamefont {Stefanucci}\ \emph {et~al.}(2023)\citenamefont
  {Stefanucci}, \citenamefont {van Leeuwen},\ and\ \citenamefont
  {Perfetto}}]{stefanucci2023and}%
  \BibitemOpen
  \bibfield  {author} {\bibinfo {author} {\bibfnamefont {G.}~\bibnamefont
  {Stefanucci}}, \bibinfo {author} {\bibfnamefont {R.}~\bibnamefont {van
  Leeuwen}},\ and\ \bibinfo {author} {\bibfnamefont {E.}~\bibnamefont
  {Perfetto}},\ }\bibfield  {title} {\bibinfo {title} {In and
  out-of-equilibrium ab initio theory of electrons and phonons},\ }\href@noop
  {} {\bibfield  {journal} {\bibinfo  {journal} {Phys. Rev. X}\ }\textbf
  {\bibinfo {volume} {13}},\ \bibinfo {pages} {031026} (\bibinfo {year}
  {2023})}\BibitemShut {NoStop}%
\bibitem [{\citenamefont {Perfetto}\ \emph {et~al.}(2024)\citenamefont
  {Perfetto}, \citenamefont {Wu},\ and\ \citenamefont
  {Stefanucci}}]{perfetto_theory_2024}%
  \BibitemOpen
  \bibfield  {author} {\bibinfo {author} {\bibfnamefont {E.}~\bibnamefont
  {Perfetto}}, \bibinfo {author} {\bibfnamefont {K.}~\bibnamefont {Wu}},\ and\
  \bibinfo {author} {\bibfnamefont {G.}~\bibnamefont {Stefanucci}},\ }\bibfield
   {title} {\bibinfo {title} {Theory of coherent phonons coupled to excitons},\
  }\href {https://doi.org/10.1038/s41699-024-00474-9} {\bibfield  {journal}
  {\bibinfo  {journal} {npj 2D mater. appl.}\ }\textbf {\bibinfo {volume}
  {8}},\ \bibinfo {pages} {40} (\bibinfo {year} {2024})}\BibitemShut {NoStop}%
\bibitem [{\citenamefont {Rethfeld}\ \emph {et~al.}(2002)\citenamefont
  {Rethfeld}, \citenamefont {Kaiser}, \citenamefont {Vicanek},\ and\
  \citenamefont {Simon}}]{rethfeld2002ultrafast}%
  \BibitemOpen
  \bibfield  {author} {\bibinfo {author} {\bibfnamefont {B.}~\bibnamefont
  {Rethfeld}}, \bibinfo {author} {\bibfnamefont {A.}~\bibnamefont {Kaiser}},
  \bibinfo {author} {\bibfnamefont {M.}~\bibnamefont {Vicanek}},\ and\ \bibinfo
  {author} {\bibfnamefont {G.}~\bibnamefont {Simon}},\ }\bibfield  {title}
  {\bibinfo {title} {Ultrafast dynamics of nonequilibrium electrons in metals
  under femtosecond laser irradiation},\ }\href@noop {} {\bibfield  {journal}
  {\bibinfo  {journal} {Phys. Rev. B}\ }\textbf {\bibinfo {volume} {65}},\
  \bibinfo {pages} {214303} (\bibinfo {year} {2002})}\BibitemShut {NoStop}%
\bibitem [{\citenamefont {Lee}\ \emph {et~al.}(2023{\natexlab{b}})\citenamefont
  {Lee}, \citenamefont {Poncé}, \citenamefont {Bushick}, \citenamefont
  {Hajinazar}, \citenamefont {Lafuente-Bartolome}, \citenamefont {Leveillee},
  \citenamefont {Lian}, \citenamefont {Lihm}, \citenamefont {Macheda},
  \citenamefont {Mori}, \citenamefont {Paudyal}, \citenamefont {Sio},
  \citenamefont {Tiwari}, \citenamefont {Zacharias}, \citenamefont {Zhang},
  \citenamefont {Bonini}, \citenamefont {Kioupakis}, \citenamefont {Margine},\
  and\ \citenamefont {Giustino}}]{lee2023electron}%
  \BibitemOpen
  \bibfield  {author} {\bibinfo {author} {\bibfnamefont {H.}~\bibnamefont
  {Lee}}, \bibinfo {author} {\bibfnamefont {S.}~\bibnamefont {Poncé}},
  \bibinfo {author} {\bibfnamefont {K.}~\bibnamefont {Bushick}}, \bibinfo
  {author} {\bibfnamefont {S.}~\bibnamefont {Hajinazar}}, \bibinfo {author}
  {\bibfnamefont {J.}~\bibnamefont {Lafuente-Bartolome}}, \bibinfo {author}
  {\bibfnamefont {J.}~\bibnamefont {Leveillee}}, \bibinfo {author}
  {\bibfnamefont {C.}~\bibnamefont {Lian}}, \bibinfo {author} {\bibfnamefont
  {J.-M.}\ \bibnamefont {Lihm}}, \bibinfo {author} {\bibfnamefont
  {F.}~\bibnamefont {Macheda}}, \bibinfo {author} {\bibfnamefont
  {H.}~\bibnamefont {Mori}}, \bibinfo {author} {\bibfnamefont {H.}~\bibnamefont
  {Paudyal}}, \bibinfo {author} {\bibfnamefont {W.~H.}\ \bibnamefont {Sio}},
  \bibinfo {author} {\bibfnamefont {S.}~\bibnamefont {Tiwari}}, \bibinfo
  {author} {\bibfnamefont {M.}~\bibnamefont {Zacharias}}, \bibinfo {author}
  {\bibfnamefont {X.}~\bibnamefont {Zhang}}, \bibinfo {author} {\bibfnamefont
  {N.}~\bibnamefont {Bonini}}, \bibinfo {author} {\bibfnamefont
  {E.}~\bibnamefont {Kioupakis}}, \bibinfo {author} {\bibfnamefont {E.~R.}\
  \bibnamefont {Margine}},\ and\ \bibinfo {author} {\bibfnamefont
  {F.}~\bibnamefont {Giustino}},\ }\bibfield  {title} {\bibinfo {title}
  {Electron-phonon physics from first principles using the {{\tt EPW}} code},\
  }\href@noop {} {\bibfield  {journal} {\bibinfo  {journal} {npj Comput.
  Mater.}\ }\textbf {\bibinfo {volume} {9}},\ \bibinfo {pages} {156} (\bibinfo
  {year} {2023}{\natexlab{b}})}\BibitemShut {NoStop}%
\bibitem [{\citenamefont {Giannozzi}\ \emph {et~al.}(2009)\citenamefont
  {Giannozzi}, \citenamefont {Baroni}, \citenamefont {Bonini}, \citenamefont
  {Calandra}, \citenamefont {Car}, \citenamefont {Cavazzoni}, \citenamefont
  {Ceresoli}, \citenamefont {Chiarotti}, \citenamefont {Cococcioni},
  \citenamefont {Dabo}, \citenamefont {Corso}, \citenamefont {de~Gironcoli},
  \citenamefont {Fabris}, \citenamefont {Fratesi}, \citenamefont {Gebauer},
  \citenamefont {Gerstmann}, \citenamefont {Gougoussis}, \citenamefont
  {Kokalj}, \citenamefont {Lazzeri}, \citenamefont {Martin-Samos},
  \citenamefont {Marzari}, \citenamefont {Mauri}, \citenamefont {Mazzarello},
  \citenamefont {Paolini}, \citenamefont {Pasquarello}, \citenamefont
  {Paulatto}, \citenamefont {Sbraccia}, \citenamefont {Scandolo}, \citenamefont
  {Sclauzero}, \citenamefont {Seitsonen}, \citenamefont {Smogunov},
  \citenamefont {Umari},\ and\ \citenamefont
  {Wentzcovitch}}]{giannozzi2009quantum}%
  \BibitemOpen
  \bibfield  {author} {\bibinfo {author} {\bibfnamefont {P.}~\bibnamefont
  {Giannozzi}}, \bibinfo {author} {\bibfnamefont {S.}~\bibnamefont {Baroni}},
  \bibinfo {author} {\bibfnamefont {N.}~\bibnamefont {Bonini}}, \bibinfo
  {author} {\bibfnamefont {M.}~\bibnamefont {Calandra}}, \bibinfo {author}
  {\bibfnamefont {R.}~\bibnamefont {Car}}, \bibinfo {author} {\bibfnamefont
  {C.}~\bibnamefont {Cavazzoni}}, \bibinfo {author} {\bibfnamefont
  {D.}~\bibnamefont {Ceresoli}}, \bibinfo {author} {\bibfnamefont {G.~L.}\
  \bibnamefont {Chiarotti}}, \bibinfo {author} {\bibfnamefont {M.}~\bibnamefont
  {Cococcioni}}, \bibinfo {author} {\bibfnamefont {I.}~\bibnamefont {Dabo}},
  \bibinfo {author} {\bibfnamefont {A.~D.}\ \bibnamefont {Corso}}, \bibinfo
  {author} {\bibfnamefont {S.}~\bibnamefont {de~Gironcoli}}, \bibinfo {author}
  {\bibfnamefont {S.}~\bibnamefont {Fabris}}, \bibinfo {author} {\bibfnamefont
  {G.}~\bibnamefont {Fratesi}}, \bibinfo {author} {\bibfnamefont
  {R.}~\bibnamefont {Gebauer}}, \bibinfo {author} {\bibfnamefont
  {U.}~\bibnamefont {Gerstmann}}, \bibinfo {author} {\bibfnamefont
  {C.}~\bibnamefont {Gougoussis}}, \bibinfo {author} {\bibfnamefont
  {A.}~\bibnamefont {Kokalj}}, \bibinfo {author} {\bibfnamefont
  {M.}~\bibnamefont {Lazzeri}}, \bibinfo {author} {\bibfnamefont
  {L.}~\bibnamefont {Martin-Samos}}, \bibinfo {author} {\bibfnamefont
  {N.}~\bibnamefont {Marzari}}, \bibinfo {author} {\bibfnamefont
  {F.}~\bibnamefont {Mauri}}, \bibinfo {author} {\bibfnamefont
  {R.}~\bibnamefont {Mazzarello}}, \bibinfo {author} {\bibfnamefont
  {S.}~\bibnamefont {Paolini}}, \bibinfo {author} {\bibfnamefont
  {A.}~\bibnamefont {Pasquarello}}, \bibinfo {author} {\bibfnamefont
  {L.}~\bibnamefont {Paulatto}}, \bibinfo {author} {\bibfnamefont
  {C.}~\bibnamefont {Sbraccia}}, \bibinfo {author} {\bibfnamefont
  {S.}~\bibnamefont {Scandolo}}, \bibinfo {author} {\bibfnamefont
  {G.}~\bibnamefont {Sclauzero}}, \bibinfo {author} {\bibfnamefont {A.~P.}\
  \bibnamefont {Seitsonen}}, \bibinfo {author} {\bibfnamefont {A.}~\bibnamefont
  {Smogunov}}, \bibinfo {author} {\bibfnamefont {P.}~\bibnamefont {Umari}},\
  and\ \bibinfo {author} {\bibfnamefont {R.~M.}\ \bibnamefont {Wentzcovitch}},\
  }\bibfield  {title} {\bibinfo {title} {{{\tt Quantum ESPRESSO}}: a modular
  and open-source software project for quantum simulations of materials},\
  }\href@noop {} {\bibfield  {journal} {\bibinfo  {journal} {J. Phys. Condens.
  Matter}\ }\textbf {\bibinfo {volume} {21}},\ \bibinfo {pages} {395502}
  (\bibinfo {year} {2009})}\BibitemShut {NoStop}%
\bibitem [{\citenamefont {Giannozzi}\ \emph {et~al.}(2017)\citenamefont
  {Giannozzi}, \citenamefont {Andreussi}, \citenamefont {Brumme}, \citenamefont
  {Bunau}, \citenamefont {Nardelli}, \citenamefont {Calandra}, \citenamefont
  {Car}, \citenamefont {Cavazzoni}, \citenamefont {Ceresoli}, \citenamefont
  {Cococcioni} \emph {et~al.}}]{giannozzi2017advanced}%
  \BibitemOpen
  \bibfield  {author} {\bibinfo {author} {\bibfnamefont {P.}~\bibnamefont
  {Giannozzi}}, \bibinfo {author} {\bibfnamefont {O.}~\bibnamefont
  {Andreussi}}, \bibinfo {author} {\bibfnamefont {T.}~\bibnamefont {Brumme}},
  \bibinfo {author} {\bibfnamefont {O.}~\bibnamefont {Bunau}}, \bibinfo
  {author} {\bibfnamefont {M.~B.}\ \bibnamefont {Nardelli}}, \bibinfo {author}
  {\bibfnamefont {M.}~\bibnamefont {Calandra}}, \bibinfo {author}
  {\bibfnamefont {R.}~\bibnamefont {Car}}, \bibinfo {author} {\bibfnamefont
  {C.}~\bibnamefont {Cavazzoni}}, \bibinfo {author} {\bibfnamefont
  {D.}~\bibnamefont {Ceresoli}}, \bibinfo {author} {\bibfnamefont
  {M.}~\bibnamefont {Cococcioni}}, \emph {et~al.},\ }\bibfield  {title}
  {\bibinfo {title} {Advanced capabilities for materials modelling with {{\tt
  Quantum ESPRESSO}}},\ }\href@noop {} {\bibfield  {journal} {\bibinfo
  {journal} {J. Phys. Condens. Matter}\ }\textbf {\bibinfo {volume} {29}},\
  \bibinfo {pages} {465901} (\bibinfo {year} {2017})}\BibitemShut {NoStop}%
\bibitem [{\citenamefont {Baroni}\ \emph {et~al.}(2001)\citenamefont {Baroni},
  \citenamefont {De~Gironcoli}, \citenamefont {Dal~Corso},\ and\ \citenamefont
  {Giannozzi}}]{baroni2001phonons}%
  \BibitemOpen
  \bibfield  {author} {\bibinfo {author} {\bibfnamefont {S.}~\bibnamefont
  {Baroni}}, \bibinfo {author} {\bibfnamefont {S.}~\bibnamefont
  {De~Gironcoli}}, \bibinfo {author} {\bibfnamefont {A.}~\bibnamefont
  {Dal~Corso}},\ and\ \bibinfo {author} {\bibfnamefont {P.}~\bibnamefont
  {Giannozzi}},\ }\bibfield  {title} {\bibinfo {title} {Phonons and related
  crystal properties from density-functional perturbation theory},\ }\href@noop
  {} {\bibfield  {journal} {\bibinfo  {journal} {Rev. Mod. Phys.}\ }\textbf
  {\bibinfo {volume} {73}},\ \bibinfo {pages} {515} (\bibinfo {year}
  {2001})}\BibitemShut {NoStop}%
\bibitem [{\citenamefont {Marzari}\ \emph {et~al.}(2012)\citenamefont
  {Marzari}, \citenamefont {Mostofi}, \citenamefont {Yates}, \citenamefont
  {Souza},\ and\ \citenamefont {Vanderbilt}}]{marzari2012maximally}%
  \BibitemOpen
  \bibfield  {author} {\bibinfo {author} {\bibfnamefont {N.}~\bibnamefont
  {Marzari}}, \bibinfo {author} {\bibfnamefont {A.~A.}\ \bibnamefont
  {Mostofi}}, \bibinfo {author} {\bibfnamefont {J.~R.}\ \bibnamefont {Yates}},
  \bibinfo {author} {\bibfnamefont {I.}~\bibnamefont {Souza}},\ and\ \bibinfo
  {author} {\bibfnamefont {D.}~\bibnamefont {Vanderbilt}},\ }\bibfield  {title}
  {\bibinfo {title} {Maximally localized $\textrm{Wannier}$ functions: Theory
  and applications},\ }\href {https://doi.org/10.1103/RevModPhys.84.1419}
  {\bibfield  {journal} {\bibinfo  {journal} {Rev. Mod. Phys.}\ }\textbf
  {\bibinfo {volume} {84}},\ \bibinfo {pages} {1419} (\bibinfo {year}
  {2012})}\BibitemShut {NoStop}%
\bibitem [{\citenamefont {Giustino}\ \emph {et~al.}(2007)\citenamefont
  {Giustino}, \citenamefont {Cohen},\ and\ \citenamefont
  {Louie}}]{Giustino2007}%
  \BibitemOpen
  \bibfield  {author} {\bibinfo {author} {\bibfnamefont {F.}~\bibnamefont
  {Giustino}}, \bibinfo {author} {\bibfnamefont {M.~L.}\ \bibnamefont
  {Cohen}},\ and\ \bibinfo {author} {\bibfnamefont {S.~G.}\ \bibnamefont
  {Louie}},\ }\bibfield  {title} {\bibinfo {title} {Electron-phonon interaction
  using $\textrm{Wannier}$ functions},\ }\href
  {https://doi.org/10.1103/PhysRevB.76.165108} {\bibfield  {journal} {\bibinfo
  {journal} {Phys. Rev. B}\ }\textbf {\bibinfo {volume} {76}},\ \bibinfo
  {pages} {165108} (\bibinfo {year} {2007})}\BibitemShut {NoStop}%
\bibitem [{\citenamefont {Pizzi}\ \emph {et~al.}(2020)\citenamefont {Pizzi},
  \citenamefont {Vitale}, \citenamefont {Arita}, \citenamefont {Blügel},
  \citenamefont {Freimuth}, \citenamefont {Géranton}, \citenamefont
  {Gibertini}, \citenamefont {Gresch}, \citenamefont {Johnson}, \citenamefont
  {Koretsune}, \citenamefont {Ibañez-Azpiroz}, \citenamefont {Lee},
  \citenamefont {Lihm}, \citenamefont {Marchand}, \citenamefont {Marrazzo},
  \citenamefont {Mokrousov}, \citenamefont {Mustafa}, \citenamefont {Nohara},
  \citenamefont {Nomura}, \citenamefont {Paulatto}, \citenamefont {Poncé},
  \citenamefont {Ponweiser}, \citenamefont {Qiao}, \citenamefont {Thöle},
  \citenamefont {Tsirkin}, \citenamefont {Wierzbowska}, \citenamefont
  {Marzari}, \citenamefont {Vanderbilt}, \citenamefont {Souza}, \citenamefont
  {Mostofi},\ and\ \citenamefont {Yates}}]{pizzi2020wannier90}%
  \BibitemOpen
  \bibfield  {author} {\bibinfo {author} {\bibfnamefont {G.}~\bibnamefont
  {Pizzi}}, \bibinfo {author} {\bibfnamefont {V.}~\bibnamefont {Vitale}},
  \bibinfo {author} {\bibfnamefont {R.}~\bibnamefont {Arita}}, \bibinfo
  {author} {\bibfnamefont {S.}~\bibnamefont {Blügel}}, \bibinfo {author}
  {\bibfnamefont {F.}~\bibnamefont {Freimuth}}, \bibinfo {author}
  {\bibfnamefont {G.}~\bibnamefont {Géranton}}, \bibinfo {author}
  {\bibfnamefont {M.}~\bibnamefont {Gibertini}}, \bibinfo {author}
  {\bibfnamefont {D.}~\bibnamefont {Gresch}}, \bibinfo {author} {\bibfnamefont
  {C.}~\bibnamefont {Johnson}}, \bibinfo {author} {\bibfnamefont
  {T.}~\bibnamefont {Koretsune}}, \bibinfo {author} {\bibfnamefont
  {J.}~\bibnamefont {Ibañez-Azpiroz}}, \bibinfo {author} {\bibfnamefont
  {H.}~\bibnamefont {Lee}}, \bibinfo {author} {\bibfnamefont {J.-M.}\
  \bibnamefont {Lihm}}, \bibinfo {author} {\bibfnamefont {D.}~\bibnamefont
  {Marchand}}, \bibinfo {author} {\bibfnamefont {A.}~\bibnamefont {Marrazzo}},
  \bibinfo {author} {\bibfnamefont {Y.}~\bibnamefont {Mokrousov}}, \bibinfo
  {author} {\bibfnamefont {J.~I.}\ \bibnamefont {Mustafa}}, \bibinfo {author}
  {\bibfnamefont {Y.}~\bibnamefont {Nohara}}, \bibinfo {author} {\bibfnamefont
  {Y.}~\bibnamefont {Nomura}}, \bibinfo {author} {\bibfnamefont
  {L.}~\bibnamefont {Paulatto}}, \bibinfo {author} {\bibfnamefont
  {S.}~\bibnamefont {Poncé}}, \bibinfo {author} {\bibfnamefont
  {T.}~\bibnamefont {Ponweiser}}, \bibinfo {author} {\bibfnamefont
  {J.}~\bibnamefont {Qiao}}, \bibinfo {author} {\bibfnamefont {F.}~\bibnamefont
  {Thöle}}, \bibinfo {author} {\bibfnamefont {S.~S.}\ \bibnamefont {Tsirkin}},
  \bibinfo {author} {\bibfnamefont {M.}~\bibnamefont {Wierzbowska}}, \bibinfo
  {author} {\bibfnamefont {N.}~\bibnamefont {Marzari}}, \bibinfo {author}
  {\bibfnamefont {D.}~\bibnamefont {Vanderbilt}}, \bibinfo {author}
  {\bibfnamefont {I.}~\bibnamefont {Souza}}, \bibinfo {author} {\bibfnamefont
  {A.~A.}\ \bibnamefont {Mostofi}},\ and\ \bibinfo {author} {\bibfnamefont
  {J.~R.}\ \bibnamefont {Yates}},\ }\bibfield  {title} {\bibinfo {title} {{\tt
  Wannier90} as a community code: new features and applications},\ }\href@noop
  {} {\bibfield  {journal} {\bibinfo  {journal} {J. Condens. Matter Phys.}\
  }\textbf {\bibinfo {volume} {32}},\ \bibinfo {pages} {165902} (\bibinfo
  {year} {2020})}\BibitemShut {NoStop}%
\bibitem [{\citenamefont {Jauernik}\ \emph {et~al.}(2018)\citenamefont
  {Jauernik}, \citenamefont {Hein}, \citenamefont {Gurgel}, \citenamefont
  {Falke},\ and\ \citenamefont {Bauer}}]{Jauernik2018}%
  \BibitemOpen
  \bibfield  {author} {\bibinfo {author} {\bibfnamefont {S.}~\bibnamefont
  {Jauernik}}, \bibinfo {author} {\bibfnamefont {P.}~\bibnamefont {Hein}},
  \bibinfo {author} {\bibfnamefont {M.}~\bibnamefont {Gurgel}}, \bibinfo
  {author} {\bibfnamefont {J.}~\bibnamefont {Falke}},\ and\ \bibinfo {author}
  {\bibfnamefont {M.}~\bibnamefont {Bauer}},\ }\bibfield  {title} {\bibinfo
  {title} {{Probing long-range structural order in $\mathrm{SnPc/Ag}$(111) by
  umklapp process assisted low-energy angle-resolved photoelectron
  spectroscopy}},\ }\href {https://doi.org/10.1103/PhysRevB.97.125413}
  {\bibfield  {journal} {\bibinfo  {journal} {Phys. Rev. B}\ }\textbf {\bibinfo
  {volume} {97}},\ \bibinfo {pages} {125413} (\bibinfo {year}
  {2018})}\BibitemShut {NoStop}%
\bibitem [{\citenamefont {Hass}\ and\ \citenamefont
  {Hadley}(1972)}]{hass1972optical}%
  \BibitemOpen
  \bibfield  {author} {\bibinfo {author} {\bibfnamefont {G.}~\bibnamefont
  {Hass}}\ and\ \bibinfo {author} {\bibfnamefont {L.}~\bibnamefont {Hadley}},\
  }\bibfield  {title} {\bibinfo {title} {Optical properties of metals},\
  }\href@noop {} {\bibfield  {journal} {\bibinfo  {journal} {American Institute
  of Physics Handbook}\ }\textbf {\bibinfo {volume} {2}},\ \bibinfo {pages} {6}
  (\bibinfo {year} {1972})}\BibitemShut {NoStop}%
\bibitem [{\citenamefont {Liu}\ and\ \citenamefont
  {Allen}(1995)}]{liu1995electronic}%
  \BibitemOpen
  \bibfield  {author} {\bibinfo {author} {\bibfnamefont {Y.}~\bibnamefont
  {Liu}}\ and\ \bibinfo {author} {\bibfnamefont {R.~E.}\ \bibnamefont
  {Allen}},\ }\bibfield  {title} {\bibinfo {title} {Electronic structure of the
  semimetals $\mathrm{Bi}$ and $\mathrm{Sb}$},\ }\href@noop {} {\bibfield
  {journal} {\bibinfo  {journal} {Phys. Rev. B}\ }\textbf {\bibinfo {volume}
  {52}},\ \bibinfo {pages} {1566} (\bibinfo {year} {1995})}\BibitemShut
  {NoStop}%
\bibitem [{\citenamefont {Zijlstra}\ \emph {et~al.}(2006)\citenamefont
  {Zijlstra}, \citenamefont {Tatarinova},\ and\ \citenamefont
  {Garcia}}]{Zijlstra2006}%
  \BibitemOpen
  \bibfield  {author} {\bibinfo {author} {\bibfnamefont {E.~S.}\ \bibnamefont
  {Zijlstra}}, \bibinfo {author} {\bibfnamefont {L.~L.}\ \bibnamefont
  {Tatarinova}},\ and\ \bibinfo {author} {\bibfnamefont {M.~E.}\ \bibnamefont
  {Garcia}},\ }\bibfield  {title} {\bibinfo {title} {Laser-induced
  phonon-phonon interactions in bismuth},\ }\href
  {https://doi.org/10.1103/PhysRevB.74.220301} {\bibfield  {journal} {\bibinfo
  {journal} {Phys. Rev. B}\ }\textbf {\bibinfo {volume} {74}},\ \bibinfo
  {pages} {220301} (\bibinfo {year} {2006})}\BibitemShut {NoStop}%
\bibitem [{sup()}]{supp}%
  \BibitemOpen
  \href@noop {} {}\bibinfo {note} {See Supplemental Material at
  URL-will-be-inserted-by-publisher, which include Supplementary Note~S1,
  Supplementary Figures~S1-S7, one Supplementary Video and citations to
  Ref.~\cite{campi2012phonons,setyawan2010high}}\BibitemShut {NoStop}%
\bibitem [{\citenamefont {Clark}\ \emph {et~al.}(2021)\citenamefont {Clark},
  \citenamefont {Freyse}, \citenamefont {Aguilera}, \citenamefont {Frolov},
  \citenamefont {Ionov}, \citenamefont {Bozhko}, \citenamefont {Yashina},\ and\
  \citenamefont {S{\'a}nchez-Barriga}}]{clark2021observation}%
  \BibitemOpen
  \bibfield  {author} {\bibinfo {author} {\bibfnamefont {O.~J.}\ \bibnamefont
  {Clark}}, \bibinfo {author} {\bibfnamefont {F.}~\bibnamefont {Freyse}},
  \bibinfo {author} {\bibfnamefont {I.}~\bibnamefont {Aguilera}}, \bibinfo
  {author} {\bibfnamefont {A.~S.}\ \bibnamefont {Frolov}}, \bibinfo {author}
  {\bibfnamefont {A.~M.}\ \bibnamefont {Ionov}}, \bibinfo {author}
  {\bibfnamefont {S.~I.}\ \bibnamefont {Bozhko}}, \bibinfo {author}
  {\bibfnamefont {L.~V.}\ \bibnamefont {Yashina}},\ and\ \bibinfo {author}
  {\bibfnamefont {J.}~\bibnamefont {S{\'a}nchez-Barriga}},\ }\bibfield  {title}
  {\bibinfo {title} {Observation of a giant mass enhancement in the ultrafast
  electron dynamics of a topological semimetal},\ }\href@noop {} {\bibfield
  {journal} {\bibinfo  {journal} {Commun. Phys.}\ }\textbf {\bibinfo {volume}
  {4}},\ \bibinfo {pages} {165} (\bibinfo {year} {2021})}\BibitemShut {NoStop}%
\bibitem [{\citenamefont {Sakamoto}\ \emph {et~al.}(2022)\citenamefont
  {Sakamoto}, \citenamefont {Gauthier}, \citenamefont {Kirchmann},
  \citenamefont {Sobota},\ and\ \citenamefont {Shen}}]{sakamoto2022connection}%
  \BibitemOpen
  \bibfield  {author} {\bibinfo {author} {\bibfnamefont {S.}~\bibnamefont
  {Sakamoto}}, \bibinfo {author} {\bibfnamefont {N.}~\bibnamefont {Gauthier}},
  \bibinfo {author} {\bibfnamefont {P.~S.}\ \bibnamefont {Kirchmann}}, \bibinfo
  {author} {\bibfnamefont {J.~A.}\ \bibnamefont {Sobota}},\ and\ \bibinfo
  {author} {\bibfnamefont {Z.-X.}\ \bibnamefont {Shen}},\ }\bibfield  {title}
  {\bibinfo {title} {Connection between coherent phonons and electron-phonon
  coupling in {Sb} (111)},\ }\href@noop {} {\bibfield  {journal} {\bibinfo
  {journal} {Phys. Rev. B}\ }\textbf {\bibinfo {volume} {105}},\ \bibinfo
  {pages} {L161107} (\bibinfo {year} {2022})}\BibitemShut {NoStop}%
\bibitem [{\citenamefont {Lax}(1964)}]{LAX1964487}%
  \BibitemOpen
  \bibfield  {author} {\bibinfo {author} {\bibfnamefont {M.}~\bibnamefont
  {Lax}},\ }\bibfield  {title} {\bibinfo {title} {Quantum relaxation, the shape
  of lattice absorption and inelastic neutron scattering lines},\ }\href
  {https://doi.org/https://doi.org/10.1016/0022-3697(64)90122-2} {\bibfield
  {journal} {\bibinfo  {journal} {J. Phys. Chem. Solids}\ }\textbf {\bibinfo
  {volume} {25}},\ \bibinfo {pages} {487} (\bibinfo {year} {1964})}\BibitemShut
  {NoStop}%
\bibitem [{\citenamefont {Hase}\ \emph {et~al.}(1998)\citenamefont {Hase},
  \citenamefont {Mizoguchi}, \citenamefont {Harima}, \citenamefont
  {Nakashima},\ and\ \citenamefont {Sakai}}]{hase1998dynamics}%
  \BibitemOpen
  \bibfield  {author} {\bibinfo {author} {\bibfnamefont {M.}~\bibnamefont
  {Hase}}, \bibinfo {author} {\bibfnamefont {K.}~\bibnamefont {Mizoguchi}},
  \bibinfo {author} {\bibfnamefont {H.}~\bibnamefont {Harima}}, \bibinfo
  {author} {\bibfnamefont {S.-i.}\ \bibnamefont {Nakashima}},\ and\ \bibinfo
  {author} {\bibfnamefont {K.}~\bibnamefont {Sakai}},\ }\bibfield  {title}
  {\bibinfo {title} {Dynamics of coherent phonons in bismuth generated by
  ultrashort laser pulses},\ }\href@noop {} {\bibfield  {journal} {\bibinfo
  {journal} {Phys. Rev. B}\ }\textbf {\bibinfo {volume} {58}},\ \bibinfo
  {pages} {5448} (\bibinfo {year} {1998})}\BibitemShut {NoStop}%
\bibitem [{\citenamefont {Ishioka}\ \emph {et~al.}(2001)\citenamefont
  {Ishioka}, \citenamefont {Hase}, \citenamefont {Kitajima},\ and\
  \citenamefont {Ushida}}]{ishioka2001ultrafast}%
  \BibitemOpen
  \bibfield  {author} {\bibinfo {author} {\bibfnamefont {K.}~\bibnamefont
  {Ishioka}}, \bibinfo {author} {\bibfnamefont {M.}~\bibnamefont {Hase}},
  \bibinfo {author} {\bibfnamefont {M.}~\bibnamefont {Kitajima}},\ and\
  \bibinfo {author} {\bibfnamefont {K.}~\bibnamefont {Ushida}},\ }\bibfield
  {title} {\bibinfo {title} {Ultrafast carrier and phonon dynamics in
  ion-irradiated graphite},\ }\href@noop {} {\bibfield  {journal} {\bibinfo
  {journal} {Appl. Phys. Lett.}\ }\textbf {\bibinfo {volume} {78}},\ \bibinfo
  {pages} {3965} (\bibinfo {year} {2001})}\BibitemShut {NoStop}%
\bibitem [{\citenamefont {Fahy}\ \emph {et~al.}(2016)\citenamefont {Fahy},
  \citenamefont {Murray},\ and\ \citenamefont {Reis}}]{fahy2016resonant}%
  \BibitemOpen
  \bibfield  {author} {\bibinfo {author} {\bibfnamefont {S.}~\bibnamefont
  {Fahy}}, \bibinfo {author} {\bibfnamefont {{\'E}.~D.}\ \bibnamefont
  {Murray}},\ and\ \bibinfo {author} {\bibfnamefont {D.~A.}\ \bibnamefont
  {Reis}},\ }\bibfield  {title} {\bibinfo {title} {Resonant squeezing and the
  anharmonic decay of coherent phonons},\ }\href@noop {} {\bibfield  {journal}
  {\bibinfo  {journal} {Phys. Rev. B}\ }\textbf {\bibinfo {volume} {93}},\
  \bibinfo {pages} {134308} (\bibinfo {year} {2016})}\BibitemShut {NoStop}%
\bibitem [{\citenamefont {Lafuente-Bartolome}\ \emph
  {et~al.}(2022{\natexlab{a}})\citenamefont {Lafuente-Bartolome}, \citenamefont
  {Lian}, \citenamefont {Sio}, \citenamefont {Gurtubay}, \citenamefont
  {Eiguren},\ and\ \citenamefont {Giustino}}]{lafuente-bartolome_unified_2022}%
  \BibitemOpen
  \bibfield  {author} {\bibinfo {author} {\bibfnamefont {J.}~\bibnamefont
  {Lafuente-Bartolome}}, \bibinfo {author} {\bibfnamefont {C.}~\bibnamefont
  {Lian}}, \bibinfo {author} {\bibfnamefont {W.~H.}\ \bibnamefont {Sio}},
  \bibinfo {author} {\bibfnamefont {I.~G.}\ \bibnamefont {Gurtubay}}, \bibinfo
  {author} {\bibfnamefont {A.}~\bibnamefont {Eiguren}},\ and\ \bibinfo {author}
  {\bibfnamefont {F.}~\bibnamefont {Giustino}},\ }\bibfield  {title} {\bibinfo
  {title} {Unified approach to polarons and phonon-induced band structure
  renormalization},\ }\href {https://doi.org/10.1103/PhysRevLett.129.076402}
  {\bibfield  {journal} {\bibinfo  {journal} {Phys. Rev. Lett.}\ }\textbf
  {\bibinfo {volume} {129}},\ \bibinfo {pages} {076402} (\bibinfo {year}
  {2022}{\natexlab{a}})}\BibitemShut {NoStop}%
\bibitem [{\citenamefont {Stefanucci}\ and\ \citenamefont
  {Perfetto}(2024)}]{stefanucci2024semiconductor}%
  \BibitemOpen
  \bibfield  {author} {\bibinfo {author} {\bibfnamefont {G.}~\bibnamefont
  {Stefanucci}}\ and\ \bibinfo {author} {\bibfnamefont {E.}~\bibnamefont
  {Perfetto}},\ }\bibfield  {title} {\bibinfo {title} {Semiconductor
  electron-phonon equations: A rung above {Boltzmann} in the many-body
  ladder},\ }\href@noop {} {\bibfield  {journal} {\bibinfo  {journal} {SciPost
  Phys.}\ }\textbf {\bibinfo {volume} {16}},\ \bibinfo {pages} {073} (\bibinfo
  {year} {2024})}\BibitemShut {NoStop}%
\bibitem [{\citenamefont {Lafuente-Bartolome}\ \emph
  {et~al.}(2022{\natexlab{b}})\citenamefont {Lafuente-Bartolome}, \citenamefont
  {Lian}, \citenamefont {Sio}, \citenamefont {Gurtubay}, \citenamefont
  {Eiguren},\ and\ \citenamefont {Giustino}}]{lafuente2022ab}%
  \BibitemOpen
  \bibfield  {author} {\bibinfo {author} {\bibfnamefont {J.}~\bibnamefont
  {Lafuente-Bartolome}}, \bibinfo {author} {\bibfnamefont {C.}~\bibnamefont
  {Lian}}, \bibinfo {author} {\bibfnamefont {W.~H.}\ \bibnamefont {Sio}},
  \bibinfo {author} {\bibfnamefont {I.~G.}\ \bibnamefont {Gurtubay}}, \bibinfo
  {author} {\bibfnamefont {A.}~\bibnamefont {Eiguren}},\ and\ \bibinfo {author}
  {\bibfnamefont {F.}~\bibnamefont {Giustino}},\ }\bibfield  {title} {\bibinfo
  {title} {Ab initio self-consistent many-body theory of polarons at all
  couplings},\ }\href@noop {} {\bibfield  {journal} {\bibinfo  {journal} {Phys.
  Rev. B}\ }\textbf {\bibinfo {volume} {106}},\ \bibinfo {pages} {075119}
  (\bibinfo {year} {2022}{\natexlab{b}})}\BibitemShut {NoStop}%
\bibitem [{\citenamefont {Ponc{\'e}}\ \emph {et~al.}(2016)\citenamefont
  {Ponc{\'e}}, \citenamefont {Margine}, \citenamefont {Verdi},\ and\
  \citenamefont {Giustino}}]{ponce2016epw}%
  \BibitemOpen
  \bibfield  {author} {\bibinfo {author} {\bibfnamefont {S.}~\bibnamefont
  {Ponc{\'e}}}, \bibinfo {author} {\bibfnamefont {E.~R.}\ \bibnamefont
  {Margine}}, \bibinfo {author} {\bibfnamefont {C.}~\bibnamefont {Verdi}},\
  and\ \bibinfo {author} {\bibfnamefont {F.}~\bibnamefont {Giustino}},\
  }\bibfield  {title} {\bibinfo {title} {{\tt EPW}: Electron-phonon coupling,
  transport and superconducting properties using maximally localized {Wannier}
  functions},\ }\href@noop {} {\bibfield  {journal} {\bibinfo  {journal}
  {Comput. Phys. Commun.}\ }\textbf {\bibinfo {volume} {209}},\ \bibinfo
  {pages} {116} (\bibinfo {year} {2016})}\BibitemShut {NoStop}%
\bibitem [{\citenamefont {Caruso}\ \emph {et~al.}(2017)\citenamefont {Caruso},
  \citenamefont {Hoesch}, \citenamefont {Achatz}, \citenamefont {Serrano},
  \citenamefont {Krisch}, \citenamefont {Bustarret},\ and\ \citenamefont
  {Giustino}}]{caruso_2017}%
  \BibitemOpen
  \bibfield  {author} {\bibinfo {author} {\bibfnamefont {F.}~\bibnamefont
  {Caruso}}, \bibinfo {author} {\bibfnamefont {M.}~\bibnamefont {Hoesch}},
  \bibinfo {author} {\bibfnamefont {P.}~\bibnamefont {Achatz}}, \bibinfo
  {author} {\bibfnamefont {J.}~\bibnamefont {Serrano}}, \bibinfo {author}
  {\bibfnamefont {M.}~\bibnamefont {Krisch}}, \bibinfo {author} {\bibfnamefont
  {E.}~\bibnamefont {Bustarret}},\ and\ \bibinfo {author} {\bibfnamefont
  {F.}~\bibnamefont {Giustino}},\ }\bibfield  {title} {\bibinfo {title}
  {Nonadiabatic {Kohn} anomaly in heavily boron-doped diamond},\ }\href
  {https://doi.org/10.1103/PhysRevLett.119.017001} {\bibfield  {journal}
  {\bibinfo  {journal} {Phys. Rev. Lett.}\ }\textbf {\bibinfo {volume} {119}},\
  \bibinfo {pages} {017001} (\bibinfo {year} {2017})}\BibitemShut {NoStop}%
\bibitem [{\citenamefont {Novko}\ \emph {et~al.}(2020)\citenamefont {Novko},
  \citenamefont {Caruso}, \citenamefont {Draxl},\ and\ \citenamefont
  {Cappelluti}}]{novko_2020}%
  \BibitemOpen
  \bibfield  {author} {\bibinfo {author} {\bibfnamefont {D.}~\bibnamefont
  {Novko}}, \bibinfo {author} {\bibfnamefont {F.}~\bibnamefont {Caruso}},
  \bibinfo {author} {\bibfnamefont {C.}~\bibnamefont {Draxl}},\ and\ \bibinfo
  {author} {\bibfnamefont {E.}~\bibnamefont {Cappelluti}},\ }\bibfield  {title}
  {\bibinfo {title} {Ultrafast hot phonon dynamics in $\mathrm{MgB_2}$ driven
  by anisotropic electron-phonon coupling},\ }\href
  {https://doi.org/10.1103/PhysRevLett.124.077001} {\bibfield  {journal}
  {\bibinfo  {journal} {Phys. Rev. Lett.}\ }\textbf {\bibinfo {volume} {124}},\
  \bibinfo {pages} {077001} (\bibinfo {year} {2020})}\BibitemShut {NoStop}%
\bibitem [{\citenamefont {Fausti}\ \emph {et~al.}(2009)\citenamefont {Fausti},
  \citenamefont {Misochko},\ and\ \citenamefont {van
  Loosdrecht}}]{fausti2009ultrafast}%
  \BibitemOpen
  \bibfield  {author} {\bibinfo {author} {\bibfnamefont {D.}~\bibnamefont
  {Fausti}}, \bibinfo {author} {\bibfnamefont {O.~V.}\ \bibnamefont
  {Misochko}},\ and\ \bibinfo {author} {\bibfnamefont {P.~H.}\ \bibnamefont
  {van Loosdrecht}},\ }\bibfield  {title} {\bibinfo {title} {Ultrafast
  photoinduced structure phase transition in antimony single crystals},\
  }\href@noop {} {\bibfield  {journal} {\bibinfo  {journal} {Phys. Rev. B}\
  }\textbf {\bibinfo {volume} {80}},\ \bibinfo {pages} {161207} (\bibinfo
  {year} {2009})}\BibitemShut {NoStop}%
\bibitem [{\citenamefont {Sakano}\ \emph {et~al.}(2020)\citenamefont {Sakano},
  \citenamefont {Hirayama}, \citenamefont {Takahashi}, \citenamefont {Akebi},
  \citenamefont {Nakayama}, \citenamefont {Kuroda}, \citenamefont {Taguchi},
  \citenamefont {Yoshikawa}, \citenamefont {Miyamoto}, \citenamefont {Okuda},
  \citenamefont {Ono}, \citenamefont {Kumigashira}, \citenamefont {Ideue},
  \citenamefont {Iwasa}, \citenamefont {Mitsuishi}, \citenamefont {Ishizaka},
  \citenamefont {Shin}, \citenamefont {Miyake}, \citenamefont {Murakami},
  \citenamefont {Sasagawa},\ and\ \citenamefont {Kondo}}]{sakano_radial_2020}%
  \BibitemOpen
  \bibfield  {author} {\bibinfo {author} {\bibfnamefont {M.}~\bibnamefont
  {Sakano}}, \bibinfo {author} {\bibfnamefont {M.}~\bibnamefont {Hirayama}},
  \bibinfo {author} {\bibfnamefont {T.}~\bibnamefont {Takahashi}}, \bibinfo
  {author} {\bibfnamefont {S.}~\bibnamefont {Akebi}}, \bibinfo {author}
  {\bibfnamefont {M.}~\bibnamefont {Nakayama}}, \bibinfo {author}
  {\bibfnamefont {K.}~\bibnamefont {Kuroda}}, \bibinfo {author} {\bibfnamefont
  {K.}~\bibnamefont {Taguchi}}, \bibinfo {author} {\bibfnamefont
  {T.}~\bibnamefont {Yoshikawa}}, \bibinfo {author} {\bibfnamefont
  {K.}~\bibnamefont {Miyamoto}}, \bibinfo {author} {\bibfnamefont
  {T.}~\bibnamefont {Okuda}}, \bibinfo {author} {\bibfnamefont
  {K.}~\bibnamefont {Ono}}, \bibinfo {author} {\bibfnamefont {H.}~\bibnamefont
  {Kumigashira}}, \bibinfo {author} {\bibfnamefont {T.}~\bibnamefont {Ideue}},
  \bibinfo {author} {\bibfnamefont {Y.}~\bibnamefont {Iwasa}}, \bibinfo
  {author} {\bibfnamefont {N.}~\bibnamefont {Mitsuishi}}, \bibinfo {author}
  {\bibfnamefont {K.}~\bibnamefont {Ishizaka}}, \bibinfo {author}
  {\bibfnamefont {S.}~\bibnamefont {Shin}}, \bibinfo {author} {\bibfnamefont
  {T.}~\bibnamefont {Miyake}}, \bibinfo {author} {\bibfnamefont
  {S.}~\bibnamefont {Murakami}}, \bibinfo {author} {\bibfnamefont
  {T.}~\bibnamefont {Sasagawa}},\ and\ \bibinfo {author} {\bibfnamefont
  {T.}~\bibnamefont {Kondo}},\ }\bibfield  {title} {\bibinfo {title} {Radial
  spin texture in elemental tellurium with chiral crystal structure},\ }\href
  {https://doi.org/10.1103/PhysRevLett.124.136404} {\bibfield  {journal}
  {\bibinfo  {journal} {Phys. Rev. Lett.}\ }\textbf {\bibinfo {volume} {124}},\
  \bibinfo {pages} {136404} (\bibinfo {year} {2020})}\BibitemShut {NoStop}%
\bibitem [{\citenamefont {Gatti}\ \emph {et~al.}(2020)\citenamefont {Gatti},
  \citenamefont {Gosálbez-Martínez}, \citenamefont {Tsirkin}, \citenamefont
  {Fanciulli}, \citenamefont {Puppin}, \citenamefont {Polishchuk},
  \citenamefont {Moser}, \citenamefont {Testa}, \citenamefont {Martino},
  \citenamefont {Roth}, \citenamefont {Bugnon}, \citenamefont {Moreschini},
  \citenamefont {Bostwick}, \citenamefont {Jozwiak}, \citenamefont {Rotenberg},
  \citenamefont {Di~Santo}, \citenamefont {Petaccia}, \citenamefont {Vobornik},
  \citenamefont {Fujii}, \citenamefont {Wong}, \citenamefont {Jariwala},
  \citenamefont {Atwater}, \citenamefont {Rønnow}, \citenamefont {Chergui},
  \citenamefont {Yazyev}, \citenamefont {Grioni},\ and\ \citenamefont
  {Crepaldi}}]{gatti_radial_2020}%
  \BibitemOpen
  \bibfield  {author} {\bibinfo {author} {\bibfnamefont {G.}~\bibnamefont
  {Gatti}}, \bibinfo {author} {\bibfnamefont {D.}~\bibnamefont
  {Gosálbez-Martínez}}, \bibinfo {author} {\bibfnamefont {S.}~\bibnamefont
  {Tsirkin}}, \bibinfo {author} {\bibfnamefont {M.}~\bibnamefont {Fanciulli}},
  \bibinfo {author} {\bibfnamefont {M.}~\bibnamefont {Puppin}}, \bibinfo
  {author} {\bibfnamefont {S.}~\bibnamefont {Polishchuk}}, \bibinfo {author}
  {\bibfnamefont {S.}~\bibnamefont {Moser}}, \bibinfo {author} {\bibfnamefont
  {L.}~\bibnamefont {Testa}}, \bibinfo {author} {\bibfnamefont
  {E.}~\bibnamefont {Martino}}, \bibinfo {author} {\bibfnamefont
  {S.}~\bibnamefont {Roth}}, \bibinfo {author} {\bibfnamefont {P.}~\bibnamefont
  {Bugnon}}, \bibinfo {author} {\bibfnamefont {L.}~\bibnamefont {Moreschini}},
  \bibinfo {author} {\bibfnamefont {A.}~\bibnamefont {Bostwick}}, \bibinfo
  {author} {\bibfnamefont {C.}~\bibnamefont {Jozwiak}}, \bibinfo {author}
  {\bibfnamefont {E.}~\bibnamefont {Rotenberg}}, \bibinfo {author}
  {\bibfnamefont {G.}~\bibnamefont {Di~Santo}}, \bibinfo {author}
  {\bibfnamefont {L.}~\bibnamefont {Petaccia}}, \bibinfo {author}
  {\bibfnamefont {I.}~\bibnamefont {Vobornik}}, \bibinfo {author}
  {\bibfnamefont {J.}~\bibnamefont {Fujii}}, \bibinfo {author} {\bibfnamefont
  {J.}~\bibnamefont {Wong}}, \bibinfo {author} {\bibfnamefont {D.}~\bibnamefont
  {Jariwala}}, \bibinfo {author} {\bibfnamefont {H.}~\bibnamefont {Atwater}},
  \bibinfo {author} {\bibfnamefont {H.}~\bibnamefont {Rønnow}}, \bibinfo
  {author} {\bibfnamefont {M.}~\bibnamefont {Chergui}}, \bibinfo {author}
  {\bibfnamefont {O.}~\bibnamefont {Yazyev}}, \bibinfo {author} {\bibfnamefont
  {M.}~\bibnamefont {Grioni}},\ and\ \bibinfo {author} {\bibfnamefont
  {A.}~\bibnamefont {Crepaldi}},\ }\bibfield  {title} {\bibinfo {title} {Radial
  spin texture of the {Weyl} fermions in chiral tellurium},\ }\href
  {https://doi.org/10.1103/PhysRevLett.125.216402} {\bibfield  {journal}
  {\bibinfo  {journal} {Phys. Rev. Lett.}\ }\textbf {\bibinfo {volume} {125}},\
  \bibinfo {pages} {216402} (\bibinfo {year} {2020})}\BibitemShut {NoStop}%
\bibitem [{\citenamefont {Yan}\ and\ \citenamefont
  {Felser}(2017)}]{yan_topological_2017}%
  \BibitemOpen
  \bibfield  {author} {\bibinfo {author} {\bibfnamefont {B.}~\bibnamefont
  {Yan}}\ and\ \bibinfo {author} {\bibfnamefont {C.}~\bibnamefont {Felser}},\
  }\bibfield  {title} {\bibinfo {title} {Topological materials: {Weyl}
  semimetals},\ }\href
  {https://doi.org/10.1146/annurev-conmatphys-031016-025458} {\bibfield
  {journal} {\bibinfo  {journal} {Annu. Rev. Condens. Matter Phys.}\ }\textbf
  {\bibinfo {volume} {8}},\ \bibinfo {pages} {337} (\bibinfo {year}
  {2017})}\BibitemShut {NoStop}%
\bibitem [{\citenamefont {Ning}\ \emph {et~al.}(2022)\citenamefont {Ning},
  \citenamefont {Mehio}, \citenamefont {Lian}, \citenamefont {Li},
  \citenamefont {Zoghlin}, \citenamefont {Zhou}, \citenamefont {Cheng},
  \citenamefont {Wilson}, \citenamefont {Wong},\ and\ \citenamefont
  {Hsieh}}]{Ning_2022_light}%
  \BibitemOpen
  \bibfield  {author} {\bibinfo {author} {\bibfnamefont {H.}~\bibnamefont
  {Ning}}, \bibinfo {author} {\bibfnamefont {O.}~\bibnamefont {Mehio}},
  \bibinfo {author} {\bibfnamefont {C.}~\bibnamefont {Lian}}, \bibinfo {author}
  {\bibfnamefont {X.}~\bibnamefont {Li}}, \bibinfo {author} {\bibfnamefont
  {E.}~\bibnamefont {Zoghlin}}, \bibinfo {author} {\bibfnamefont
  {P.}~\bibnamefont {Zhou}}, \bibinfo {author} {\bibfnamefont {B.}~\bibnamefont
  {Cheng}}, \bibinfo {author} {\bibfnamefont {S.~D.}\ \bibnamefont {Wilson}},
  \bibinfo {author} {\bibfnamefont {B.~M.}\ \bibnamefont {Wong}},\ and\
  \bibinfo {author} {\bibfnamefont {D.}~\bibnamefont {Hsieh}},\ }\bibfield
  {title} {\bibinfo {title} {Light-induced {Weyl} semiconductor-to-metal
  transition mediated by {Peierls} instability},\ }\href
  {https://doi.org/10.1103/PhysRevB.106.205118} {\bibfield  {journal} {\bibinfo
   {journal} {Phys. Rev. B}\ }\textbf {\bibinfo {volume} {106}},\ \bibinfo
  {pages} {205118} (\bibinfo {year} {2022})}\BibitemShut {NoStop}%
\bibitem [{\citenamefont {Guan}\ \emph {et~al.}(2022)\citenamefont {Guan},
  \citenamefont {Liu}, \citenamefont {Chen}, \citenamefont {Li}, \citenamefont
  {Qi}, \citenamefont {Yang}, \citenamefont {You},\ and\ \citenamefont
  {Meng}}]{guan2022optical}%
  \BibitemOpen
  \bibfield  {author} {\bibinfo {author} {\bibfnamefont {M.-X.}\ \bibnamefont
  {Guan}}, \bibinfo {author} {\bibfnamefont {X.-B.}\ \bibnamefont {Liu}},
  \bibinfo {author} {\bibfnamefont {D.-Q.}\ \bibnamefont {Chen}}, \bibinfo
  {author} {\bibfnamefont {X.-Y.}\ \bibnamefont {Li}}, \bibinfo {author}
  {\bibfnamefont {Y.-P.}\ \bibnamefont {Qi}}, \bibinfo {author} {\bibfnamefont
  {Q.}~\bibnamefont {Yang}}, \bibinfo {author} {\bibfnamefont {P.-W.}\
  \bibnamefont {You}},\ and\ \bibinfo {author} {\bibfnamefont {S.}~\bibnamefont
  {Meng}},\ }\bibfield  {title} {\bibinfo {title} {Optical control of
  multistage phase transition via phonon coupling in $\mathrm{MoTe_2}$},\
  }\href@noop {} {\bibfield  {journal} {\bibinfo  {journal} {Phys. Rev. Lett.}\
  }\textbf {\bibinfo {volume} {128}},\ \bibinfo {pages} {015702} (\bibinfo
  {year} {2022})}\BibitemShut {NoStop}%
\bibitem [{\citenamefont {Gauthier}\ \emph {et~al.}(2020)\citenamefont
  {Gauthier}, \citenamefont {Sobota}, \citenamefont {Gauthier}, \citenamefont
  {Xu}, \citenamefont {Pfau}, \citenamefont {Rotundu}, \citenamefont {Shen},\
  and\ \citenamefont {Kirchmann}}]{gauthier2020tuning}%
  \BibitemOpen
  \bibfield  {author} {\bibinfo {author} {\bibfnamefont {A.}~\bibnamefont
  {Gauthier}}, \bibinfo {author} {\bibfnamefont {J.~A.}\ \bibnamefont
  {Sobota}}, \bibinfo {author} {\bibfnamefont {N.}~\bibnamefont {Gauthier}},
  \bibinfo {author} {\bibfnamefont {K.-J.}\ \bibnamefont {Xu}}, \bibinfo
  {author} {\bibfnamefont {H.}~\bibnamefont {Pfau}}, \bibinfo {author}
  {\bibfnamefont {C.~R.}\ \bibnamefont {Rotundu}}, \bibinfo {author}
  {\bibfnamefont {Z.-X.}\ \bibnamefont {Shen}},\ and\ \bibinfo {author}
  {\bibfnamefont {P.~S.}\ \bibnamefont {Kirchmann}},\ }\bibfield  {title}
  {\bibinfo {title} {Tuning time and energy resolution in time-resolved
  photoemission spectroscopy with nonlinear crystals},\ }\href@noop {}
  {\bibfield  {journal} {\bibinfo  {journal} {J. Appl. Phys.}\ }\textbf
  {\bibinfo {volume} {128}} (\bibinfo {year} {2020})}\BibitemShut {NoStop}%
\bibitem [{\citenamefont {Dal~Corso}(2014)}]{dal2014pseudopotentials}%
  \BibitemOpen
  \bibfield  {author} {\bibinfo {author} {\bibfnamefont {A.}~\bibnamefont
  {Dal~Corso}},\ }\bibfield  {title} {\bibinfo {title} {Pseudopotentials
  periodic table: From $\mathrm{H}$ to $\mathrm{Pu}$},\ }\href@noop {}
  {\bibfield  {journal} {\bibinfo  {journal} {Comput. Mater. Sci.}\ }\textbf
  {\bibinfo {volume} {95}},\ \bibinfo {pages} {337} (\bibinfo {year}
  {2014})}\BibitemShut {NoStop}%
\bibitem [{\citenamefont {Perdew}\ \emph {et~al.}(1996)\citenamefont {Perdew},
  \citenamefont {Burke},\ and\ \citenamefont
  {Ernzerhof}}]{perdew1996generalized}%
  \BibitemOpen
  \bibfield  {author} {\bibinfo {author} {\bibfnamefont {J.~P.}\ \bibnamefont
  {Perdew}}, \bibinfo {author} {\bibfnamefont {K.}~\bibnamefont {Burke}},\ and\
  \bibinfo {author} {\bibfnamefont {M.}~\bibnamefont {Ernzerhof}},\ }\bibfield
  {title} {\bibinfo {title} {Generalized gradient approximation made simple},\
  }\href@noop {} {\bibfield  {journal} {\bibinfo  {journal} {Phys. Rev. Lett.}\
  }\textbf {\bibinfo {volume} {77}},\ \bibinfo {pages} {3865} (\bibinfo {year}
  {1996})}\BibitemShut {NoStop}%
\bibitem [{\citenamefont {Sangalli}\ \emph {et~al.}(2019)\citenamefont
  {Sangalli}, \citenamefont {Ferretti}, \citenamefont {Miranda}, \citenamefont
  {Attaccalite}, \citenamefont {Marri}, \citenamefont {Cannuccia},
  \citenamefont {Melo}, \citenamefont {Marsili}, \citenamefont {Paleari},
  \citenamefont {Marrazzo}, \citenamefont {Prandini}, \citenamefont {Bonfà},
  \citenamefont {Atambo}, \citenamefont {Affinito}, \citenamefont {Palummo},
  \citenamefont {Molina-Sánchez}, \citenamefont {Hogan}, \citenamefont
  {Grüning}, \citenamefont {Varsano},\ and\ \citenamefont
  {Marini}}]{sangalli2019many}%
  \BibitemOpen
  \bibfield  {author} {\bibinfo {author} {\bibfnamefont {D.}~\bibnamefont
  {Sangalli}}, \bibinfo {author} {\bibfnamefont {A.}~\bibnamefont {Ferretti}},
  \bibinfo {author} {\bibfnamefont {H.}~\bibnamefont {Miranda}}, \bibinfo
  {author} {\bibfnamefont {C.}~\bibnamefont {Attaccalite}}, \bibinfo {author}
  {\bibfnamefont {I.}~\bibnamefont {Marri}}, \bibinfo {author} {\bibfnamefont
  {E.}~\bibnamefont {Cannuccia}}, \bibinfo {author} {\bibfnamefont
  {P.}~\bibnamefont {Melo}}, \bibinfo {author} {\bibfnamefont {M.}~\bibnamefont
  {Marsili}}, \bibinfo {author} {\bibfnamefont {F.}~\bibnamefont {Paleari}},
  \bibinfo {author} {\bibfnamefont {A.}~\bibnamefont {Marrazzo}}, \bibinfo
  {author} {\bibfnamefont {G.}~\bibnamefont {Prandini}}, \bibinfo {author}
  {\bibfnamefont {P.}~\bibnamefont {Bonfà}}, \bibinfo {author} {\bibfnamefont
  {M.~O.}\ \bibnamefont {Atambo}}, \bibinfo {author} {\bibfnamefont
  {F.}~\bibnamefont {Affinito}}, \bibinfo {author} {\bibfnamefont
  {M.}~\bibnamefont {Palummo}}, \bibinfo {author} {\bibfnamefont
  {A.}~\bibnamefont {Molina-Sánchez}}, \bibinfo {author} {\bibfnamefont
  {C.}~\bibnamefont {Hogan}}, \bibinfo {author} {\bibfnamefont
  {M.}~\bibnamefont {Grüning}}, \bibinfo {author} {\bibfnamefont
  {D.}~\bibnamefont {Varsano}},\ and\ \bibinfo {author} {\bibfnamefont
  {A.}~\bibnamefont {Marini}},\ }\bibfield  {title} {\bibinfo {title}
  {Many-body perturbation theory calculations using the {\tt yambo} code},\
  }\href@noop {} {\bibfield  {journal} {\bibinfo  {journal} {J. Condens. Matter
  Phys.}\ }\textbf {\bibinfo {volume} {31}},\ \bibinfo {pages} {325902}
  (\bibinfo {year} {2019})}\BibitemShut {NoStop}%
\bibitem [{\citenamefont {Li}\ \emph {et~al.}(2014)\citenamefont {Li},
  \citenamefont {Carrete}, \citenamefont {Katcho},\ and\ \citenamefont
  {Mingo}}]{ShengBTE_2014}%
  \BibitemOpen
  \bibfield  {author} {\bibinfo {author} {\bibfnamefont {W.}~\bibnamefont
  {Li}}, \bibinfo {author} {\bibfnamefont {J.}~\bibnamefont {Carrete}},
  \bibinfo {author} {\bibfnamefont {N.~A.}\ \bibnamefont {Katcho}},\ and\
  \bibinfo {author} {\bibfnamefont {N.}~\bibnamefont {Mingo}},\ }\bibfield
  {title} {\bibinfo {title} {{\tt ShengBTE:} a solver of the {B}oltzmann
  transport equation for phonons},\ }\href
  {https://doi.org/10.1016/j.cpc.2014.02.015} {\bibfield  {journal} {\bibinfo
  {journal} {Comp. Phys. Commun.}\ }\textbf {\bibinfo {volume} {185}},\
  \bibinfo {pages} {1747} (\bibinfo {year} {2014})}\BibitemShut {NoStop}%
\bibitem [{\citenamefont {Moser}(2017)}]{moser2017experimentalist}%
  \BibitemOpen
  \bibfield  {author} {\bibinfo {author} {\bibfnamefont {S.}~\bibnamefont
  {Moser}},\ }\bibfield  {title} {\bibinfo {title} {An experimentalist's guide
  to the matrix element in angle-resolved photoemission},\ }\href@noop {}
  {\bibfield  {journal} {\bibinfo  {journal} {J. Electron Spectros. Relat.
  Phenomena}\ }\textbf {\bibinfo {volume} {214}},\ \bibinfo {pages} {29}
  (\bibinfo {year} {2017})}\BibitemShut {NoStop}%
\bibitem [{\citenamefont {Perfetto}\ and\ \citenamefont
  {Stefanucci}(2023)}]{PerfettoStefanucci2023}%
  \BibitemOpen
  \bibfield  {author} {\bibinfo {author} {\bibfnamefont {E.}~\bibnamefont
  {Perfetto}}\ and\ \bibinfo {author} {\bibfnamefont {G.}~\bibnamefont
  {Stefanucci}},\ }\bibfield  {title} {\bibinfo {title} {Real-time
  {GW}-{Ehrenfest}-{Fan}-{Migdal} method for nonequilibrium {2D} materials},\
  }\href {https://doi.org/10.1021/acs.nanolett.3c01772} {\bibfield  {journal}
  {\bibinfo  {journal} {Nano Lett.}\ }\textbf {\bibinfo {volume} {23}},\
  \bibinfo {pages} {7029} (\bibinfo {year} {2023})}\BibitemShut {NoStop}%
\bibitem [{\citenamefont {Karlsson}\ \emph {et~al.}(2021)\citenamefont
  {Karlsson}, \citenamefont {van Leeuwen}, \citenamefont {Pavlyukh},
  \citenamefont {Perfetto},\ and\ \citenamefont {Stefanucci}}]{Karlsson2021}%
  \BibitemOpen
  \bibfield  {author} {\bibinfo {author} {\bibfnamefont {D.}~\bibnamefont
  {Karlsson}}, \bibinfo {author} {\bibfnamefont {R.}~\bibnamefont {van
  Leeuwen}}, \bibinfo {author} {\bibfnamefont {Y.}~\bibnamefont {Pavlyukh}},
  \bibinfo {author} {\bibfnamefont {E.}~\bibnamefont {Perfetto}},\ and\
  \bibinfo {author} {\bibfnamefont {G.}~\bibnamefont {Stefanucci}},\ }\bibfield
   {title} {\bibinfo {title} {Fast {Green}'s function method for ultrafast
  electron-{Boson} dynamics},\ }\href
  {https://doi.org/10.1103/PhysRevLett.127.036402} {\bibfield  {journal}
  {\bibinfo  {journal} {Phys. Rev. Lett.}\ }\textbf {\bibinfo {volume} {127}},\
  \bibinfo {pages} {036402} (\bibinfo {year} {2021})}\BibitemShut {NoStop}%
\bibitem [{\citenamefont {Molina-Sánchez}\ \emph {et~al.}(2017)\citenamefont
  {Molina-Sánchez}, \citenamefont {Sangalli}, \citenamefont {Wirtz},\ and\
  \citenamefont {Marini}}]{Sanchez2017}%
  \BibitemOpen
  \bibfield  {author} {\bibinfo {author} {\bibfnamefont {A.}~\bibnamefont
  {Molina-Sánchez}}, \bibinfo {author} {\bibfnamefont {D.}~\bibnamefont
  {Sangalli}}, \bibinfo {author} {\bibfnamefont {L.}~\bibnamefont {Wirtz}},\
  and\ \bibinfo {author} {\bibfnamefont {A.}~\bibnamefont {Marini}},\
  }\bibfield  {title} {\bibinfo {title} {Ab initio calculations of ultrashort
  carrier dynamics in two-dimensional materials: Valley depolarization in
  single-layer $\mathrm{WSe_2}$},\ }\href
  {https://doi.org/10.1021/acs.nanolett.7b00175} {\bibfield  {journal}
  {\bibinfo  {journal} {Nano Lett.}\ }\textbf {\bibinfo {volume} {17}},\
  \bibinfo {pages} {4549} (\bibinfo {year} {2017})}\BibitemShut {NoStop}%
\bibitem [{\citenamefont {Campi}\ \emph {et~al.}(2012)\citenamefont {Campi},
  \citenamefont {Bernasconi},\ and\ \citenamefont
  {Benedek}}]{campi2012phonons}%
  \BibitemOpen
  \bibfield  {author} {\bibinfo {author} {\bibfnamefont {D.}~\bibnamefont
  {Campi}}, \bibinfo {author} {\bibfnamefont {M.}~\bibnamefont {Bernasconi}},\
  and\ \bibinfo {author} {\bibfnamefont {G.}~\bibnamefont {Benedek}},\
  }\bibfield  {title} {\bibinfo {title} {Phonons and electron-phonon
  interaction at the {Sb} (111) surface},\ }\href@noop {} {\bibfield  {journal}
  {\bibinfo  {journal} {Phys. Rev. B}\ }\textbf {\bibinfo {volume} {86}},\
  \bibinfo {pages} {075446} (\bibinfo {year} {2012})}\BibitemShut {NoStop}%
\bibitem [{\citenamefont {Setyawan}\ and\ \citenamefont
  {Curtarolo}(2010)}]{setyawan2010high}%
  \BibitemOpen
  \bibfield  {author} {\bibinfo {author} {\bibfnamefont {W.}~\bibnamefont
  {Setyawan}}\ and\ \bibinfo {author} {\bibfnamefont {S.}~\bibnamefont
  {Curtarolo}},\ }\bibfield  {title} {\bibinfo {title} {High-throughput
  electronic band structure calculations: Challenges and tools},\ }\href@noop
  {} {\bibfield  {journal} {\bibinfo  {journal} {Comput. Mater. Sci.}\ }\textbf
  {\bibinfo {volume} {49}},\ \bibinfo {pages} {299} (\bibinfo {year}
  {2010})}\BibitemShut {NoStop}%
\end{thebibliography}

%

\end{document}